\documentclass[lettersize,journal]{IEEEtran}
\usepackage{amsmath,amsfonts}
\usepackage{algorithmic}
\usepackage{array}
\usepackage{textcomp}
\usepackage{stfloats}
\usepackage{url}
\usepackage{verbatim}
\usepackage{graphicx}

\def\BibTeX{{\rm B\kern-.05em{\sc i\kern-.025em b}\kern-.08em
    T\kern-.1667em\lower.7ex\hbox{E}\kern-.125emX}}
\usepackage{balance}

\usepackage{tcolorbox}
\usepackage{mathtools}
\usepackage{subfigure}

\usepackage{mathrsfs}

\newcommand{\revOne}[2]{#2}
\newcommand{\revTwo}[2]{#2}

\newcommand{\commentout}[1]{}
\newcommand{\hardcoded}[2]{{#1}} 

\usepackage{fancyhdr}
\usepackage{extramarks}

\begin{document}

\title{Quantifying the Performance Gap for Simple Versus Optimal Dynamic Server Allocation Policies}


\IEEEaftertitletext{\vspace{-2\baselineskip}}

\author{\IEEEauthorblockN{Niklas Carlsson}
  \IEEEauthorblockA{\textit{Link\"oping University}, Sweden}
  \\
  \IEEEauthorblockN{Derek Eager}
  \IEEEauthorblockA{\textit{University of Saskatchewan}, Canada}
   }



\maketitle

\begin{abstract}
Cloud computing enables the dynamic provisioning of server resources.  To exploit this opportunity, a policy is needed for dynamically allocating (and deallocating) servers in response to the current load conditions.
In this paper we describe several simple policies for dynamic server allocation and develop analytic models for their analysis.  We also design semi-Markov decision models that enable determination of the performance achieved with optimal policies, allowing us to quantify the performance gap between simple, easily implemented policies, and optimal policies.  Finally, we apply our models to study the potential performance benefits of state-dependent routing in multi-site systems when using dynamic server allocation at each site. 
Insights from our results are valuable to service providers wanting to balance cloud service costs and delays. 
\end{abstract}

\fancypagestyle{firststyle}
               {
                 \fancyhf{}
                 \fancyfoot[CF]{\tiny
                   \copyright IEEE (2026). This is the author's version of the work. It is posted here by permission of IEEE for your personal use. Not for redistribution. 
                   \\The definitive version will be published in 
                   {\em  IEEE Transactions on Cloud Computing},
                   accepted Mar. 2026,
                   \url{https://doi.org/10.1109/TCC.2026.3672446}.}
               }
               \thispagestyle{firststyle}

\begin{IEEEkeywords}
Cloud computing, Edge cloud, Dynamic server allocation, Performance gap, Optimal policies
\end{IEEEkeywords}

\section{Introduction}\label{sec:intro}

\IEEEPARstart{W}{ith}
the advent and continued evolution of cloud computing systems, dynamic allocation of servers (e.g., in VMs) has become common, and is becoming increasingly lightweight.\footnote{\revOne{}{
In this paper, “server allocation” and “deallocation” refer to acquiring (activating) and releasing (deactivating) server instances in response to load, as is common in cloud resource-provisioning settings. This contrasts with usage in some scheduling contexts where “server allocation” refers instead to assigning individual jobs to specific servers.}}  
\revOne{Dynamic server allocation is an attractive strategy when cost is incurred by the service provider only when a server is allocation but not when that server is deallocated.}
{Dynamic server allocation is an attractive strategy for efficiently balancing service performance and resource cost when there is significant additional cost incurred by the service provider when a server is allocated (and either actively serving requests or idle) compared to when it is deallocated.}
In such a context, an important problem is that of how to best dynamically allocate (and deallocate) server instances based on the current system load, cloud provider pricing, and the desired balance between service delay and service cost.  

A dynamic server allocation policy must balance several potentially conflicting objectives, including low request response time and low service provider's cost.  For example, allocating more server instances may reduce request response time, but will increase the service provider's cost.  Furthermore, there is some overhead cost associated with launching a new server instance.  Therefore, while quick reactions to changes in request load through allocation/deallocation actions may achieve a better matching of server resources to load, there will also be an increase in the overhead costs. 

In this paper, we derive and quantitatively compare the above-mentioned performance tradeoffs for both simple and optimal dynamic server allocation policies, leading to insights about the performance gap between simple and optimal policies.  To capture the above cloud-relevant tradeoffs, typically ignored in the queuing and performance evaluation literature, all policies are compared under our novel system model that incorporates the key tradeoffs between response times (allocated over all requests) and server allocation costs 
(Sec.~\ref{sec:metrics}).  

For our analysis, we first describe a variety of dynamic server allocation policies and develop corresponding analytic models (Sec.~\ref{sec:policiesandmodels}). Our policy selection includes \revTwo{both
simple natural-to-implement policies that use
a limited number of servers and somewhat less practical policies that do not put a cap on the number of servers.}{simple natural-to-implement policies that use single, dual, or arbitrary numbers of servers.}
We find simple response time and cost expressions for a policy for dynamic allocation/deallocation of a single server, and simple product forms for the state probabilities of two dynamic allocation policies for systems with no cap on the number of allocated server instances.  More complex solutions are obtained for analytic models of dual server policies.

Second, we employ semi-Markov decision models (Sec.~\ref{sec:optimal}) to quantify the performance gaps between the simple dynamic server allocation policies
\revTwo{, the less practical policies without a cap on the number of servers, }{that we consider}
and the optimal allocation policies (Sec.~\ref{sec:performancecopmarisons}).
We find that simple policies can often yield close to optimal performance, even in scenarios with potentially many servers and highly-complex optimal policies.  Performance gaps widen, however, as the cost of server allocation increases.

Finally, we consider systems with multiple geographically distributed server sites and populations of clients, and the problem of routing each client service request to an appropriate site (Sec.~\ref{sec:routing}).  
Request routing policies broadly fall into two classes: state-dependent routing policies, which utilize current system state for decisions, and state-oblivious policies, relying only on average request rates.
The use of dynamic server allocation adds a new important aspect to the performance comparison between these policy categories. 
\revOne{One}{On} 
the one hand, one might anticipate increased benefits for state-dependent routing due to its potential utilization of server allocation and request queues' state at each site. However, on the other hand, dynamic server allocation empowers individual sites to better handle local load fluctuations, potentially reducing the necessity for state-dependent routing.
We take a first look at this issue by comparing the performance of optimal state-oblivious routing of requests to that of optimal state-dependent routing for a scenario with two server sites each using dynamic server allocation, and find only modest performance gaps.

\revTwo{}{These results yield important practical insights for cloud computing system design.  Policies for dynamic server allocation and request routing have major performance impacts in cloud computing systems, and have been a focus of considerable design effort.  A challenge when designing any form of resource management policy is knowing what potential room for improvement there may be, beyond the performance provided by some existing simple heuristic or some simpler policy class, such as state-independent policies in the case of request routing.  Through comparisons between simple and optimal dynamic server allocation policies, and between optimal state-independent and state-dependent routing policies, we address this challenge for these key cloud computing policies.}  

\textit{Outline:}
Related work is described in Sec.~\ref{sec:relatedwork}.
Sec.~\ref{sec:metrics} describes the type of system 
considered
and the metrics and objective function used for policy evaluation.  
Several
simple policies for dynamic server allocation are introduced in Sec.~\ref{sec:policiesandmodels}, together with corresponding analytic models.  Sec.~\ref{sec:optimal} describes semi-Markov decision models that allow determination of optimal policies and their performance.  Performance comparisons among the simple policies of Sec.~\ref{sec:policiesandmodels} and with optimal policy performance are presented in Sec.~\ref{sec:performancecopmarisons}.  Sec.~\ref{sec:routing} considers the potential benefits of state-dependent routing in a system with multiple server sites, each using dynamic server provisioning. 
Finally, Sec.~\ref{sec:conclusions} concludes the paper.

\section{Related Work}\label{sec:relatedwork}

A number of prior works have developed analytic models of systems with server setup delays.  An important early work is that by Welch~\cite{Welc64}, who considered a generalization of an M/G/1 single server system in which a customer arriving when the server is idle has a different service time distribution.  This analysis was applied by Meisner et al., for example, in a queueing model-based analysis of a server energy conservation approach using rapid transitions between active and near-zero-power idle states~\cite{Meis11}.  Later studies analyze multiple server systems with exponentially distributed setup delays~\cite{AELH05, GHBA10, GDHB14, Phun17}, 
\revOne{}{\cite{maccio2018structural}}.  
\revOne{}{Among these later studies, Maccio and Down~\cite{maccio2018structural} show that for a broad class of optimization functions, the optimal policy for such systems is a (in general hysteretic) threshold policy.}
Williams et al. consider the perhaps more realistic case where setup delays are deterministic rather than exponential~\cite{WHBW22}.  
\revOne{Other work considers}{Other work considers single server systems with setup delays but where, when “on”, the server is able to instantaneously switch service rate among some finite set of possible rates at zero cost~\cite{badian2021optimal}, multi-server systems in which some of the servers are “dynamic” servers able to instantaneously switch on or off based on a threshold on the number of queued tasks~\cite{xie2024integrating}, and} 
multi-server systems in which there is a ``delayed-off'' delay before a server is deallocated after becoming idle~\cite{GGHB10, GHBR12, GDHB14}.    

Our work differs in a number of respects from these prior studies.  First, our focus is the performance of simple server allocation policies for single, dual, and unlimited server scenarios, the performance gap with respect to optimal policies, and the impact of use of dynamic server allocation on the performance comparison between state-dependent and state-oblivious routing in a multi-site system.   
Second, our evaluation captures the fundamental tradeoff between maintaining low response times and low service provider costs, as calculated over all requests.
Here, we adopt a performance metric combining response time and server cost that can be naturally extended to systems with multiple server sites.
Third,
with respect to the queuing modeling results,
our work differs with respect to the particular simple allocation policies we consider and the analytic models we derive for them.  Specifically:
(1) we find simple response time and cost expressions for a single server policy that incorporates an Erlang distributed ``delayed-off'' delay or optionally batching of requests before server allocation is initiated, as well for the optimal single server policy given our performance metric;  
\revOne{(2) with respect to dual server policies, we analyze a new model (Figure~\ref{fig:statetransitiondiagramforMM1Scaling}) in which one of the two servers is kept allocated;}{(2) with respect to dual server policies, we analyze a new model (Figure~\ref{fig:statetransitiondiagramforMM1Scaling}) in which one of the two servers is kept allocated, and the other server is allocated/deallocated according to upper and lower thresholds;} 
(3) we provide explicit closed-form expressions for mean response time and cost for the M/M/2 with exponential setup times model in which two server allocations can be in progress at once, in contrast to the alternative solution method outlined in \cite{GDHB14} that requires a system of linear equations to be solved; and
(4)
we find new, simple product form solutions for 
\revTwo{server allocation}{``unlimited server''} 
models (Figures~\ref{fig:StateTransitionDiagramforUnlimitedServer} and~\ref{fig:StateTransitionDiagramforProactiveUnlimitedServer}) in which there is no bound on the number of servers that can be allocated.  
\revTwo{}{Unlimited server models are important as they can yield insight into cloud computing scenarios in which the limit on the number of allocated servers is large enough that it is rarely if ever reached. Our new product form solutions for this important case are particularly noteworthy, since they were unexpected (such solutions are relatively rare), and they yield simple expressions for the performance measures of interest.}

\revOne{}{Other prior analytic modelling work has concerned the combination of load balancing with dynamic server provisioning in a setting where each server has its own queue.  For example, Anselmi~\cite{anselmi2024asynchronous}  develops an “Asynchronous Load Balancing  and Auto-scaling” Markovian framework incorporating “Join-the-Idle-Queue” and “Power-of-d” load balancing, and identifies scaling rules that yield delay and relative energy optimality (in the context of the model).  Mukherjee et al.~\cite{mukherjee2017optimal} propose a load balancing and auto-scaling scheme that is asymptotically optimal (in the context of their model) as the total traffic volume and nominal number of servers grow large in proportion.  Eshwar et al.~\cite{eshwar2024online} formulates the problem of load balancing and auto-scaling as a finite horizon weakly-coupled Markov Decision Process and proposes an online learning algorithm to find the optimal load balancing and auto-scaling policy.  Down~\cite{down2022optimal} describes a queueing model with multiple servers that can be switched off and on (with setup delay), each with its own queue, and the joint problem of determining job routing and when servers should be switched off and on.   In contrast, in our work, we assume that the servers dynamically provisioned at a site share a common queue, so that no server is on but idle when there is waiting work, and load balancing is not an issue.}

There is also considerable prior work on more complex dynamic server provisioning policies, evaluated through simulation and/or prototype implementation (e.g., \cite{GHBR12, USCG08, Qu18, Stra22}).  A variety of contexts have been considered, including physical servers and ``deallocation'' in the sense of moving the server to a low power state, as well as VM scenarios of varying agility. 
While we do not explicitly address practical implementation issues in this work, we expect that the simple policies presented here are relatively easy to implement and that our modelling results concerning these policies and their performance gaps versus optimal policies yield useful insight to their practical value.

Finally, there has been much work on low-overhead sandboxed execution environments, with a common objective being to achieve the security of traditional VMs but at much lower performance cost~\cite{Agac20, gVis18, Manc17, Kata, Broo23}.  Lowering the cost of server allocation can enable more aggressive server allocation/deallocation policies, such as those considered here.        

\section{System Description and Metrics}\label{sec:metrics}

We consider a system, such as in a cloud computing environment, where server instances can be dynamically allocated and deallocated.  Initially, we assume just a single server site.  Requests for service arrive at a rate $\lambda$.  The service provider wishes to achieve a low request response time, but also a low cost, where the cost at each point in time is assumed proportional to the total service rate of the allocated servers.  Here the time that a server is considered to be ``allocated'' includes the setup delay from when a server allocation is initiated until the server is ready to process requests.  We assume homogeneous servers each with service rate $\mu$ (i.e., $1/\mu$ is the average service time of a request), and so the cost at each point in time is proportional to $\mu$ times the number of currently allocated servers.

There is a tradeoff between low response time and low cost, but also with respect to the frequency of server allocations and deallocations.
A factor favoring low frequency is that during the server setup delay cost is being incurred with no direct benefit.  However, cost is also incurred with no direct benefit when server(s) are left allocated when there are no requests for them to process.  Of interest are server allocation/deallocation policies that efficiently mediate these tradeoffs in a manner that yields a desired balance between low response time and low cost.

We adopt the following objective function that we wish to minimize:
\begin{align}\label{eq:objfunction}
\omega (\lambda R) + C,
\end{align}
where $R$ is the average request response time, $C$ is the cost as measured by the average over time of the total service rate of the allocated servers (= $1/\mu$ times the average number of allocated servers), and $\omega$ is a factor chosen according to the desired balance in importance between the two objective function terms.  
\revOne{Note that $\lambda R$ can be thought of as the average rate at which}{From Little’s Law, $\lambda R$ is equal to the average number of requests in the system.  It can also be thought of as the average rate at which}
``request time'' (i.e., request waiting and service time) is being accumulated.  We can also define a notion of ``server time''.  If units are normalized so that the service of a single request uses, on average, 1 unit of server time, then when there is one allocated server, server time is being accumulated at rate $\mu$, and $C$ gives the average rate at which server time is being accumulated.  With these definitions, the objective function can be viewed as a weighted sum of the rates of accumulation of ``request time'' and ``server time''. 

\revOne{}{Note that the two objective function terms will tend to conflict.   For example, for fixed $\lambda$, increasing the service rate $\mu$ will decrease $R$ but typically increase $C$.  The factor $\omega$ reflects the desired balance between the two objective-function terms,
acting as a weighting factor that quantifies the relative emphasis placed on cost versus delay, with larger values corresponding to greater priority on minimizing response time.
Server allocation/deallocation policies differ not only in the importance they attach to the two terms, but also in how effectively the tradeoff is mediated; for example, an efficient policy may be able to achieve a decrease in $R$ with a lower increase in $C$ than can be achieved with other policies.  We seek efficient policies that yield a minimal or close to minimal objective function value compared to that of other candidate policies across a wide range of $\omega$ values.}

The objective function~(\ref{eq:objfunction}) 
can also be used at each site in a service provider system with multiple sites.  Using the subscript $i$ to index the metrics of site $i$, note that the overall average response time would be given by $(\sum_i \lambda_i R_i)/(\sum_j \lambda_j)$, and so it would be appropriate to weight each $R_i$ by $\lambda_i$ in an objective function for each site, while using a single system-wide value of $\omega$. 

\section{Policies and Models}\label{sec:policiesandmodels}

We begin in Sec.~\ref{sec:singleserver} with the simplest case of dynamic allocation/deallocation of a single server.  In Sec.~\ref{sec:dualserver}, we consider a more flexible dual-server scenario.  Sec.~\ref{sec:unlimmitedserver} presents and analyses policies for systems with no cap on the number of allocated servers.  

In all cases, the models that we develop assume a Poisson request arrival process at rate $\lambda$, exponentially distributed request service times with mean $1/\mu$, and exponentially distributed server setup delays with mean $\Delta$. 

\revOne{}{While our model assumes a centralized queue and Markovian dynamics for analytic tractability, these assumptions, common in foundational queueing analyses, are made for analytic tractability. Our objective is not to reproduce any specific cloud deployment in full detail, but rather to gain insight into the performance gap between simple and optimal dynamic allocation policies. Although real systems may exhibit non-stationary demand and parameter uncertainty, fixed-rate models provide a useful abstraction over short timescales, and our numerical results (Sec.~VI) span wide parameter ranges to assess robustness across operating regimes.}

\subsection{Single Server}\label{sec:singleserver}

\revTwo{With just a single server, two}{Two} 
basic policy options are to always leave the server allocated, or to deallocate the server when the system empties of requests, possible after some ``holding-on'' / ``delayed-off'' delay \revTwo{that is motivated}{motivated}
by the 
\revTwo{possibility}{chance} 
that a new request 
\revTwo{might arrive shortly after the server has become idle.  If the server is always left allocated, the resulting model is the standard M/M/1 model.}{arrives soon.}

\begin{figure}[t]
\centering
\includegraphics[width=0.46\textwidth]{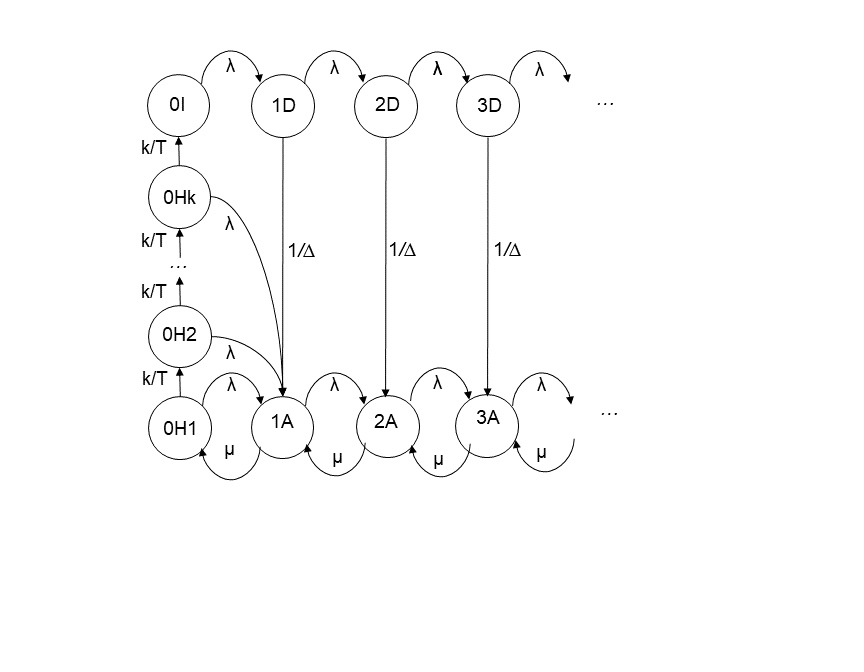}
\vspace{-50pt}
\caption{State-Transition Diagram for Single Server \revTwo{Allocation/Deallocation Policy}{System}.}
\label{fig:mm1statetransitiondiagram}
\vspace{-2pt}
\end{figure}

A state-transition diagram for the second policy option, assuming an Erlang-distributed holding-on delay with shape parameter $k$ and rate parameter $k/T$ (yielding a mean delay of $T$), and assuming that server allocation is triggered whenever a request arrives to an empty system with deallocated server, is shown in Figure~\ref{fig:mm1statetransitiondiagram}.  (We subsequently consider a variant of this 
\revTwo{policy option with request batching prior to the triggering of server allocation.}{policy with request batching.})  
\revTwo{Here each}{Each} 
state is labeled by the number of 
\revTwo{client requests}{requests} 
present at the server, followed by "A" (server is active processing requests), "D" 
(\revTwo{server is in setup}{setup} 
delay), "I" (\revTwo{system is idle and the}{idle,} 
server is deallocated), or "H$j$" 
\revTwo{for integer $j$ between 1 and $k$ (server is}{($1 \leq j \leq k$, server is in}  
the $j$'th stage of the Erlang-distributed holding-on period).\revTwo{  The feasible states are states $i$A ($i \geq 1$), $i$D ($i \geq 1$), 0I, and 0H$j$ ($1 \leq j \leq k$), with state transition rates as shown in the figure.}{\footnote{See \revTwo{the supplemental material}{Appendix~B} for the analysis of an extension of this model incorporating state-dependent service rates $\mu_i$.}}

Denote the steady-state probability of 
\revTwo{the state with label $s$}{state $s$} 
by $p_s$.
Assuming $\mu > \lambda$, 
\revTwo{}{the server utilization is $\lambda/\mu$,} 
and we must have
\begin{align}\label{eq:util}
    \sum_{i=1}^{\infty} p_{i\textrm{A}} = \frac{\lambda}{\mu} \Leftrightarrow \sum_{j=1}^{k} p_{0\textrm{H}j} + p_{0\textrm{I}} + \sum_{i=1}^{\infty} p_{i\textrm{D}} = 1 - \frac{\lambda}{\mu}.
\end{align}
\revOne{}{\revTwo{The leftmost equality in Equation~(\ref{eq:util}) follows from the fact that the sum of the probabilities of the “A” states gives the fraction of time the server is actively processing requests.  This is the server utilization, which is equal to $\lambda/\mu$.}}{}

\revTwo{We can express the $p_{0\textrm{H}j}$ and $p_{i\textrm{D}}$ probabilities in terms of $p_{0\textrm{I}}$, allowing the solution for $p_{0\textrm{I}}$ using Equation~(\ref{eq:util}).  From flow balance, we have
\begin{align}
p_{0\textrm{H}k} \left( \frac{k}{T} \right) = p_{0\textrm{I}} \lambda
\end{align}
and
\begin{align}
p_{0\textrm{H}j} \left( \frac{k}{T} \right) = p_{0\textrm{H}(j+1)} (\lambda + k/T) \;\;\;\;\;\; 1 \leq j < k,
\end{align}
yielding
\begin{align}\label{eq:p0Hj}
p_{0\textrm{H}j} = p_{0\textrm{I}} \left( \frac{\lambda T}{k} \right) \left( \frac{\lambda + k/T}{k/T} \right)^{k-j} \;\;\;\;\;\; 1 \leq j \leq k.
\end{align}
From Equation~(\ref{eq:p0Hj}) we get:
\begin{align}\label{eq:sump0Hj}
\sum_{j=1}^{k} p_{0\textrm{H}j} 
=  p_{0\textrm{I}} \left( (\lambda T / k + 1 )^k - 1 \right).
\end{align}
\revOne{}{Note that adding $p_{0I}$ to both sides of this equation and then dividing both sides by $p_{0I}$ yields the expected result that the ratio of the probability of being in a state with zero requests to the probability of being in state $0I$, is equal to the probability that no arrival occurs during a holding-on period.}
\revOne{We also have}{We also have, from flow balance,}
\begin{align}\label{eq:p1D}
p_{1\textrm{D}} ( \lambda + 1/\Delta ) = p_{0\textrm{I}} \lambda
\end{align}
and
\begin{align}
p_{(i+1)\textrm{D}} ( \lambda + 1/\Delta ) = p_{i\textrm{D}} \lambda  \;\;\;\;\;\; i \geq 1,
\end{align}
yielding
\begin{align}\label{eq:piDoffwhenidle}
p_{i\textrm{D}} = p_{0\textrm{I}} \left( \frac{\lambda}{\lambda + 1/\Delta} \right)^i \;\;\;\;\;\; i \geq 1.
\end{align}
From Equation~(\ref{eq:piDoffwhenidle}) we get:
\begin{align}\label{eq:sumpiD}
\sum_{i=1}^{\infty} p_{i\textrm{D}} =
p_{0\textrm{I}} \frac{\lambda / (\lambda + 1/\Delta)}{1 - \lambda / (\lambda + 1/\Delta)} = p_{0\textrm{I}}\lambda \Delta.
\end{align}
\revOne{}{Note that this equation yields the expected result that the ratio of the probability of being in a state in which the server is in setup delay to the probability of being in state $0I$, is equal to the ratio of the mean setup delay ($\Delta$) to the mean interarrival time ($1/\lambda$).}
Using Equations~(\ref{eq:sump0Hj}) and~(\ref{eq:sumpiD}) to substitute into~(\ref{eq:util}) gives:
\begin{align}\label{eq:p0I}
p_{0\textrm{I}} = \frac{1 - \lambda/\mu}{(\lambda T / k + 1 )^k + \lambda \Delta}. 
\end{align}

Denote by $p_n$ ($n \geq 1$) the steady-state probability of $n$ client requests being present at the server, i.e. the sum of $p_{n\textrm{A}}$ and $p_{n\textrm{D}}$.  From flow balance,
\begin{align}\label{eq:p1fb}
\mu (p_1 - p_{1\textrm{D}}) = \lambda \left( \sum_{j=1}^{k} p_{0\textrm{H}j} + p_{0I} \right).
\end{align}
Substitution from~(\ref{eq:sump0Hj}),~(\ref{eq:piDoffwhenidle}), and~(\ref{eq:p0I}) gives
\begin{align}\label{eq:p1}
p_1 =  \left( \frac{\lambda}{\mu} (\lambda T / k + 1 )^k + \frac{\lambda}{\lambda + 1/\Delta} \right) \left(\frac{1 - \lambda/\mu}{(\lambda T / k + 1 )^k + \lambda \Delta}\right).
\end{align}
Again applying flow balance,
\begin{align}\label{eq:pifb}
\mu (p_i - p_{i\textrm{D}}) = \lambda p_{i-1} \;\;\;\;\;\; i \geq 2,
\end{align}
yielding, for all $i \geq 1$,}
{We can express the $p_{0\textrm{H}j}$ and $p_{i\textrm{D}}$ in terms of $p_{0\textrm{I}}$, allowing the solution for $p_{0\textrm{I}}$ using Equation~(\ref{eq:util}). The model in Figure~\ref{fig:mm1statetransitiondiagram} can then be solved by applying the respective flow balance equations.\footnote{\revTwo{}{These and other derivation details can be found in Appendix~A.}}  Denoting by $p_i$ the probability of $i$ client requests at the server, i.e. the sum of $p_{i\textrm{A}}$ and $p_{i\textrm{D}}$, we obtain, for $i \geq 1$,}

{\footnotesize\begin{align}\label{eq:pi}
p_i =  \left( ( \frac{\lambda}{\mu} )^i (\frac{\lambda T}{k} + 1 )^k + \sum_{l=1}^{i} ( \frac{\lambda}{\lambda + \frac{1}{\Delta}} )^l ( \frac{\lambda}{\mu})^{i-l}  \right) \frac{1 - \lambda/\mu}{(\frac{\lambda T}{k} + 1 )^k + \lambda \Delta}.
\end{align}}

\revTwo{Considering now the mean number of requests in the system $\sum_{i=1}^{\infty} i p_i$, where $p_i$ is given by Equation~(\ref{eq:pi}), note that
\begin{align}\label{eq:sumsoln}
\sum_{i=1}^{\infty} i (\lambda / \mu)^i = \frac{\lambda / \mu}{(1 - \lambda / \mu)^2}
\end{align}
and
{\footnotesize \begin{align}\label{eq:regroupingeq}
\sum_{i=1}^{\infty} i \sum_{l=1}^{i} \left( \frac{\lambda}{\lambda + 1/\Delta} \right)^l \left( \frac{\lambda}{\mu}\right)^{i-l} & = \sum_{i=1}^{\infty}  \left( \frac{\lambda}{\lambda + 1/\Delta} \right)^i \sum_{l=0}^{\infty} (l+i) \left( \frac{\lambda}{\mu} \right)^l \nonumber \\
& = \frac{\lambda / \mu}{(1 - \lambda / \mu)^2} (\lambda \Delta) + \frac{\lambda \Delta (1 + \lambda \Delta)}{1 - \lambda/\mu}.
\end{align}
}
Note that in the first line of (\ref{eq:regroupingeq}), the original double summation is rewritten to group together all of the resulting terms that include the same power of $\lambda / (\lambda + 1/\Delta)$ as one of the factors. Applying~(\ref{eq:sumsoln}) and~(\ref{eq:regroupingeq}) with~(\ref{eq:pi}), the mean number of requests in the system is given by
{\footnotesize\begin{align}
\sum_{i=1}^{\infty} i p_i & = \left( \frac{\lambda / \mu}{(1 - \lambda / \mu)^2} \left( (\frac{\lambda T}{k} + 1 )^k + \lambda \Delta \right) + \frac{\lambda \Delta (1 + \lambda \Delta)}{1 - \lambda/\mu} \right) \times \nonumber \\
& \;\;\;\;\;\; \;\;\;\;\;\; \;\;\;\;\;\; \;\;\;\;\;\; \;\;\;\;\;\; \;\;\;\;\;\; \;\;\;\;\;\; \;\;\;\;\;\; \;\;\;\;
\left(\frac{1 - \lambda/\mu}{(\frac{\lambda T}{k} + 1 )^k + \lambda \Delta}\right) \nonumber \\
& = \frac{\lambda / \mu}{1 -\lambda / \mu} + \frac{\lambda \Delta (1 + \lambda \Delta)} {(\frac{\lambda T}{k} + 1 )^k + \lambda \Delta}.
\end{align}
}
From Little's Law, the mean request response time $R$ is given by
\revOne{}{the mean number of requests in the system divided by $\lambda$:}}{The mean number of requests in the system is given by $\sum_{i=1}^{\infty} i p_i$.  Applying Little's Law and simplifying yields:} 
\begin{align}
R = \frac{1/\mu}{1 -\lambda /\mu} + \frac{\Delta (1 + \lambda \Delta)} {(\frac{\lambda T}{k} + 1 )^k + \lambda \Delta}.
\end{align}
The cost $C$ is given by $\mu$ times the probability that the server is active or in setup or holding-on delay, i.e., by $\mu (1-p_{0I})$\revTwo{, yielding}{:}
\begin{align}
C = \mu  - \frac{\mu -\lambda }{(\lambda T/k + 1)^k + \lambda  \Delta} .
\end{align}
Two special cases of interest are: \\
1) The holding-on delay $T$ is exponentially distributed ($k=1$):
\begin{align}\label{eq:exponential}
R = \frac{1/\mu}{1 - \lambda /\mu} + \frac{\Delta (1 + \lambda \Delta)}{ \lambda T + 1 + \lambda  \Delta},~~~~
C = \mu - \frac{\mu -\lambda }{\lambda T + 1 + \lambda  \Delta}.
\end{align}
2) The holding on delay $T$ is deterministic ($k \rightarrow \infty$):
\begin{align}\label{eq:deterministic}
R = \frac{1/\mu}{1 - \lambda /\mu} + \frac{\Delta (1 + \lambda \Delta)}{ e^{\lambda T} + \lambda  \Delta},~~~~
C = \mu - \frac{\mu-\lambda }{e^{\lambda T} + \lambda  \Delta}.
\end{align}

\revTwo{The above model can be extended to incorporate state-dependent service rates $\mu_i$, potentially providing the ability to more accurately model behavior when multiple requests can be in service simultaneously. 
(See supplemental material.)}{}

\revTwo{Consider now the extremes of $T=0$ and $T\rightarrow\infty$.  For $T\rightarrow\infty$, the server is always left allocated, resulting in the standard M/M/1 model as noted above.  For $T=0$, the server is immediately deallocated when the system empties of requests.
It is easy to see that under fixed rate Poisson (memoryless) arrivals an optimal policy will either always leave the server on, or immediately deallocate.}{It is easy to see that under fixed rate Poisson (memoryless) arrivals an optimal policy will either always leave the server on ($T\rightarrow\infty$), or immediately deallocate ($T=0$).}
\revOne{}{This follows since as time elapses during a system idle period, the distribution of time until the next arrival \revTwo{(which determines the optimal choice between the server being left on or being deallocated) }{}does not change.  Note, however, that this optimal choice depends on a priori knowledge of the request rate\revTwo{.  
This motivates}{, motivating} the use of intermediate “holding-on” policies, \revTwo{which are}{as} often employed in practice\revTwo{, and motivates our analysis of such policies}.{}}

\revTwo{Considering the case of immediate deallocation ($T=0$), we}{Consider now the case of immediate deallocation ($T=0$).  We} 
have assumed that server allocation is triggered whenever a request arrives to an empty system with deallocated server, but another option would be to 
\revTwo{not trigger server allocation until multiple request arrivals have occurred.}{wait for multiple arrivals.}  
\revTwo{Under fixed-rate}{Given} 
Poisson arrivals, an optimal policy will trigger 
\revTwo{the initiation of server}{server}
allocation 
\revTwo{when $b$ request arrivals have occurred}{after $b$ arrivals} 
for some integer\revTwo{ policy parameter}{} $b \geq 1$.  \revTwo{It is straightforward to generalize Equations~(\ref{eq:exponential})/(\ref{eq:deterministic})\revTwo{ (which are for $b=1$) to also handle the case of $b > 1$}{}.  Applying}{For this case, applying} \revTwo{the PASTA property}{PASTA}, we can write \revTwo{an expression for }{}$R$ as the sum of 1) the mean request service time, 2) the product of the mean\revTwo{ request}{} service time and the mean number of requests waiting for or receiving service, and 3) the mean time\revTwo{ spent}{} waiting for server allocation:
\begin{align}
R = \frac{1}{\mu} + \frac{R \lambda}{\mu} + (1 - \frac{\lambda}{\mu}) \left(\frac{(\lambda\Delta)\Delta+b \left(\Delta+\frac{b-1}{2}\frac{1}{\lambda}\right)}{\lambda\Delta+b}\right). 
\end{align}
\revTwo{Here the mean time spent waiting for server allocation is given by the probability that a server is not allocated ($1 - \lambda/\mu$), times the average total request waiting time for server allocation summed over those requests arriving during a period with no allocated server, divided by the average number of such requests.
}{}Solving for $R$ yields, for integer $b \geq 1$,
\begin{align}\label{eq:Rbatching}
R = \frac{1/\mu}{1-\lambda/\mu} + \Delta + \frac{b(b-1)}{2\lambda(\lambda\Delta+b)}. 
\end{align}

Denote the mean time that the server is allocated before being deallocated again by $A$.\revTwo{  The cost $C$ is given by $\mu$ times the fraction of time that the server is allocated or allocation is in progress:
\begin{align}\label{eq:CintermsofA}
C = \mu \left( \frac{\Delta+A}{b/\lambda + \Delta + A}\right). 
\end{align}
}{}
\revTwo{Since, with $T=0$, the server is immediately deallocated when the system empties of requests}{When $T=0$}, $A$ is given by the mean time required to serve the requests that are waiting at the time server allocation completes as well as the requests that arrive while the server is allocated, yielding
\begin{align}
A = \frac{\lambda(A+\Delta) + b}{\mu}. 
\end{align}
Solving for $A$\revTwo{and substituting into Equation~(\ref{eq:CintermsofA}) yields, for integer $b \geq 1$,
\begin{align}\label{eq:Cbatching}
C = \mu - \frac{b (\mu - \lambda)}{\lambda \Delta+b}. 
\end{align}}{, we can obtain the cost $C$ as
\begin{align}\label{eq:Cbatching}
C = \mu \left( \frac{\Delta+A}{b/\lambda + \Delta + A}\right) = \mu - \frac{b (\mu - \lambda)}{\lambda \Delta+b}. 
\end{align}}

From Equations~(\ref{eq:exponential}) or~(\ref{eq:deterministic}) for $T\rightarrow\infty$, and Equations~(\ref{eq:Rbatching}) and~(\ref{eq:Cbatching}), an optimal policy for a fixed request rate $\lambda$ has a value for the objective function~(\ref{eq:objfunction}) of

{\footnotesize\begin{align}\label{eq:optmetricvalue}
\omega  \frac{\lambda/\mu}{1 - \lambda /\mu} + \mu + \min\left[0, \min_b \left[\omega \left( \lambda \Delta + \frac{b(b-1)}{2(\lambda\Delta+b)}\right)  - \frac{b (\mu-\lambda)}{\lambda \Delta + b}\right] \right],
\end{align}
}where the interior minimum is taken over integers $b \geq 1$.  It is straightforward to verify that for $\omega \geq (\mu - \lambda)\lambda\Delta/(\lambda\Delta+1)$, the minimum is achieved with $b=1$, i.e. when server allocation is triggered whenever a request arrives to an empty system, but for smaller $\omega$ a larger $b$ is optimal.

\subsection{Dual Server}\label{sec:dualserver}

A policy for single server allocation/deallocation can accommodate periods \revTwo{when there are no request}{with no} arrivals and the system would otherwise be idle and 
incur
cost with no benefit.  Dual server policies, as considered in this section, can also accommodate periods with high load by allocating a second server.

The basic dual server policy options are (1) always leave both servers allocated, (2) always leave one server allocated and dynamically allocate/deallocate the second server, and (3) dynamically allocate/deallocate both servers.  
\revOne{}{For simplicity, we do not consider policies with a “holding-on” delay in the dual-server case. However, similar behavior can be achieved with the second of the above policy options by setting the deallocation threshold for the second server lower than the allocation threshold, thereby creating an implicit hysteresis effect.}
In all cases, we assume a shared request queue, and so option (1) results in the standard M/M/2 model.  Here we develop models for policy options (2) and (3), beginning with policy option (2).

\subsubsection{Dual Servers, One Server Always Allocated}

\begin{figure}[t]
\centering
\includegraphics[width=0.46\textwidth]{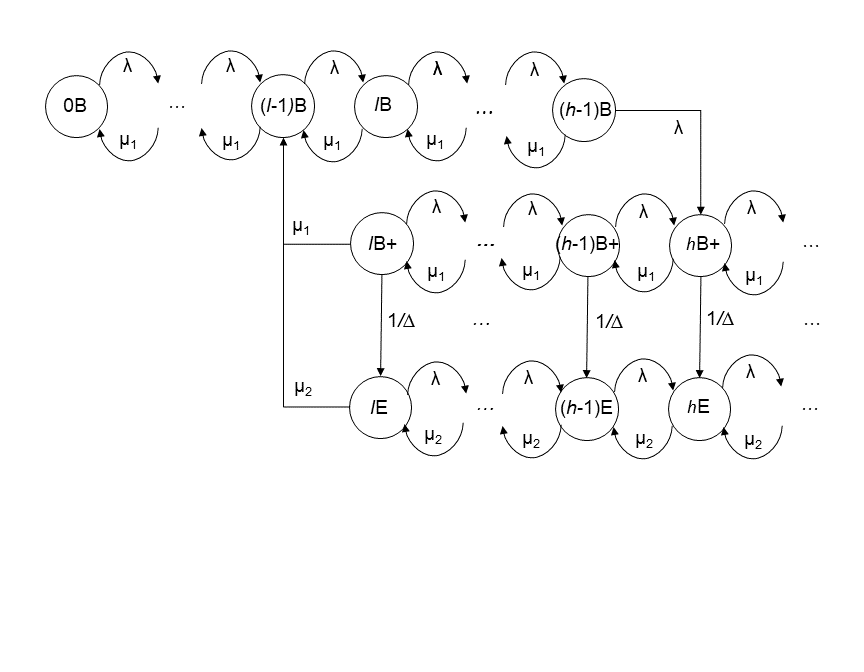}
\vspace{-60pt}
\caption{State-Transition Diagram for System with Dual Servers, One Server Always Allocated.}
\label{fig:statetransitiondiagramforMM1Scaling}
\vspace{-2pt}
\end{figure}

Our model for this policy option is somewhat more general, with state-transition diagram shown in Figure~\ref{fig:statetransitiondiagramforMM1Scaling}.  In the case of a dual server scenario with identical servers each with service rate $\mu$, $\mu_2 = 2 \mu$ and $l \geq 2$.
Each state in Figure~\ref{fig:statetransitiondiagramforMM1Scaling} is labeled by the number of client requests present at the server, followed by "B" (baseline processing resources), "B+" (baseline processing resources are in use and allocation of extra processing resources has been initiated), or "E" (extra processing resources are in use). 
\revOne{The feasible states are states $i$B for $0 \leq i < h$, where $h$ is the threshold at which allocation of extra processing resources is initiated, $i$B+ for $i \geq l$, where $l$ is the threshold below which the additional processing resources are released (or allocation of extra processing resources is abandoned in the case of the transition from state $l\textrm{B+}$ to state $(l-1)\textrm{B}$), and $i$E for $i \geq l$, with state transition rates as shown in the figure.  We assume $\mu_2 > \lambda$.}{Here, $h$ is the threshold at which allocation of extra processing resources is initiated, and $l$ is the threshold below which the additional processing resources are released (in the case of the transition from state $l\textrm{E}$ to state $(l - 1)\textrm{B}$) or the allocation of extra processing resources is abandoned (in the case of the transition from state $l\textrm{B+}$ to state $(l - 1)\textrm{B}$).  The feasible states are states $i\textrm{B}$ for $0 \leq i < h$, $i\textrm{B+}$ for $i \geq l$, and $i\textrm{E}$ for $i \geq l$, with state transition rates as shown in the figure.} 
In the following, we consider the case of $l = h$; 
\revTwo{Appendix~A}{Appendix~C} treats the general case.

\begin{figure}[t]
\centering
\includegraphics[width=0.46\textwidth]{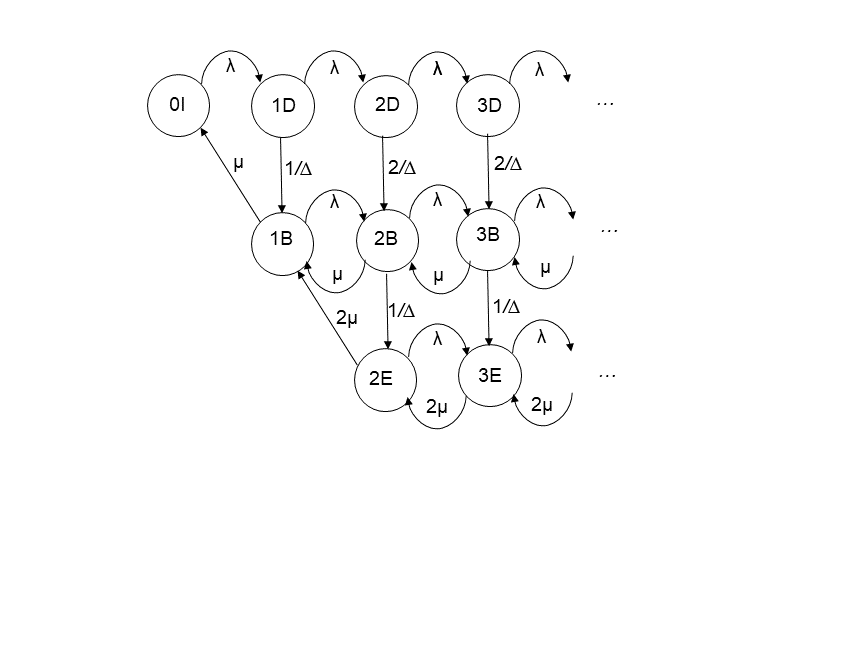}
\vspace{-60pt}
\caption{State-Transition Diagram for System with Dual Servers, both Dynamically Allocated/Deallocated.}
\label{fig:statetransitiondiagramforDualServer}
\vspace{-2pt}
\end{figure}

Our analysis proceeds by expressing all state probabilities in terms of $p_{h\textrm{B+}}$, and then solving for $p_{h\textrm{B+}}$ using the constraint that the state probabilities must sum to one.
Note that the state probabilities $p_{i\textrm{B+}}$ for $i > l = h$ depend only on the portion of the state-transition diagram consisting of these states and the value of $p_{h\textrm{B+}}$.  We seek expressions for these probabilities, in terms of $p_{h\textrm{B+}}$, such that the flow balance equation
\begin{align}\label{eq:scalingflowbalance}
p_{i\textrm{B+}} \mu_1 = p_{(i-1)\textrm{B+}} (\lambda + \mu_1 + 1/\Delta) - p_{(i-2)\textrm{B+}} \lambda
\end{align}
is satisfied for all $i \geq h+2$.  The general form of solution of this recurrence relation
\revOne{}{(e.g.,~\cite{rosen2011discrete})}
is given by
\begin{align}\label{eq:solutionform}
p_{(h+k)\textrm{B+}} = \alpha r_1^k + \beta r_2^k \;\;\;\;\;\; k \geq 0,
\end{align}
where $\alpha$ and $\beta$ are independent of $k$, and $r_1$ and $r_2$ are the roots of the equation
\begin{align}\label{eq:characteristiceq}
x^2 - \left( \frac{\lambda + \mu_1 + 1/\Delta}{\mu_1} \right) x + \frac{\lambda}{\mu_1} = 0. 
\end{align}
These roots are given by 
{\footnotesize\begin{align}\label{eq:r1andr2}
r_1 = \frac{ (\lambda + \mu_1 + 1/\Delta)/\mu_1 - \sqrt{( (\lambda + \mu_1 + 1/\Delta)/\mu_1)^2  - 4 \lambda/\mu_1}}{2},
\nonumber \\
r_2 = \frac{ (\lambda + \mu_1 + 1/\Delta)/\mu_1 + \sqrt{( (\lambda + \mu_1 + 1/\Delta)/\mu_1)^2  - 4 \lambda/\mu_1}}{2}.
\end{align}
}
It is straightforward to verify that $r_2 > 1$ and $0 < r_1 < \textrm{min}[\lambda/\mu_1, 1]$ (assuming $1/\Delta > 0$), and that therefore $\beta$ must be zero if valid state probabilities are to be obtained. Thus, we get
\begin{align}\label{eq:piD}
p_{i\textrm{B+}} = p_{h\textrm{B+}} r_1^{i-h} \;\;\;\;\;\; i \geq h.
\end{align}
Note that
\begin{align}\label{eq:sumofiD}
\sum_{i=h}^{\infty} p_{i\textrm{B+}} = \frac{p_{h\textrm{B+}}}{1 - r_1}.
\end{align}

Consider now the state probabilities $p_{i\textrm{E}}$ for $i \geq l = h$.  The state probability $p_{h\textrm{E}}$ satisfies the flow balance equation
\begin{align}
p_{h\textrm{E}} \mu_2 + p_{(h+1)\textrm{B+}} \mu_1 = p_{h\textrm{B+}} (\lambda + 1/\Delta)
\end{align}
yielding
\begin{align}
p_{h\textrm{E}} = p_{h\textrm{B+}} \frac{(\lambda + 1/\Delta - r_1 \mu_1)}{\mu_2}.
\end{align}
Each state probability $p_{i\textrm{E}}$, $i > h$, satisfies the flow balance equation
\begin{align}\label{eq:piEfb}
p_{i\textrm{E}} \mu_2 + p_{i\textrm{B+}} \mu_1 = (p_{(i-1)\textrm{E}} + p_{(i-1)\textrm{B+}} ) \lambda
\end{align}
yielding
\begin{align}\label{eq:piEorig}
& p_{i\textrm{E}} = p_{h\textrm{B+}} \left( \frac{(\lambda + 1/\Delta - r_1 \mu_1)}{\mu_2} \left( \frac{\lambda}{\mu_2} \right)^{i-h} + \right. \nonumber \\
& \left. \left( \frac{\lambda - r_1 \mu_1}{\mu_2} \right) \sum_{m=0}^{i-h-1} r_1^m \left( \frac{\lambda}{\mu_2} \right)^{i-h-1-m} \right) \;\;\;\;\;\; i \geq h .
\end{align}
Since $r_1$ satisfies Equation~(\ref{eq:characteristiceq}),
\begin{align}\label{eq:r1}
r_1 = (\lambda + \mu_1 + 1/\Delta)/\mu_1 - \lambda / (\mu_1 r_1).
\end{align}
Substitution for $r_1$ in the first term within the outer brackets on the right-hand side of Equation~(\ref{eq:piEorig}) yields, for $i \geq h$,
{\footnotesize\begin{align}\label{eq:piE}
p_{i\textrm{E}} = p_{h\textrm{B+}} \left( \frac{\frac{\lambda}{r_1} - \mu_1}{\mu_2} \right) \left( \left( \frac{\lambda}{\mu_2} \right)^{i-h} \! \! + r_1 \! \! 
\sum_{m=0}^{i-h-1} r_1^{m} \left( \frac{\lambda}{\mu_2} \right)^{i-h-1-m} \right).
\end{align}
}
Note that
{\footnotesize\begin{align}\label{eq:sumofiE}
\sum_{i=h}^{\infty} p_{i\textrm{E}} & = \! p_{h\textrm{B+}} \left( \frac{\frac{\lambda}{r_1} - \mu_1}{\mu_2 - \lambda} + \frac{\lambda - r_1 \mu_1}{\mu_2} \sum_{i=h}^{\infty} \sum_{m=0}^{i-h-1} r_1^m \left( \frac{\lambda}{\mu_2} \right)^{i-h-1-m} \right) \nonumber \\
& = p_{h\textrm{B+}} \left( \frac{\lambda/r_1 - \mu_1}{\mu_2 - \lambda} + \frac{\lambda - r_1 \mu_1}{\mu_2} \sum_{m=0}^{\infty} r_1^m \sum_{i=0}^{\infty}  \left( \frac{\lambda}{\mu_2} \right)^i \right) \nonumber \\
& = \frac{p_{h\textrm{B+}}}{\mu_2 - \lambda} \left( \lambda/r_1 - \mu_1 +  \frac{\lambda - r_1 \mu_1}{1 - r_1}  \right) \nonumber \\
& = 
p_{h\textrm{B+}}\left(\frac{\lambda/r_1 - \mu_1}{(\mu_2 - \lambda)(1 - r_1)} \right).
\end{align}
}

Finally, consider the state probabilities $p_{i\textrm{B}}$ for $0 \leq i < h$.  The state probability $p_{(h-1)\textrm{B}}$ satisfies the flow balance equation
\begin{align}
p_{(h-1)\textrm{B}} \lambda = p_{h\textrm{B+}} \mu_1 + p_{h\textrm{E}} \mu_2
\end{align}
yielding
\begin{align}
p_{(h-1)\textrm{B}} = p_{h\textrm{B+}}  (1 + (1/\Delta +(1 -  r_1) \mu_1)/\lambda).
\end{align}
Applying Equation~(\ref{eq:r1}) and simplifying yields
\begin{align}\label{eq:l=hp(h-1)B}
p_{(h-1)\textrm{B}} = p_{h\textrm{B+}}\left(\frac{1}{r_1}\right).
\end{align}
Each state probability $p_{i\textrm{B}}$, $0 \leq i < h-1$, satisfies the flow balance equation
\begin{align}
p_{i\textrm{B}} \lambda = p_{(i+1)\textrm{B}} \mu_1
\end{align}
yielding
\begin{align}\label{eq:piB}
p_{i\textrm{B}} = p_{h\textrm{B+}}  \left( \frac{1}{r_1} \right) \left( \frac{\mu_1}{\lambda} \right)^{h-1-i} \;\;\;\;\;\; 0 \leq i \leq h-1.
\end{align}
Note that
\begin{align}\label{eq:sumofiB}
\sum_{i=0}^{h-1} p_{i\textrm{B}} = p_{h\textrm{B+}}  \left(\frac{1}{r_1} \right) \left( \frac{1 - (\mu_1/\lambda)^{h}}{1 - \mu_1/\lambda} \right).
\end{align}
(In the case of $\mu_1 = \lambda$, the rightmost factor in~(\ref{eq:sumofiB}) is replaced by $h$.)

Applying the constraint that
\begin{align}
\sum_{i=h}^{\infty} p_{i\textrm{B+}} + \sum_{i=h}^{\infty} p_{i\textrm{E}} + \sum_{i=0}^{h-1} p_{i\textrm{B}} = 1,
\end{align}
Equations~(\ref{eq:sumofiD}),~(\ref{eq:sumofiE}), and~(\ref{eq:sumofiB}) yield
\begin{align}\label{eq:phD}
p_{h\textrm{B+}} = \frac{1-r_1}{1 + \frac{\lambda/r_1 - \mu_1}{\mu_2 - \lambda} + \left(\frac{1-r_1}{r_1} \right) \left( \frac{1 - (\mu_1/\lambda)^{h}}{1 - \mu_1/\lambda} \right)}.
\end{align}

Consider now the mean number of requests in the system $\sum_{i=1}^{\infty} i p_i$, where $p_i$ is given by Equations~(\ref{eq:piD}) and~(\ref{eq:piE}) for $i \geq h$ and by Equation~(\ref{eq:piB}) for $i < h$.  Note that
\begin{align}\label{eq:sumofipiD}
\sum_{i=h}^{\infty} i p_{i\textrm{B+}} = p_{h\textrm{B+}} r_1^{-h}  \sum_{i=h}^{\infty} i r_1^i = p_{h\textrm{B+}} \left( \frac{h}{1 - r_1} + \frac{r_1}{(1 - r_1 )^2} \right),
\end{align}
{\footnotesize\begin{align}\label{eq:sumofipiE}
\sum_{i=h}^{\infty} i p_{i\textrm{E}} & = p_{h\textrm{B+}} \left( \frac{\lambda/r_1 - \mu_1}{\mu_2} \right)\left(  \left( \frac{\lambda}{\mu_2} \right)^{-h} \sum_{i=h}^{\infty} i \left( \frac{\lambda}{\mu_2} \right)^{i} + \right. \nonumber \\
& \left. r_1 \sum_{i=h}^{\infty} i \sum_{m=0}^{i-h-1} r_1^m \left( \frac{\lambda}{\mu_2} \right)^{i-h-1-m} \right) \nonumber \\
& = p_{h\textrm{B+}} \left( \frac{\lambda/r_1 - \mu_1}{\mu_2} \right)\left(  \frac{h - (h - 1) \lambda/\mu_2 }{(1 - \lambda/\mu_2 )^2 } + \right. \nonumber \\
& \left. r_1 \sum_{m=0}^{\infty} r_1^m \sum_{i=0}^{\infty}  (i + h + 1 + l) \left( \frac{\lambda}{\mu_2} \right)^i \right) \nonumber \\
& = p_{h\textrm{B+}} \left( \frac{\lambda/r_1 - \mu_1}{(\mu_2 - \lambda)(1-r_1)} \right) \left(h + \frac{r_1}{1-r_1} + \frac{\lambda/\mu_2}{1 - \lambda/\mu_2} \right),
\end{align}
}

and
\begin{align}\label{eq:sumofipiB}
\sum_{i=0}^{h-1} i p_{i\textrm{B}} & = p_{h\textrm{B+}}  \left( \frac{1}{r_1} \right) \sum_{i=0}^{h-1} i \left( \frac{\mu_1}{\lambda} \right)^{h-1-i} \nonumber \\
& = p_{h\textrm{B+}}  \left( \frac{1}{r_1} \right) \left( \frac{h(1 - \mu_1/\lambda) - (1 - (\mu_1/\lambda)^h)}{(1 - \mu_1/\lambda)^2} \right).
\end{align}
(In the case of $\mu_1 = \lambda$, the rightmost factor in~(\ref{eq:sumofipiB}) is replaced by $h(h-1)/2$.)

Equations~(\ref{eq:sumofipiD}),~(\ref{eq:sumofipiE}), and~(\ref{eq:sumofipiB}), together with equation~(\ref{eq:phD}), yield, after simplification,
\begin{align}
& \sum_{i=1}^{\infty} i p_i = h + \frac{r_1}{1-r_1} + \nonumber \\
& \frac{\lambda \frac{\frac{\lambda}{r_1} - \mu_1}{(\mu_2 - \lambda)^2} +\frac{1-r_1}{r_1} \left( \frac{(h(1 - \frac{\mu_1}{\lambda}) +1)(\mu_1/\lambda)^{h}- 1}{(1 - \mu_1/\lambda)^2} \right) - \frac{1 - (\mu_1/\lambda)^{h}}{1 - \mu_1/\lambda}}{1 + \frac{\lambda/r_1 - \mu_1}{\mu_2 - \lambda} + \left(\frac{1-r_1}{r_1} \right) \left( \frac{1 - (\mu_1/\lambda)^{h}}{1 - \mu_1/\lambda} \right)}.
\end{align}
From Little’s Law, division by $\lambda$ yields the mean request response time.
The cost $C$ is given by $\mu_1$ times $\sum_{i=0}^{h-1} p_{i\textrm{B}}$ (given by Equation~(\ref{eq:sumofiB})) plus $\mu_2$ times $\sum_{i=h}^\infty ( p_{i\textrm{B+}} + p_{i\textrm{E}} )$ (given by Equations~(\ref{eq:sumofiD}) and~(\ref{eq:sumofiE})).

\subsubsection{Dual Servers, Each Server Deallocated When Idle}

We consider now a dual server system in which both servers are dynamically allocated/deallocated.  In the policy we consider, a server is never kept allocated (or its allocation allowed to continue) if it would be idle, and conversely, $is$ kept allocated (or its allocation initiated or continued), whenever it would be busy. We model this policy using the state-transition diagram shown in Figure~\ref{fig:statetransitiondiagramforDualServer}.  Each state is labeled by the number of client requests queued or in service, followed by "I" (idle with no server allocation initiated), "D" (within setup delay from when allocation of one or both servers has been initiated until one server allocation is complete and server is in use), "B" (baseline of one active server), or "E" (extra server is also in use). The feasible states are states 0I, $i$B for $i \geq 1$, $i$D for $i \geq 0$, and $i$E for $i \geq 2$, with state transition rates as shown in the figure.  Analysis for explicit expressions for $R$ and $C$ follows along similar lines as that of the dual server, one server always allocated policy.  
(See \revTwo{Appendix~B}{Appendix~D}.)

\subsection{Unlimited Server}\label{sec:unlimmitedserver}

In this section we consider three dynamic server allocation policies in which there is no cap on the number of servers that may be allocated.  \revTwo{}{The analysis of such policies can
yield insight applicable to cloud computing scenarios in which the limit
on the number of allocated servers is large enough that it is
rarely if ever reached.  Also, in conjunction with results for single server and dual server policies, better, more complete intuition can be obtained regarding simple versus optimal server allocation policy performance than if considering just single or dual or unlimited server cases in isolation.}

\subsubsection{Separate server per request}
In this policy a new server instance is allocated for each incoming request and released when the request is completed, corresponding to a function-as-a-service type of approach without any caching of server instances.
Here the average request response time $R$ is simply given by $1/\mu + \Delta$, and the cost $C$ by $\mu ( \lambda (1/\mu + \Delta) ) = \lambda(1+\Delta\mu)$.

\subsubsection{Reactive unlimited server}
\begin{figure}[t]
\centering
\includegraphics[trim = 34mm 0mm 0mm 0mm, clip, width=0.42\textwidth]{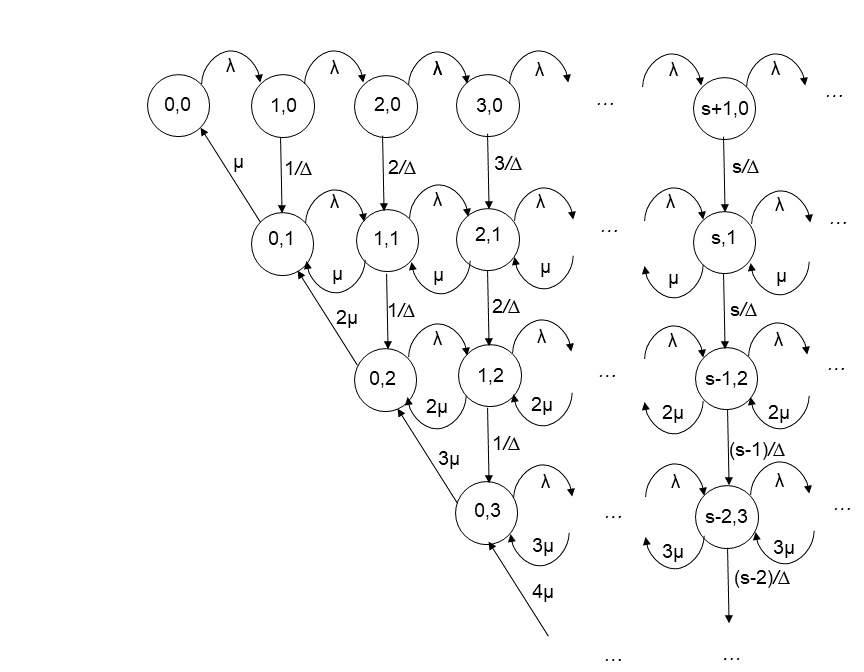}
\caption{State-Transition Diagram for Reactive Unlimited Server System.}
\label{fig:StateTransitionDiagramforUnlimitedServer}
\end{figure}

Allocation of a new server instance is initiated whenever a new request arrives, subject to the constraint that at most $s$ server allocations are allowed to be in progress at once where $s$ is a policy parameter. By de-allocating servers and canceling in-progress allocations as needed, the number of servers active or in the process of being allocated is constrained to be at most equal to the number of requests currently in the system.  A state-transition diagram for this policy is shown in Figure~\ref{fig:StateTransitionDiagramforUnlimitedServer}, where each state is labelled by a pair ($i, k$) giving the number of waiting requests in that state ($i$) and the number of requests in service ($k$).

The average request response time $R$ is given by $1 / \mu$ plus the delay until either a new server is allocated for the request, or an existing request completes service and the new request enters service using the now-free already allocated server.  Denote by $p_{i,k}$ the probability of the state with $i$ waiting requests and $k$ requests in service.  Perhaps surprisingly, we find a product form for these probabilities.

Flow balance equations for this system are, for $i, k > 0$, given by:
\begin{align}
p_{i,0} \left(\lambda + \frac{\textrm{min}[i,s]}{\Delta}\right) = p_{i-1,0},
\end{align}
\begin{align}
p_{0,k} (\lambda + k\mu) = p_{1,k-1} \left(\frac{1}{\Delta}\right) + p_{1,k} k\mu + p_{0,k+1} (k+1)\mu,
\end{align}
and
{\footnotesize\begin{align}
p_{i,k} (\lambda + \! k \mu + \! \frac{\textrm{min}[i,s]}{\Delta})  =  p_{i+1,k-1} \frac{\textrm{min}[i+1,s]}{\Delta} \! + p_{i-1,k} \lambda + p_{i+1,k} k \mu.
\end{align}
}
It is straightforward to verify that these equations are satisfied by probabilities $p_{i,k} = p_{i,*}p_{*,k}$ where $p_{i,*}$ denotes the marginal probability of $i$ waiting requests and $p_{*,k}$ denotes the marginal probability of $k$ requests in service, with $p_{i,*}$ for $i \geq 0$ given by
\begin{align}\label{eq:unlimitedstateprobs}
p_{i,*} = \frac{\prod\limits_{m=0}^i \frac{\lambda}{\lambda+\textrm{min}[m, s]/\Delta}}{\sum\limits_{n=0}^{s-1} \prod\limits_{m=0}^n \frac{\lambda}{\lambda+m/\Delta} + \left(\prod\limits_{m=0}^s\frac{\lambda}{\lambda+m/\Delta} \right) \left(\frac{1}{1 - \lambda/(\lambda+s/\Delta)} \right)}
\end{align}
and $p_{*,k}$ for $k \geq 0$ given by
\begin{align}
p_{*,k} = \frac{\left(\frac{\lambda}{\mu}\right)^k e^{-\lambda/\mu}}{k!}.
\end{align}
From Equation~(\ref{eq:unlimitedstateprobs}),
{\footnotesize\begin{align}
& \sum_{i=0}^\infty i p_{i,*} = \nonumber \\
& \frac{ \sum\limits_{i=0}^{s-1} i \prod\limits_{m=0}^i \frac{\lambda}{\lambda+m/\Delta} + \left( \prod\limits_{m=0}^s \frac{\lambda}{\lambda+m/\Delta}\right) \left( \frac{s}{1 - \frac{\lambda}{\lambda+s/\Delta}} + \frac{\lambda/(\lambda + s/\Delta)}{\left(1 - \frac{\lambda}{\lambda+s/\Delta}\right)^2}\right) }{\sum\limits_{n=0}^{s-1} \prod\limits_{m=0}^n \frac{\lambda}{\lambda+m/\Delta} + \left(\prod\limits_{m=0}^s\frac{\lambda}{\lambda+m/\Delta} \right) \left(\frac{1}{1 - \lambda/(\lambda+s/\Delta)} \right)}.
\end{align}
}
Applying Little's Law, and incorporating the service time once a request acquires a server, $R$ is given by
{\footnotesize\begin{align}
& R = 1/\mu + \nonumber \\
& \frac{ \sum\limits_{i=0}^{s-1} i \prod\limits_{m=0}^i \frac{\lambda}{\lambda+m/\Delta} + \left( \prod\limits_{m=0}^s \frac{\lambda}{\lambda+m/\Delta}\right) \left( \frac{s}{1 - \frac{\lambda}{\lambda+s/\Delta}} + \frac{\lambda/(\lambda + s/\Delta)}{\left(1 - \frac{\lambda}{\lambda+s/\Delta}\right)^2}\right) } { \lambda \left(\sum\limits_{n=0}^{s-1} \prod\limits_{m=0}^n \frac{\lambda}{\lambda+m/\Delta} + \left(\prod\limits_{m=0}^s\frac{\lambda}{\lambda+m/\Delta} \right) \left(\frac{1}{1 - \lambda/(\lambda+s/\Delta)} \right) \right)} .
\end{align}
}
The cost $C$ is given by
{\footnotesize\begin{align}
 C & = \mu \left( \lambda (1/\mu) +\left(\sum_{i=0}^\infty \textrm{min}[i, s] p_{i,*} \right) \right) \nonumber \\
 & = \lambda + \mu \left( \frac{ \sum\limits_{i=0}^{s-1} i \prod\limits_{m=0}^i \frac{\lambda}{\lambda+m/\Delta} + \left( \prod\limits_{m=0}^s \frac{\lambda}{\lambda+m/\Delta}\right) \left( \frac{s}{1 - \frac{\lambda}{\lambda+s/\Delta}}\right) }{\sum\limits_{n=0}^{s-1} \prod\limits_{m=0}^n \frac{\lambda}{\lambda+m/\Delta} + \left(\prod\limits_{m=0}^s\frac{\lambda}{\lambda+m/\Delta} \right) \left(\frac{1}{1 - \lambda/(\lambda+s/\Delta)} \right)} \right) .
\end{align}
}

For the special case of $s=1$,
these expressions reduce to a cost $C$ of $\lambda (1 + \mu /(\lambda + 1/\Delta))$ and a mean response time $R$ of $\frac{1}{\mu} + \Delta$.
Note that with $s=1$, this policy has the same mean response time as with a separate server per request, but lower cost.  This can be explained by the efficiency that results from taking a newly-free existing server for a waiting request instead of always requiring a new server.

\subsubsection{Proactive unlimited server policy}
\begin{figure}[t]
\centering
\includegraphics[width=0.46\textwidth]{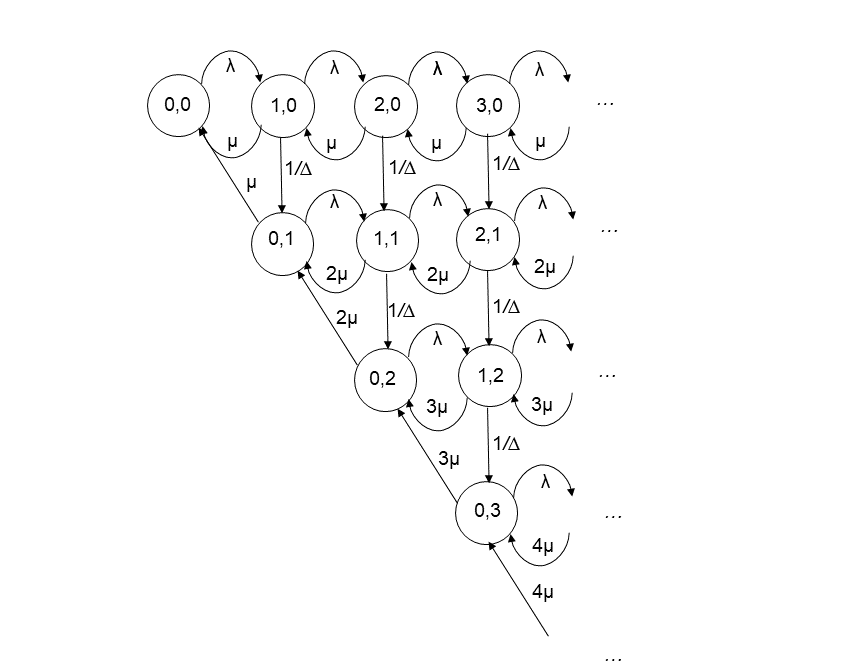}
\vspace{-14pt}
\caption{State-Transition Diagram for Proactive Unlimited Server System.}
\label{fig:StateTransitionDiagramforProactiveUnlimitedServer}
\end{figure}

With this policy, there is always at least one allocated server.  Allocation of a new server instance is initiated whenever all currently allocated servers become busy serving requests.  The number of servers active or in the process of being allocated is constrained to be at most one more than the number of requests currently in the system.   A state-transition diagram for this policy is shown in Figure~\ref{fig:StateTransitionDiagramforProactiveUnlimitedServer}.
For our analysis it is convenient to use a different state labelling here than that used in Figure~\ref{fig:StateTransitionDiagramforUnlimitedServer}. In Figure~\ref{fig:StateTransitionDiagramforProactiveUnlimitedServer}, each state is labelled by a pair ($i, k$) where $k+1$ is the number of allocated servers, and $k+i$ is the number of requests currently in the system.  Again, we find a product form for the state probabilities.

Flow balance equations for this system are, for $i, k > 0$, given by:
\begin{align}\label{eq:unlimitedserverfbeq}
p_{i,0} \left(\lambda + \mu + \frac{1}{\Delta}\right) = p_{i-1,0} \lambda + p_{i+1,0} \mu,
\end{align}
{\footnotesize\begin{align}
p_{0,k} (\lambda + k\mu) = p_{1,k-1} \left(\frac{1}{\Delta}\right) + p_{1,k} (k+1)\mu + p_{0,k+1} (k+1)\mu,
\end{align}
}
and
{\footnotesize\begin{align}
p_{i,k} \left(\lambda + (k+1) \mu + \frac{1}{\Delta}\right)  = & p_{i+1,k-1} \left(\frac{1}{\Delta}\right) + p_{i-1,k} \lambda + p_{i+1,k} (k+1).
\end{align}
}
Define $r$ as
\begin{align}
r = \frac{ (\lambda + \mu + 1/\Delta)/\mu - \sqrt{( (\lambda + \mu + 1/\Delta)/\mu)^2  - 4 \lambda/\mu}}{2}.
\end{align}
(Note the similar forms of Equations~ (\ref{eq:scalingflowbalance}) and~(\ref{eq:unlimitedserverfbeq}).)  It is straightforward to verify that the flow balance equations are satisfied by probabilities $p_{i,k} = p_{i,*}p_{*,k}$ with $p_{i,*}$ given by
\begin{align}\label{eq:unlimitedproactivestateprobi}
p_{i,*} = r^i (1 - r)\;\;\;\;\;\; i \geq 0 
\end{align}
and $p_{*,k}$ given by
\begin{align}\label{eq:unlimitedproactivestateprobk}
p_{*,k} = \frac{\left(\frac{r}{\Delta\mu(1-r)}\right)^k e^{-\frac{r}{\Delta\mu(1-r)}}}{k!}\;\;\;\;\;\; k \geq 0.
\end{align}
Applying Little's Law, $R$ is given  by
\begin{align}
R & = \frac{\sum_{i=0}^\infty \sum_{k=0}^\infty (k+i) p_{i,k}}{\lambda}
= \frac{\sum_{i=0}^\infty i p_{i,*} + \sum_{k=0}^\infty k p_{*,k}}{\lambda} \nonumber \\
& = \frac{1}{\mu} \left(\frac{\mu+ 1/\Delta}{\lambda}\right)\left(\frac{r}{1-r}\right).
\end{align}
The cost $C$ is given by
\begin{align}
C & = \mu\left(1 + \sum_{k=0}^\infty k p_{*,k} + (1 - p_{0,*}) \right) \nonumber \\
& = \mu (1+r) + \frac{1}{\Delta}\left(\frac{r}{1-r}\right).
\end{align}

\section{Optimal Policies}\label{sec:optimal}
Sec.~\ref{sec:policiesandmodels} has described and modeled a number of server allocation policies.  We address here the problem of determining \emph{optimal} policies.  From the performance of such policies, we can evaluate how much room for improvement there may be through the design of dynamic server allocation policies more sophisticated than those described in Sec.~\ref{sec:policiesandmodels}.  Although we were able to determine the optimal single server allocation policy performance in Sec.~\ref{sec:singleserver} analytically, a different approach, which we develop here, is needed for systems with multiple potential servers.  

Under our assumption of Poisson arrivals and exponentially distributed service and server allocation times, an optimal policy will take some server allocation action only when a new request arrives, a request completes service, or a server allocation completes.  We can describe the system operation using a semi-Markov decision model \cite{Tijm86}.  To obtain numerical results for an optimal server allocation policy, we truncate the state space by removing states with negligibly small probabilities and employ a version of the policy iteration algorithm.\footnote{\revOne{}{Although we apply policy iteration with state spaces that are sufficiently large to obtain accurate baselines for the infinite state space models of Sec. IV, an alternative approach would have been to apply policy iteration, and numerically evaluate finite state space variants of the models of Sec. IV, on the same truncated state spaces.}}

We consider systems where homogeneous rate $\mu$ servers can be dynamically allocated, with a shared queue so that when there are $n$ requests in the system and $m$ allocated servers, the total service rate is min[$n,m$]$\mu$.  Attention is restricted to policies in which at each decision point at most one server has its allocation initiated, is deallocated, or has its in-progress allocation terminated.  The set of system states is defined as $\{ (n,m,a) | n \geq 0, m \geq 0, a \geq 0\}$ where $n$ gives the number of requests in the system (waiting or in service), $m$ gives the number of allocated servers prior to any action taken at the point of entering the state, and $a$ gives the number of server allocations in progress at the time the state is entered.

When applying policy iteration, we cap the maximum value of $n$.  We report results with a cap that is large enough to accommodate those states with non-negligible probability and verify that larger caps do not yield different results.   The value of $m+a$ can be capped to the same value as used for $n$ (since there can be no benefit to having more servers than requests), or to a smaller value if a scenario with only a small number of potential servers is of interest. A separate cap on $a$ (limiting the number of server allocations that can be in progress at once) can also be used.  

Potential actions in a state $(n,m,a)$ are (1) $action~IA$: initiate allocation of a server; (2) $action~CA$: cancel an in-progress server allocation (possible when $a>0$); (3) $action~D$: deallocate a server (possible when $m > 0$ and $a=0$), and (4) $action~NC$: make no changes to the server allocation state.  Note that an optimal policy would not deallocate an existing server while continuing an in-progress server allocation and therefore such an action need not be considered.

We denote the rate of moving next to state $S^\prime$ when action $\mathscr{A}$ is taken at the time of entering state $S$ by $q^\mathscr{A}[S, S^\prime$] and the associated rate at which reward is earned during the sojourn time in state $S$ by $R^\mathscr{A}[S]$.  The reward rates are defined such that the undiscounted average reward rate for a particular policy is equal to the negative of the metric given by Expression~(\ref{eq:objfunction}), i.e. 
$-(\omega (\lambda R) + C)$.
The state transition and reward rates, when no caps are placed on the state variables, are given as follows.  For action $IA$:
\begin{align}
q^{IA}[(n,m,a), (n+1,m,a+1)] & = \lambda,
\nonumber \\
q^{IA}[(n,m,a), (n-1,m,a+1)] & = \textrm{min}[n,m]\mu
\nonumber \\
q^{IA}[(n,m,a), (n,m+1,a)] & = (a+1)/\Delta,
\nonumber \\
R^{IA}[(n,m,a)] &= - (\omega n + (m+a+1) \mu)
\end{align}
For action $CA$ (only possible when $a > 0)$:
\begin{align}
q^{CA}[(n,m,a), (n+1,m,a-1)] & = \lambda,
\nonumber \\
q^{CA}[(n,m,a), (n-1,m,a-1)] & = \textrm{min}[n,m]\mu,
\nonumber \\
q^{CA}[(n,m,a), (n,m+1,a-2)] & = (a-1)/\Delta,
\nonumber \\
R^{CA}[(n,m,a)] &= - (\omega n + (m+a-1) \mu).
\end{align}
For action $D$ (only possible when $m>0$ and $a=0$):
\begin{align}
q^{D}[(n,m,0), (n+1,m-1,0)] & = \lambda,
\nonumber \\
q^{D}[(n,m,0), (n-1,m-1,0)] & = \textrm{min}[n,m-1]\mu,
\nonumber \\
R^{D}[(n,m,0)] &= - (\omega n + (m-1) \mu).
\end{align}
For action $NC$:
\begin{align}
q^{NC}[(n,m,a), (n+1,m,a)] & = \lambda,
\nonumber \\
q^{NC}[(n,m,a), (n-1,m,a)] & = \textrm{min}[n,m]\mu,
\nonumber \\
q^{NC}[(n,m,a), (n,m+1,a-1)] & = a/\Delta,
\nonumber \\
R^{NC}[(n,m,a)] &= - (\omega n + (m + a) \mu).
\end{align}

A cap on $n$ requires zeroing the arrival rate in any state with $n$ at its maximum value.  States in which $m+a$ exceeds its cap are also removed from the model, and action $IA$ cannot be allowed when entering a state with $m+a$ equal to its cap.  A separate cap on $a$ can be implemented in a similar fashion by removing states and restricting the use of action $IA$.

Action selections need also be restricted to ensure that there is always a non-zero transition rate from a state following the action taken upon entry.  In particular,
action $CA$ cannot be allowed when entering a state with $n$ equal to the maximum allowed value and with $a=1$ and $m=0$, action $D$ cannot be allowed when entering a state with $n$ equal to the maximum allowed value and $m=1$, and action $NC$ cannot be allowed when entering a state with $n$ equal to the maximum allowed value and $m=a=0$.     

We carry out policy iteration by iterating the following two steps until convergence:

~\newline
\emph{Step 1}:  Value-Determination 

For the current policy $\pi$, solve the system of linear equations
\begin{align}
\bar{R}^\pi = R^{\mathscr{A}^\pi [S]} [S] + \! \! \! \! \sum_{S^\prime \in N(\mathscr{A}^\pi[S], S)} \! \! \! \! q^{\mathscr{A}^\pi [S]} [S, S^\prime]  \left( v^{\pi}[S^\prime] - v^{\pi}[S]\right) \nonumber \\
\forall S \in Z
\end{align}
for the average reward rate $\bar{R}^\pi$ and the state relative values $v^{\pi}[S]$ for all states $S$ except state (0,0,0), with $v^{\pi}[(0,0,0)]$ chosen as 0.  
\revOne{Here $Z$ denotes the set of all states, $\mathscr{A}^\pi [S]$ denotes the action selected in state $S$ when using policy $\pi$, and $N(\mathscr{A}, S)$ denotes the set of states to which there are non-zero transition rates when action $\mathscr{A}$ is taken when entering state $S$.}{Here, $Z$ is the set of all states, $\mathscr{A}^\pi [S]$ is the action selected in state $S$ under policy $\pi$, and $N(\mathscr{A}, S)$ denotes the set of states to which there are non-zero transition rates when action $\mathscr{A}$ is taken when entering state $S$.}

~\newline
\emph{Step 2}: Policy Improvement

For each state $S$, find an allowed action $\mathscr{A} [S]$ that maximizes the following expression, choosing the action used in policy $\pi$ if it is a maximizer:
\begin{align}
\frac{R^{\mathscr{A}} [S] - \bar{R}^\pi + \sum_{S^\prime \in N(\mathscr{A}, S)} q^{\mathscr{A}} [S, S^\prime]   v^{\pi}[S^\prime]}{\sum_{S^\prime \in N(\mathscr{A}, S)} q^{\mathscr{A}} [S, S^\prime]}.
\end{align}
Form the policy $\pi^\prime$ that makes the action selection of $\mathscr{A} [S]$ in each state $S$.  If  the new policy $\pi^\prime$ is identical to the policy $\pi$, stop.  Otherwise, go to step 1 with $\pi$ replaced by $\pi^\prime$.
~\newline

\section{Performance Comparisons}\label{sec:performancecopmarisons}

\subsection{Single Server Policies}

\begin{figure*}[t]
\centering
\subfigure[$\Delta = 0.5$]{\includegraphics[trim = 14mm 26mm 18mm 0mm, clip, width=0.23\textwidth]{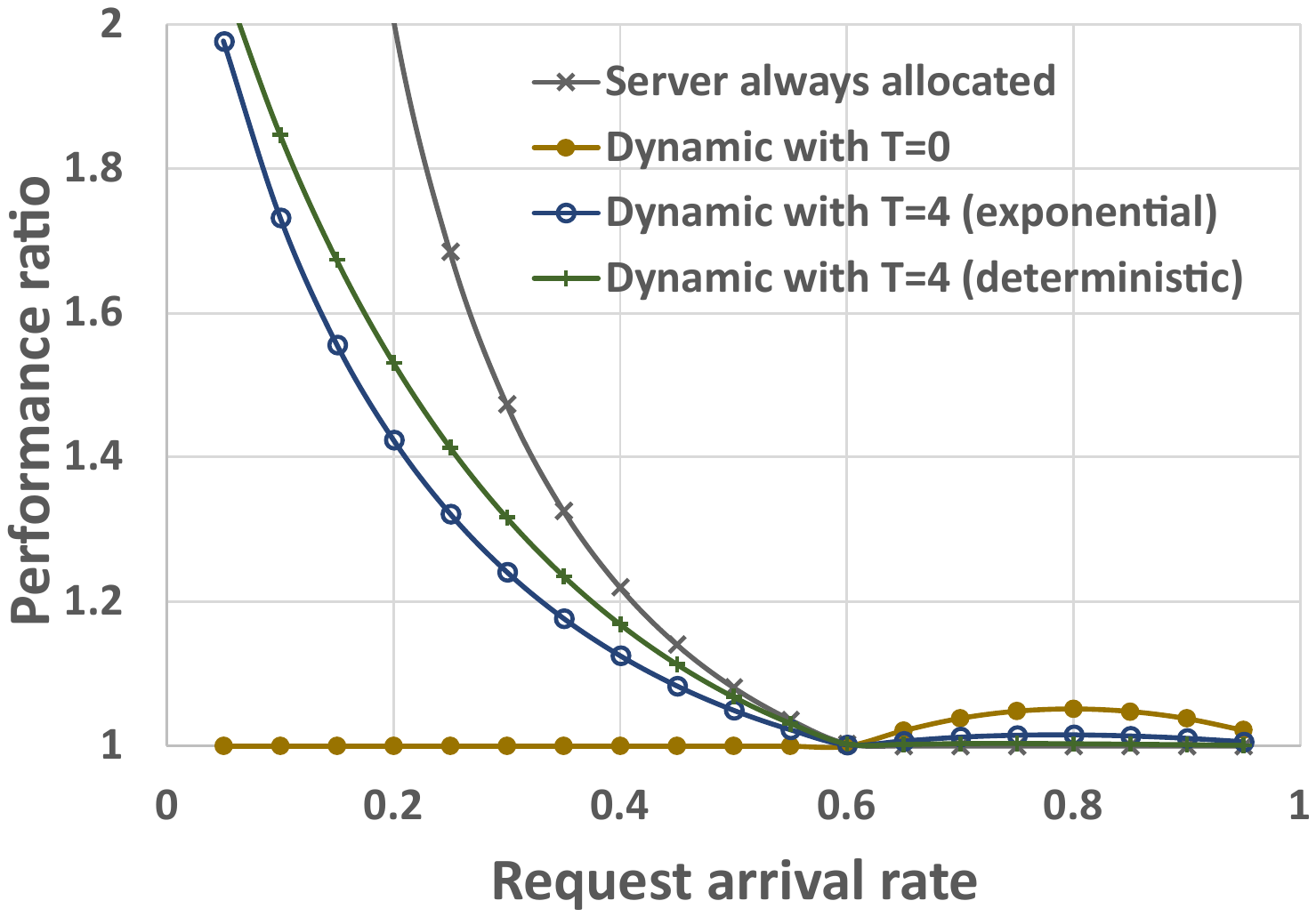}}
\subfigure[$\Delta = 1$]{\includegraphics[trim = 14mm 26mm 18mm 0mm, clip, width=0.23\textwidth]{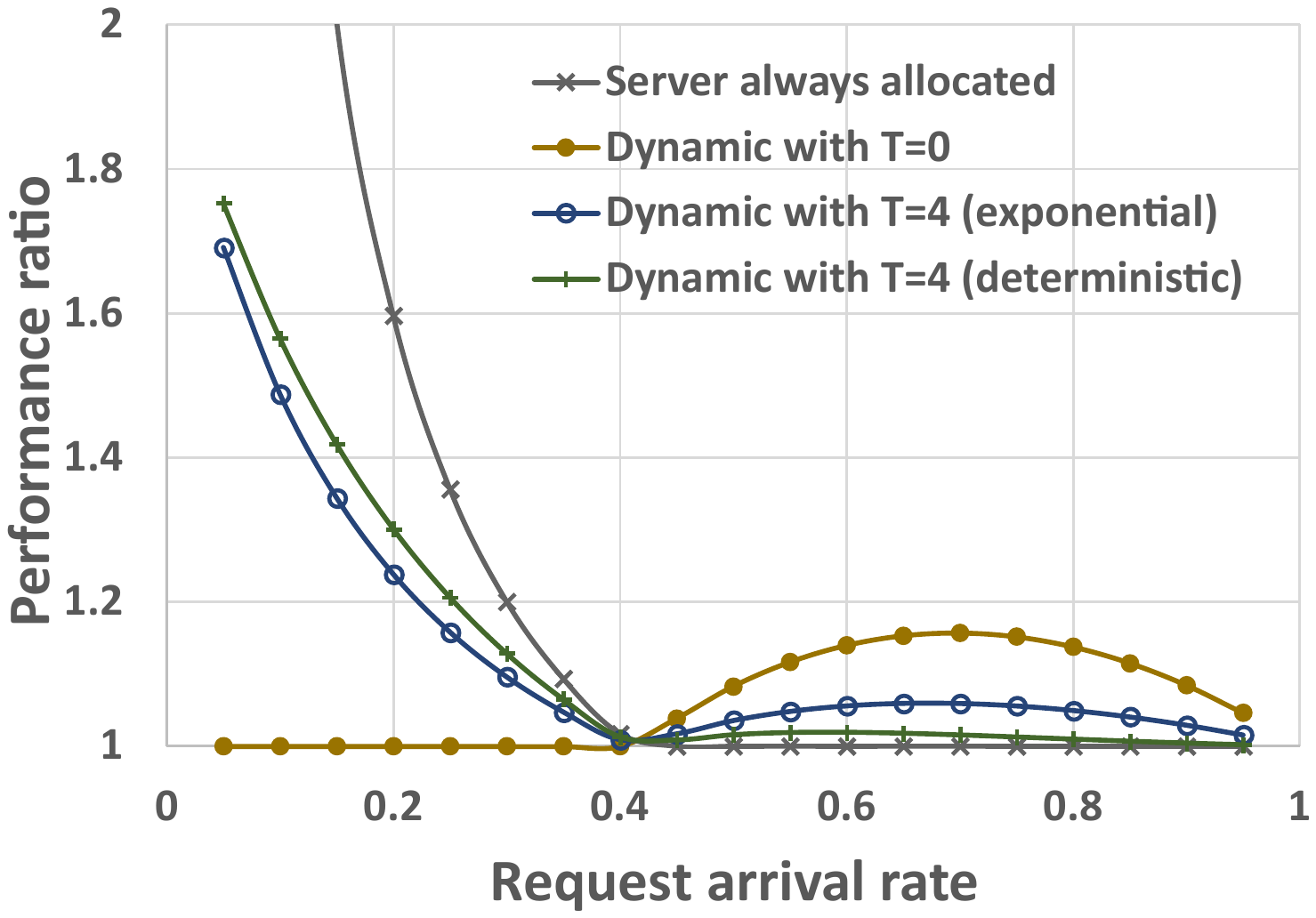}}
\subfigure[$\Delta = 2$]{\includegraphics[trim = 14mm 26mm 18mm 0mm, clip, width=0.23\textwidth]{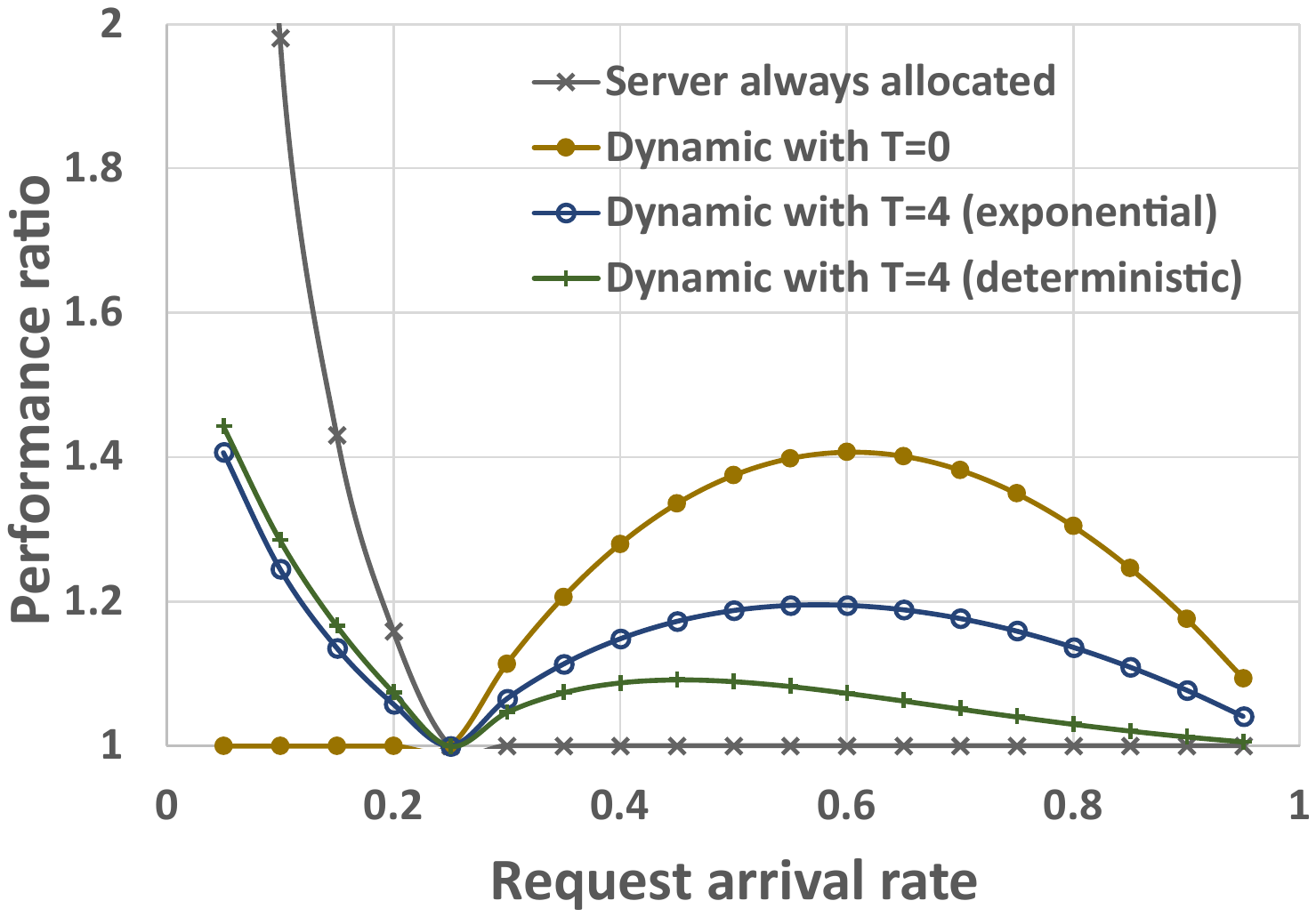}}
\subfigure[$\Delta = 4$]{\includegraphics[trim = 14mm 26mm 18mm 0mm, clip, width=0.23\textwidth]{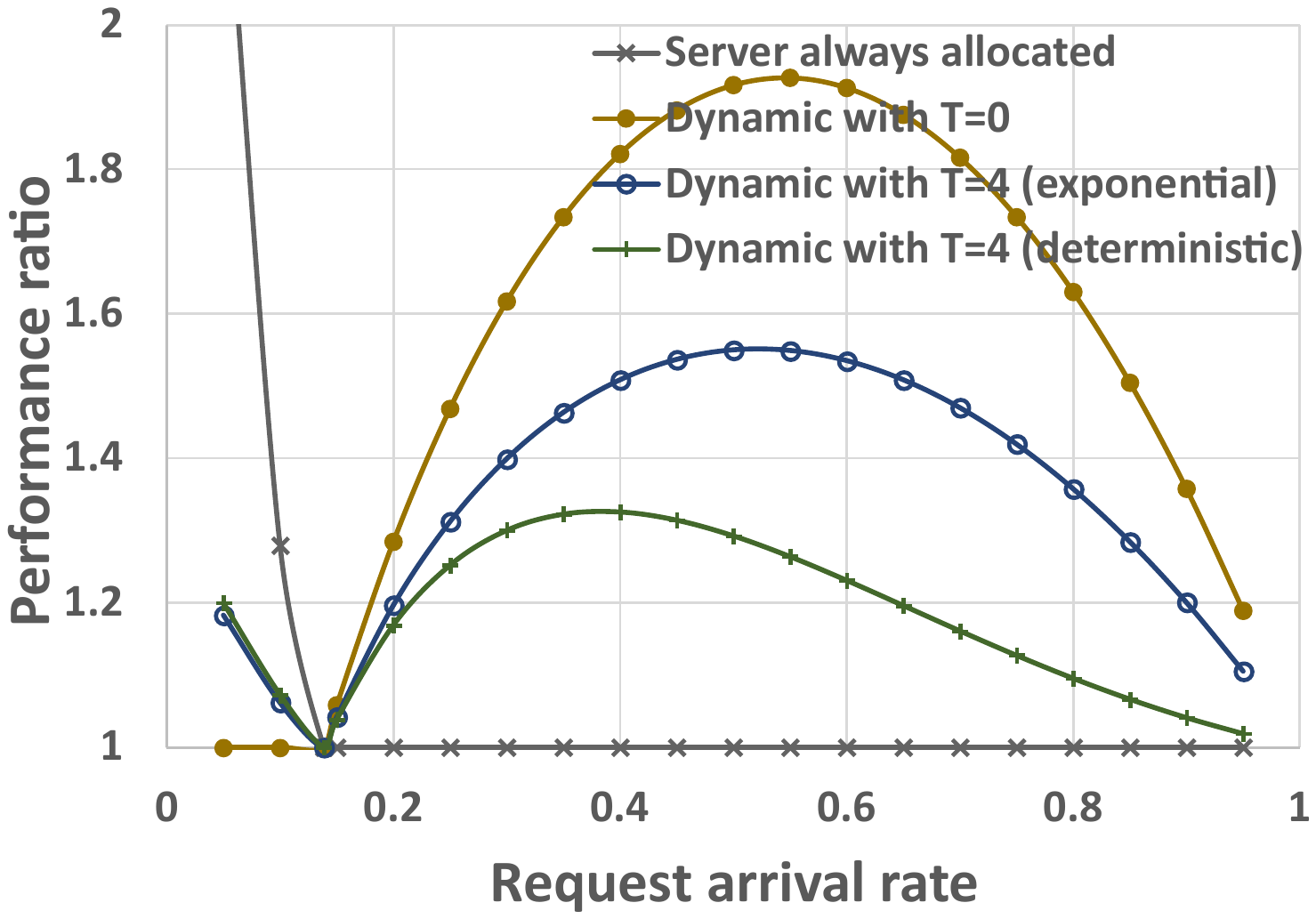}}
\vspace{-4pt}
\caption{Performance ratio for single server allocation policies ($\mu = 1$, $\omega=1$).}
\label{fig:singleserverpolicies}
\vspace{-4pt}
\end{figure*}

Figure~\ref{fig:singleserverpolicies} shows the ratio of the performance with the dynamic server allocation policy defined and analyzed in Sec.~\ref{sec:singleserver},  both for when there is an exponential or deterministic ``holding on''/``delayed-off'' time and when there is no such time, to that of an optimal single server allocation policy.  Also shown is the performance ratio for a baseline policy in which the server is never deallocated.  
\revOne{Here we measure performance by the metric given  by expression~(\ref{eq:objfunction}), with small values corresponding to better performance (lower delay and/or cost).}{We measure performance, and calculate performance ratios, using the objective function given in expression~(\ref{eq:objfunction}), which incorporates both response time and cost.  Small values are preferred and correspond to lower delay and/or cost.}  

Note that the optimal single server allocation policy is \emph{request rate dependent}.  For each request rate of interest, the performance with an optimal policy for that request rate can be obtained using policy iteration as described in Sec.~\ref{sec:optimal} (with a cap of one on the number of servers), or, in this single server context, using expression 
(\ref{eq:optmetricvalue}).
Given the parameter value choices in Figure~\ref{fig:singleserverpolicies}, $\omega$ satisfies the condition given after expression (\ref{eq:optmetricvalue}) and an optimal policy will trigger server allocation when the first request arrives to an empty system ($b=1$).  In this case, for each value of $\lambda$ the optimal policy is either the dynamic server allocation policy of   
Sec.~\ref{sec:singleserver} with $T=0$, or to never deallocate the server ($T\rightarrow\infty$).
At a value of $\lambda$ such that $\omega \lambda \Delta (1+\lambda \Delta) = \mu-\lambda$, the performance metrics of these policies are identical, yielding the convergence points evident in Figure~\ref{fig:singleserverpolicies}.  For smaller values of $\lambda$, it is optimal to immediately deallocate the server whenever it becomes idle, while for larger values of $\lambda$, it is optimal to never deallocate the server.  Finally, as $\lambda$ approaches $\mu$, the server becomes idle increasingly rarely, and the policy differences narrow.

Without requiring knowledge of request rate, a dynamic server allocation policy with a fixed intermediate value of $T$ is often able to yield performance that is not far from optimal over a wide range of request rates.  For example, as seen in Figure~\ref{fig:singleserverpolicies}(c) for $\mu=1$, $\omega=1$, and $\Delta=2$, with $T=4$ performance is within 20\% of optimal for $\lambda \geq 0.15$. With respect to the impact of the distribution of the holding-on time, for values of $\lambda$ smaller than the value at which immediate deallocation yields the same performance as never deallocating, an exponential distribution with the same mean can yield slightly better performance than a deterministic value.  For larger values of $\lambda$, however, substantially better performance can be achieved with a deterministic value.

\subsection{Dual Server Policies}

\begin{figure*}[t]
\centering
\subfigure[$\Delta = 0.5$]{\includegraphics[trim = 14mm 26mm 18mm 0mm, clip, width=0.23\textwidth]{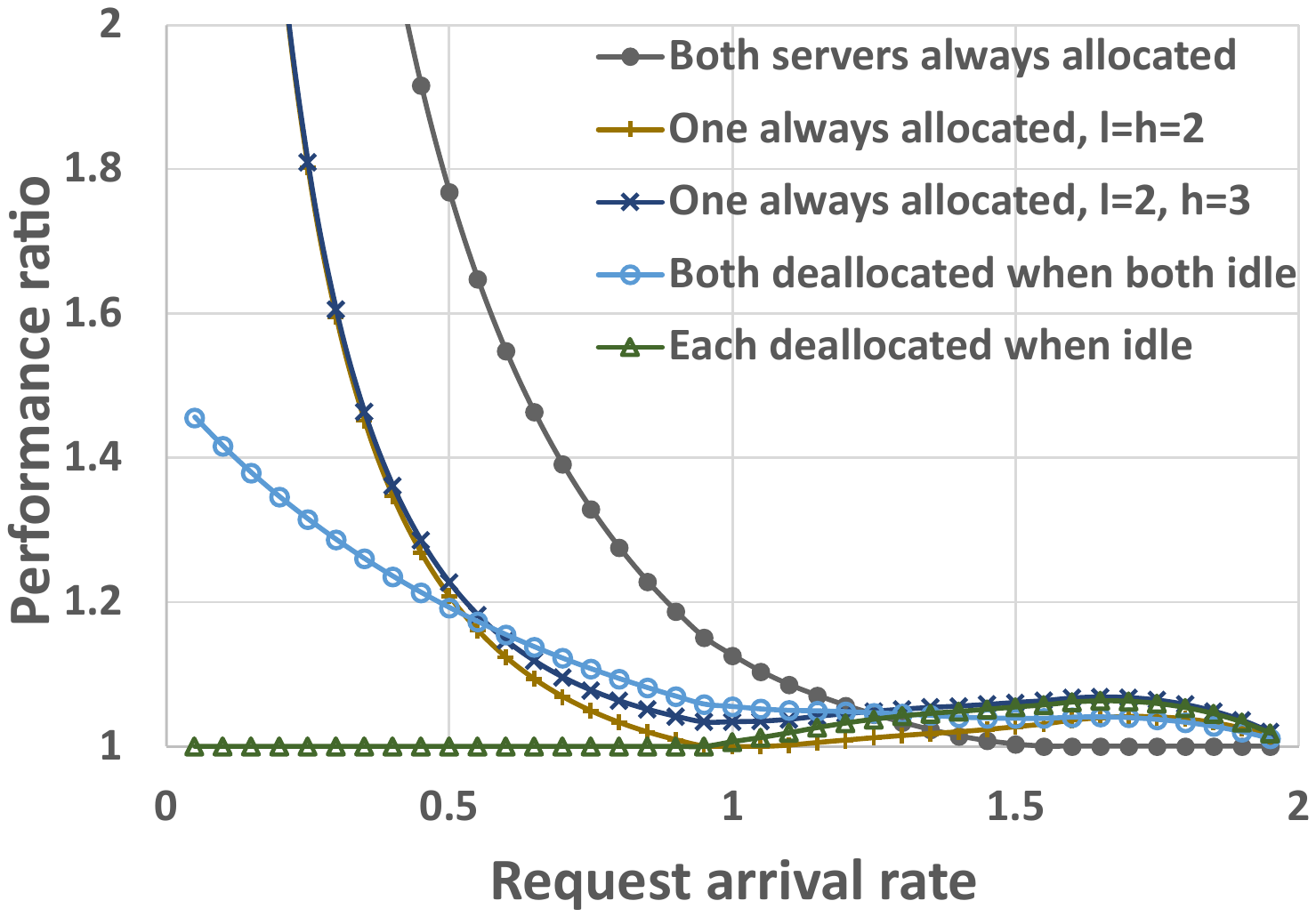}}
\subfigure[$\Delta = 1$]{\includegraphics[trim = 14mm 26mm 18mm 0mm, clip, width=0.23\textwidth]{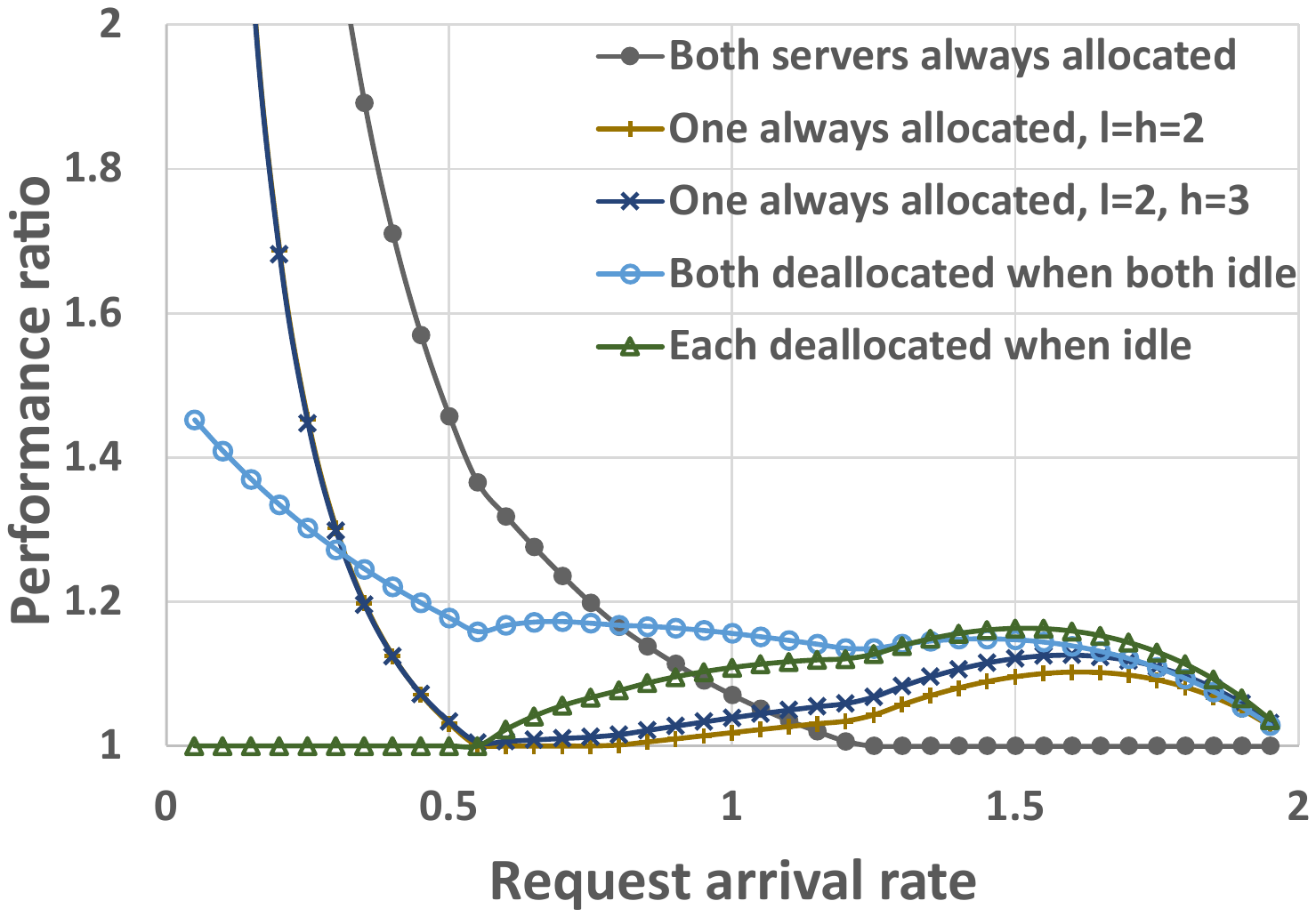}}
\subfigure[$\Delta = 2$]{\includegraphics[trim = 14mm 26mm 18mm 0mm, clip, width=0.23\textwidth]{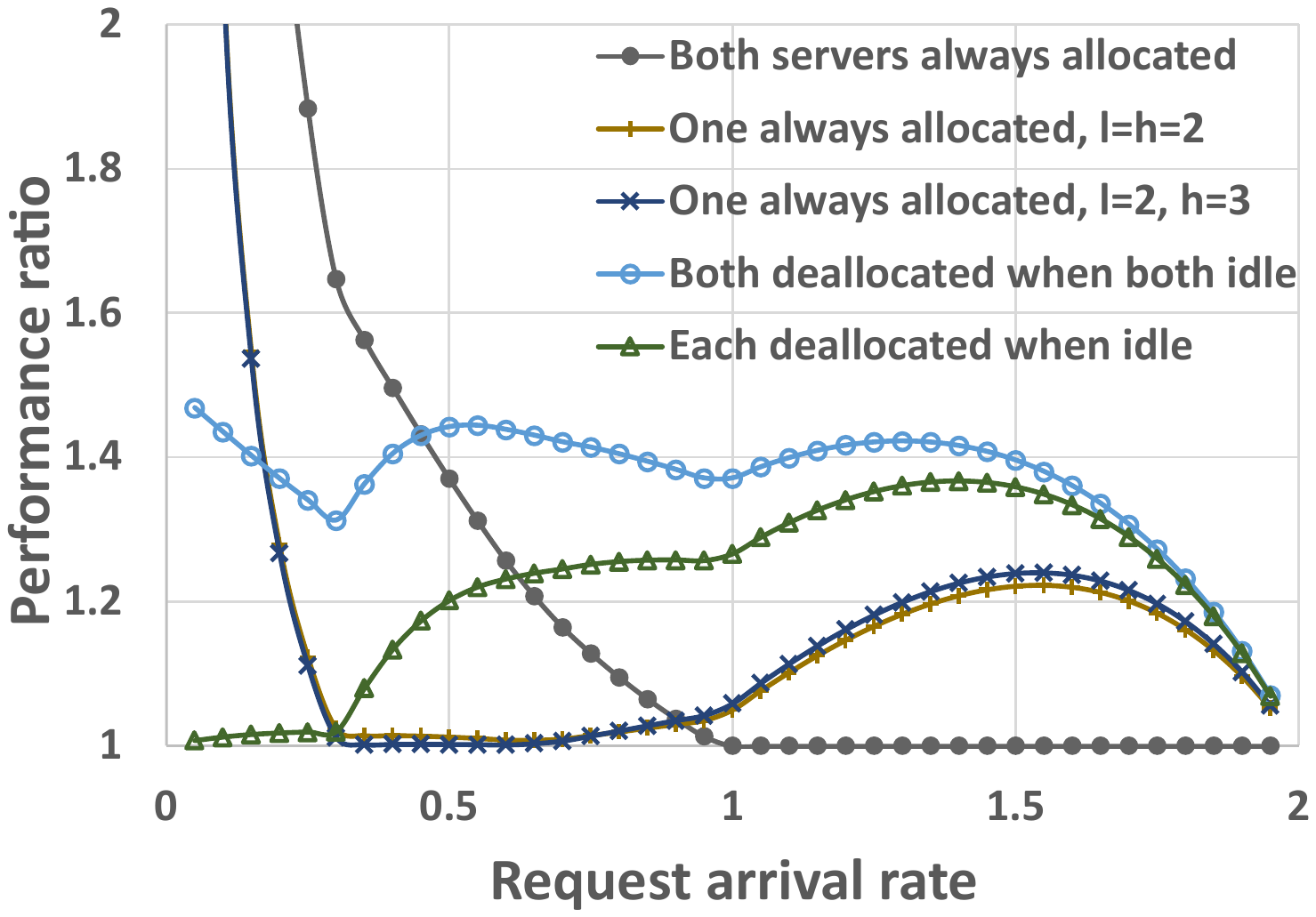}}
\subfigure[$\Delta = 4$]{\includegraphics[trim = 14mm 26mm 18mm 0mm, clip, width=0.23\textwidth]{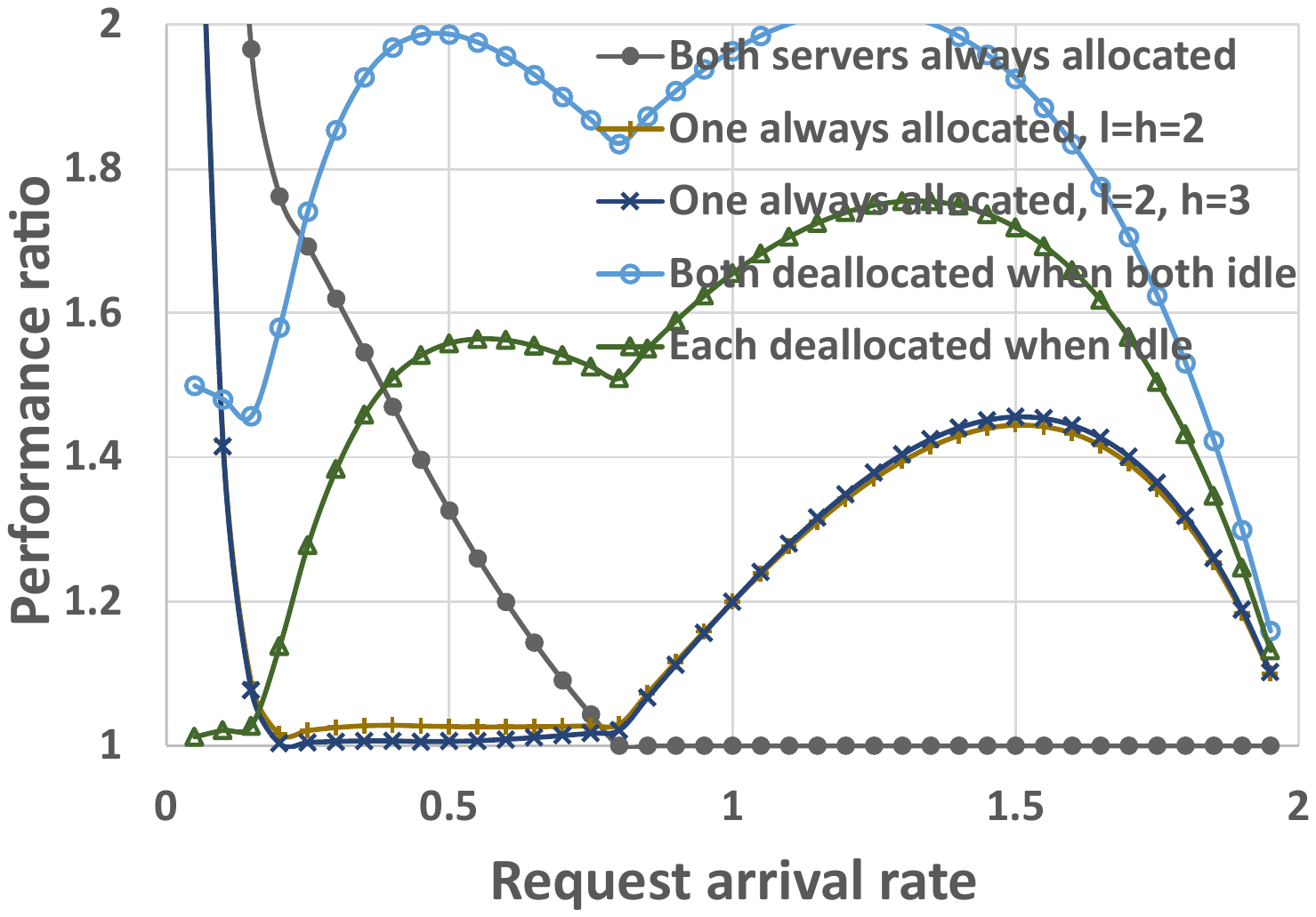}}
\vspace{-4pt}
\caption{Performance ratio for dual server allocation policies ($\mu = 1$, $\omega=1$).}
\label{fig:dualserverpolicies}
\vspace{-4pt}
\end{figure*}

Figure~\ref{fig:dualserverpolicies} shows the performance with various simple dual server policies.  As in Figure~\ref{fig:singleserverpolicies}, performance is measured using expression~(\ref{eq:objfunction}) and expressed as the ratio of the performance of the policy under consideration to the optimal performance, in this case as obtained from using policy iteration for each request arrival rate.  Results are shown for the ``Dual Servers, One Server Always Allocated'' (both with $l=h=2$, and with $l=2$, $h=3$) and the ``Dual Servers, Each Server Deallocated When Idle'' policies from Sec.~\ref{sec:dualserver}.  The figure also shows results for a policy in which both servers are always allocated/deallocated together rather than individually (obtained using the single server model with state-dependent
service rates analysis in 
\revTwo{the supplemental material}{Appendix~B}
with $c=2$, $\mu_1=\mu$, $\mu_c=2\mu$), and for a baseline policy in which both servers are always kept allocated (yielding an M/M/2 model).

Similarly, as in the case of a single server, for a sufficiently large request arrival rate, it is optimal to never deallocate the servers, and as the request rate approaches $2\mu$ (capacity load) policy differences narrow.  For smaller values of $\lambda$, there are two regions, one where it is optimal to never deallocate one of the two servers (but dynamically allocate/deallocate the other), and one region (the smallest arrival rates) where sometimes it is optimal to deallocate both servers.

Note that in each region of the parameter space covered by the graphs in Figure~\ref{fig:dualserverpolicies}, one of the simple dual server policies yields close to optimal performance.  However, different policies are best in different regions, and none of the simple dual server policies yields close to optimal performance across the entire parameter \revTwo{space.}{space when $\Delta$ is large.}  The 
``Dual Servers, Each Server Deallocated When Idle'' policy is seen to be the most robust of the policies in the sense of its maximum performance ratio, but can be quite far from optimal over a large range of request rates when $\Delta$ is substantially larger than the average request service time, as seen in Figure~\ref{fig:dualserverpolicies}(d).  

With respect to the performance comparison between
``Dual Servers, One Server Always Allocated'' with $l=h=2$, and this policy with $l=2$ and $h=3$, note that when $\Delta$ is large as in Figure~\ref{fig:dualserverpolicies}(d) reducing the frequency of initiating server allocations by using a higher queue length threshold for allocation ($h$) than for deallocation ($l$) can yield some (modest in this case) performance improvement.  Finally, note that allocating/deallocating both servers together rather than individually often (but not always) yields substantially worse performance than individually allocating/deallocating the servers. 

\subsection{Unlimited Server Policies}

\begin{figure*}[t]
\centering
\subfigure[$\Delta = 0.5$]{\includegraphics[trim = 14mm 26mm 18mm 0mm, clip, width=0.23\textwidth]{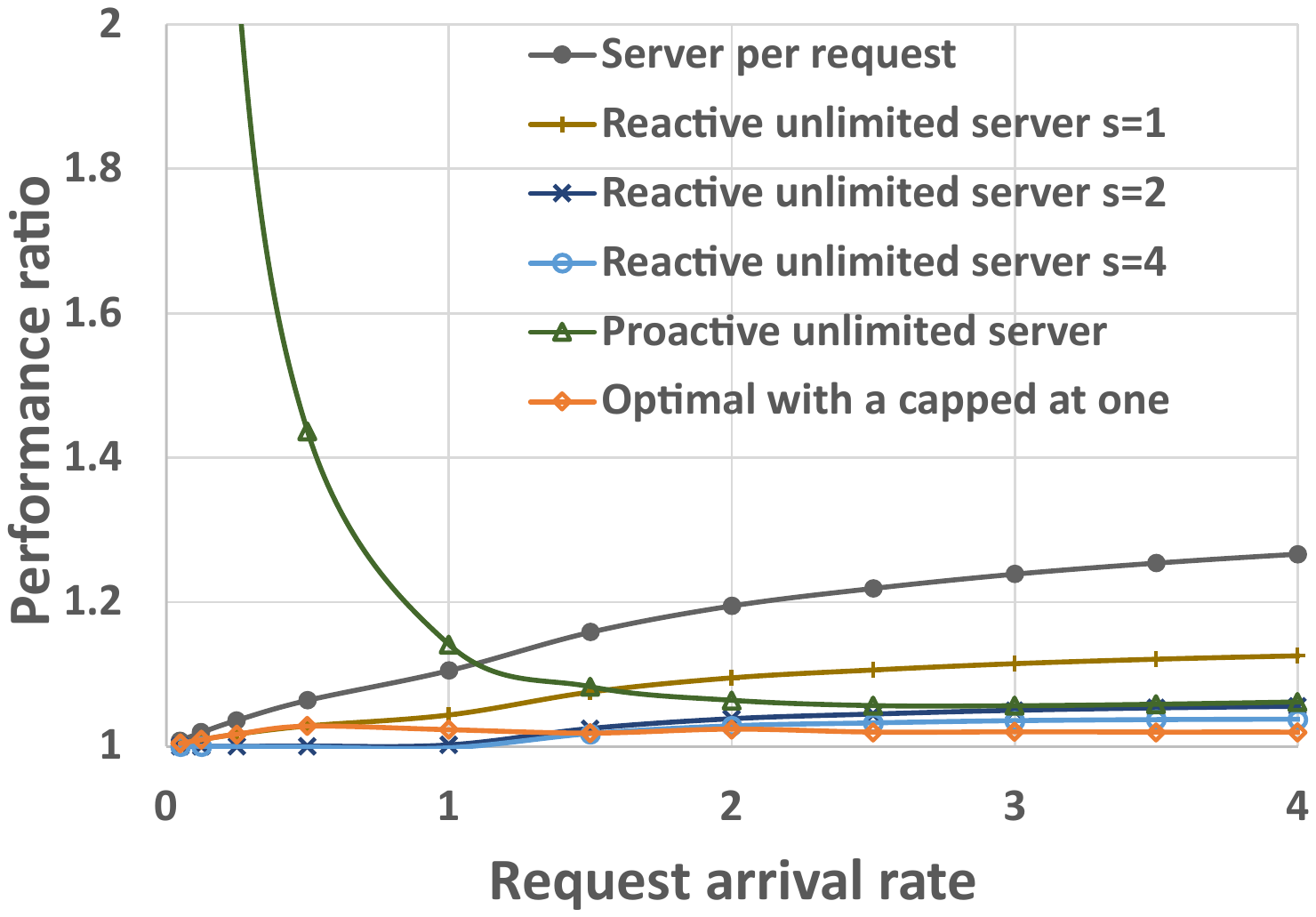}}
\subfigure[$\Delta = 1$]{\includegraphics[trim = 14mm 26mm 18mm 0mm, clip, width=0.23\textwidth]{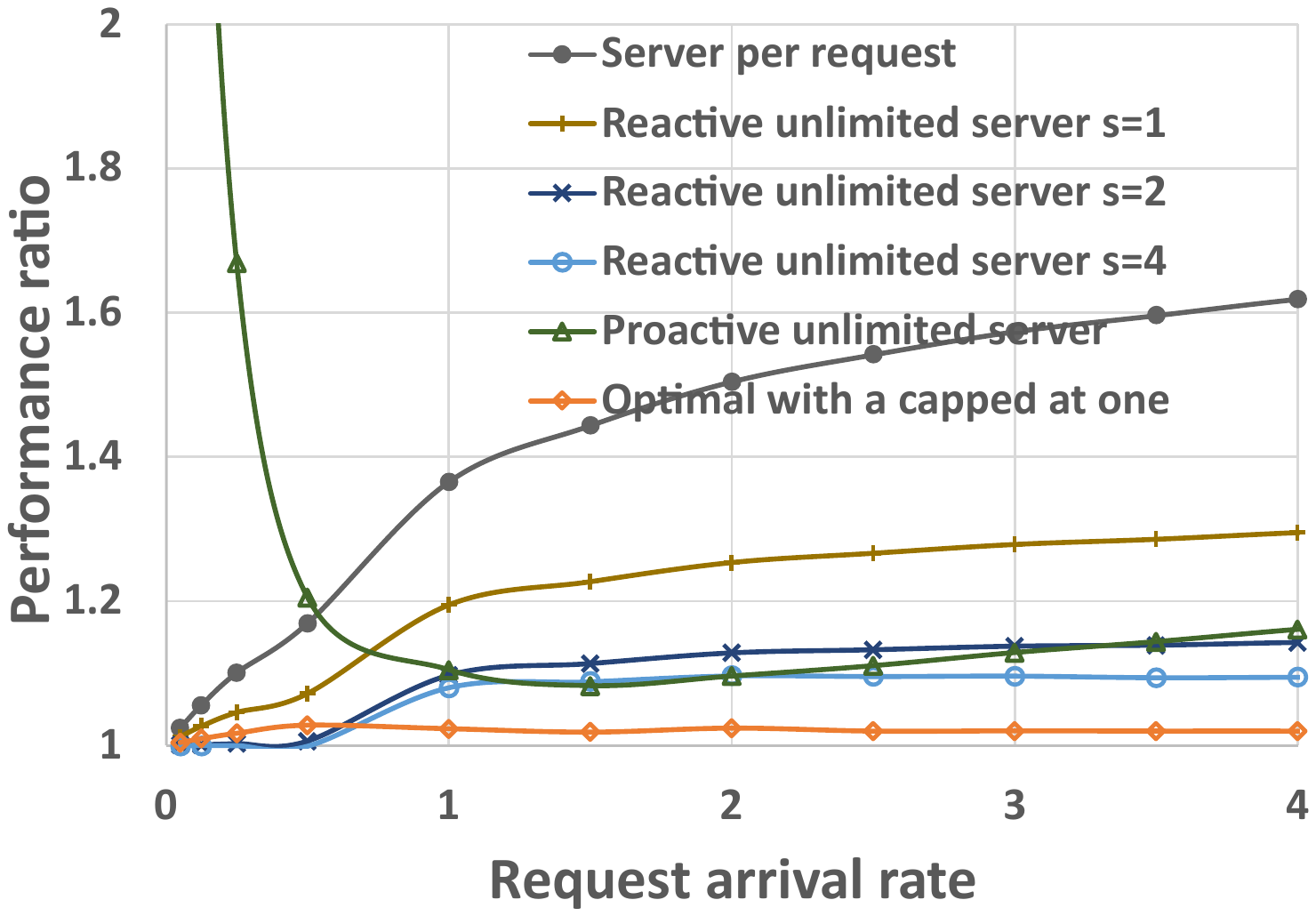}}
\subfigure[$\Delta = 2$]{\includegraphics[trim = 14mm 26mm 18mm 0mm, clip, width=0.23\textwidth]{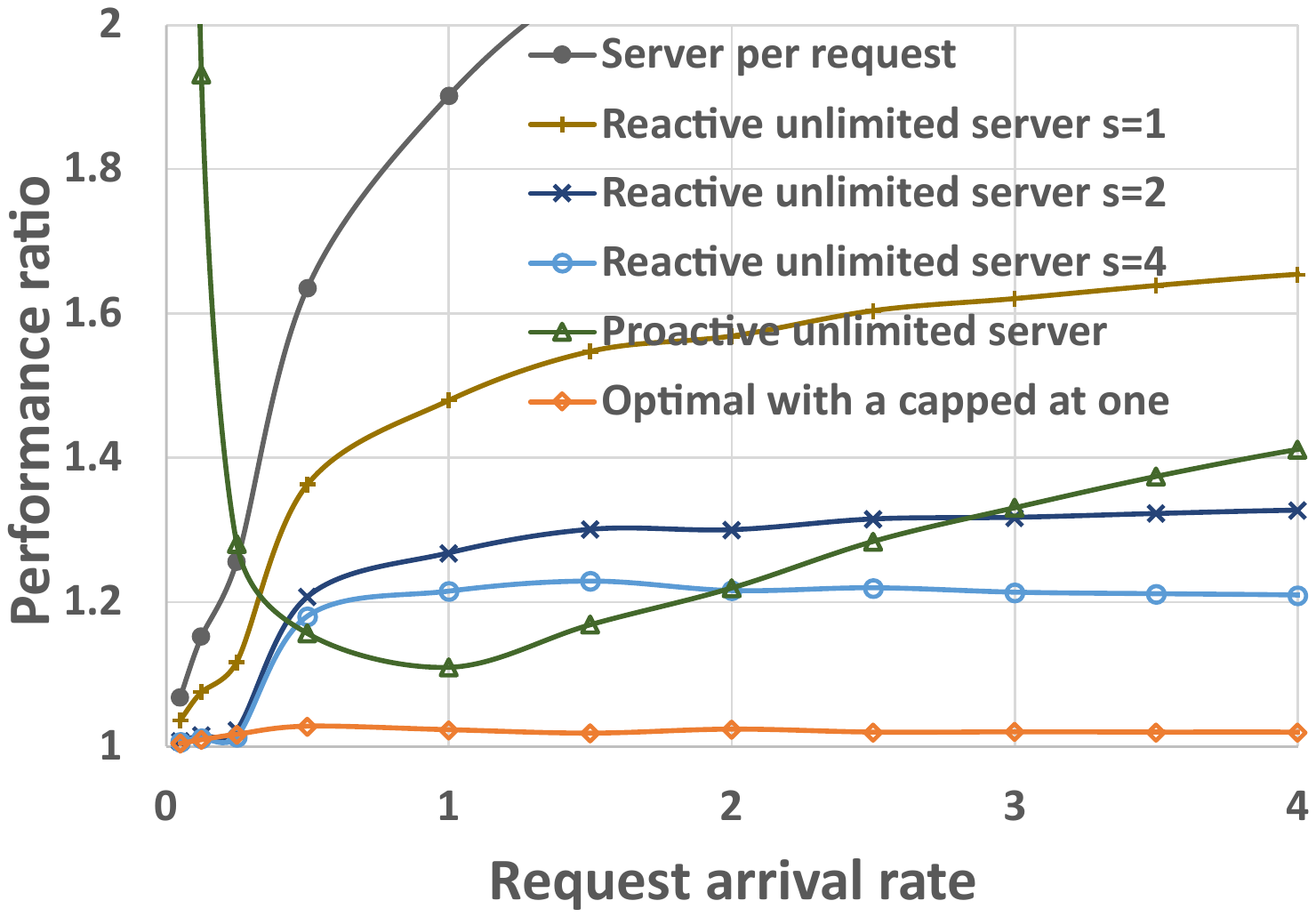}}
\subfigure[$\Delta = 4$]{\includegraphics[trim = 14mm 26mm 18mm 0mm, clip, width=0.23\textwidth]{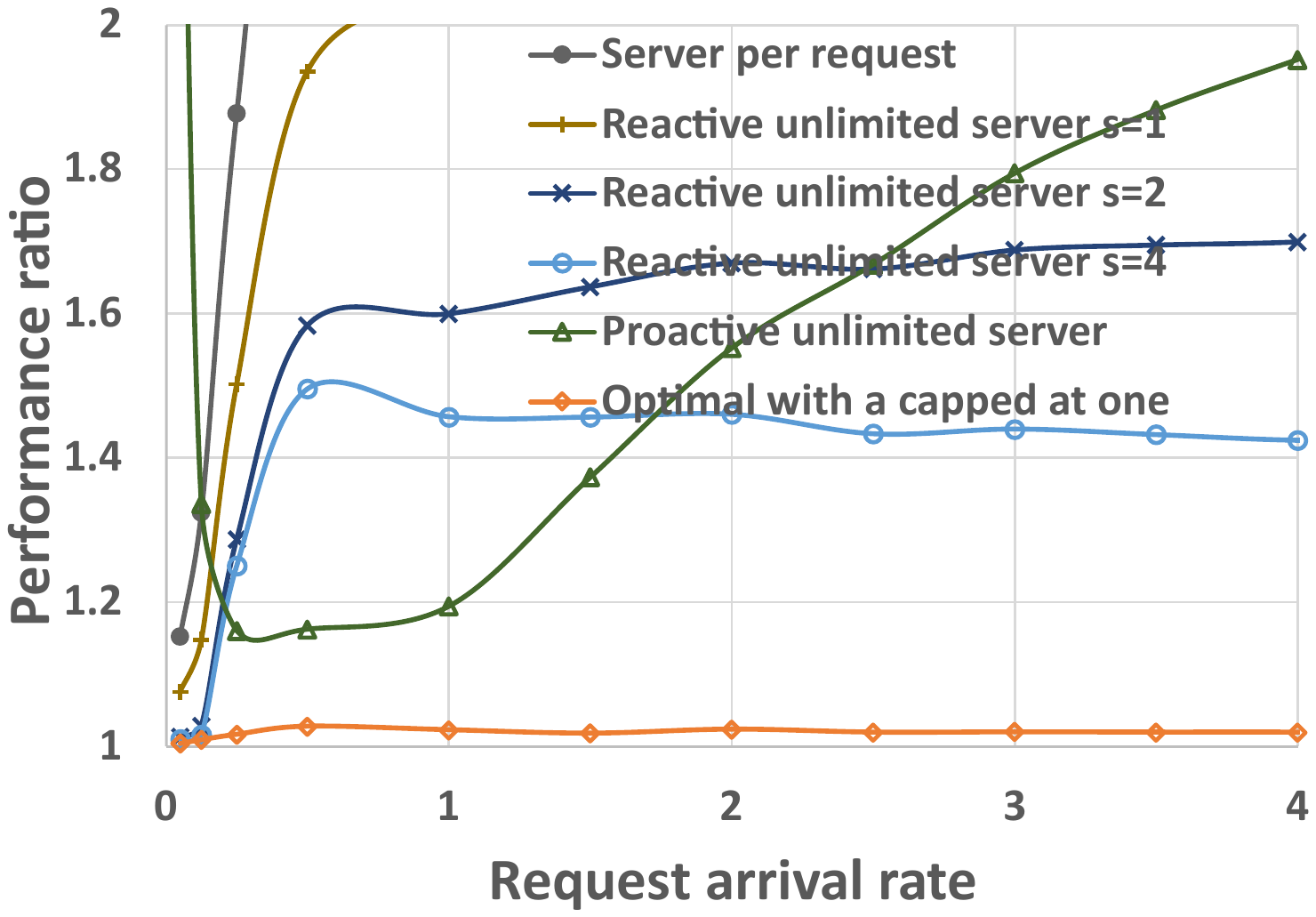}}
\vspace{-4pt}
\caption{Performance ratio for unlimited server allocation policies ($\mu = 1$, $\omega=1$).}
\label{fig:unlimitedserverpolicies}
\vspace{-4pt}
\end{figure*}

As the number of servers that can potentially be allocated increases, optimal server allocation/deallocation policies can become increasingly complex.  Interestingly, however, we find that a single simple policy, namely the ``reactive unlimited server'' policy from Sec.~\ref{sec:unlimmitedserver}, can yield performance close to optimal over the full range of request rates, as long as $\Delta$ is not too large (e.g., $\Delta \leq 2$ in the case of $\mu=1$, $\omega = 1$), and multiple server allocations can be in progress at once (the policy parameter $s$ is greater than one).  This is illustrated by the results shown in Figures~\ref{fig:unlimitedserverpolicies}(a)-(c).  For larger $\Delta$ (Figure~\ref{fig:unlimitedserverpolicies}(d)), the performance gap widens, as the optimal (request rate dependent) policy will bias allocation/deallocation decisions towards keeping some (request rate dependent) number of servers consistently allocated, and the performance benefits of such a bias increase with increasing $\Delta$.

Figure~\ref{fig:unlimitedserverpolicies} also shows results for a policy that allocates a new server for each request and deallocates the server when the request has finished service, the ``proactive unlimited server'' policy of Sec.~\ref{sec:unlimmitedserver}, and an optimal (request-rate dependent) policy under the constraint that at most one server allocation can be in progress at once (the state variable $a$ is capped at one).  Not surprisingly, the server per request policy yields relatively poor performance since there is no reuse of allocated servers across multiple requests.  Note that when the reactive policy is constrained to have at most one server allocation in progress at once ($s=1$), it yields worse performance than the proactive policy excepting in a low request rate region of size dependent on $\Delta$.  The reactive policy's performance improves, however, narrowing the region where the proactive policy yields better performance, when multiple server allocations can be in progress at once.  In this case, the reactive policy is able to more quickly increase the number of allocated servers when needed in reaction to increased numbers of queued requests, and the benefits of proactivity are reduced.

Finally, note that performance with an optimal policy under the constraint that at most one server allocation can be in progress at once is very similar to that with an unconstrained optimal policy.  This is quite different from what is seen with the reactive policy, and can be explained by the fact that an optimal policy will ensure that the number of allocated servers is never far from a target (request rate dependent) number of servers.      

\subsection{\revTwo{}{Single vs. Dual vs. Unlimited}}

\revTwo{}{Results for the single server, dual server, and unlimited server cases are often similar with respect to the magnitudes of the observed performance gaps.  For $\Delta =1$, in each of the three cases, there is a simple policy with a performance ratio within 1.2 across the full range of request arrival rates.  As $\Delta$ increases, in all three cases the corresponding upper bound on the performance ratio achievable with a fixed simple policy, across the full range of request rates, also increases, but remains under 2 even for the extreme case of $\Delta = 4$.  These similarities across diverse cases yield increased confidence with respect to the robustness of our findings.}

\revTwo{}{
There are also significant insights that come from one or two of the three cases, but not the other(s).  Perhaps most notably, the unlimited server results show the benefit that may be achieved with a simple reactive policy, and yet not with an optimal policy, when multiple server allocations can be in progress at once rather than being limited to just one.}

\subsection{Impact of Weighting of Objective Function Terms}

\begin{figure}[t]
\centering
\includegraphics[trim = 0mm 28mm 0mm 0mm, clip, width=0.38\textwidth]{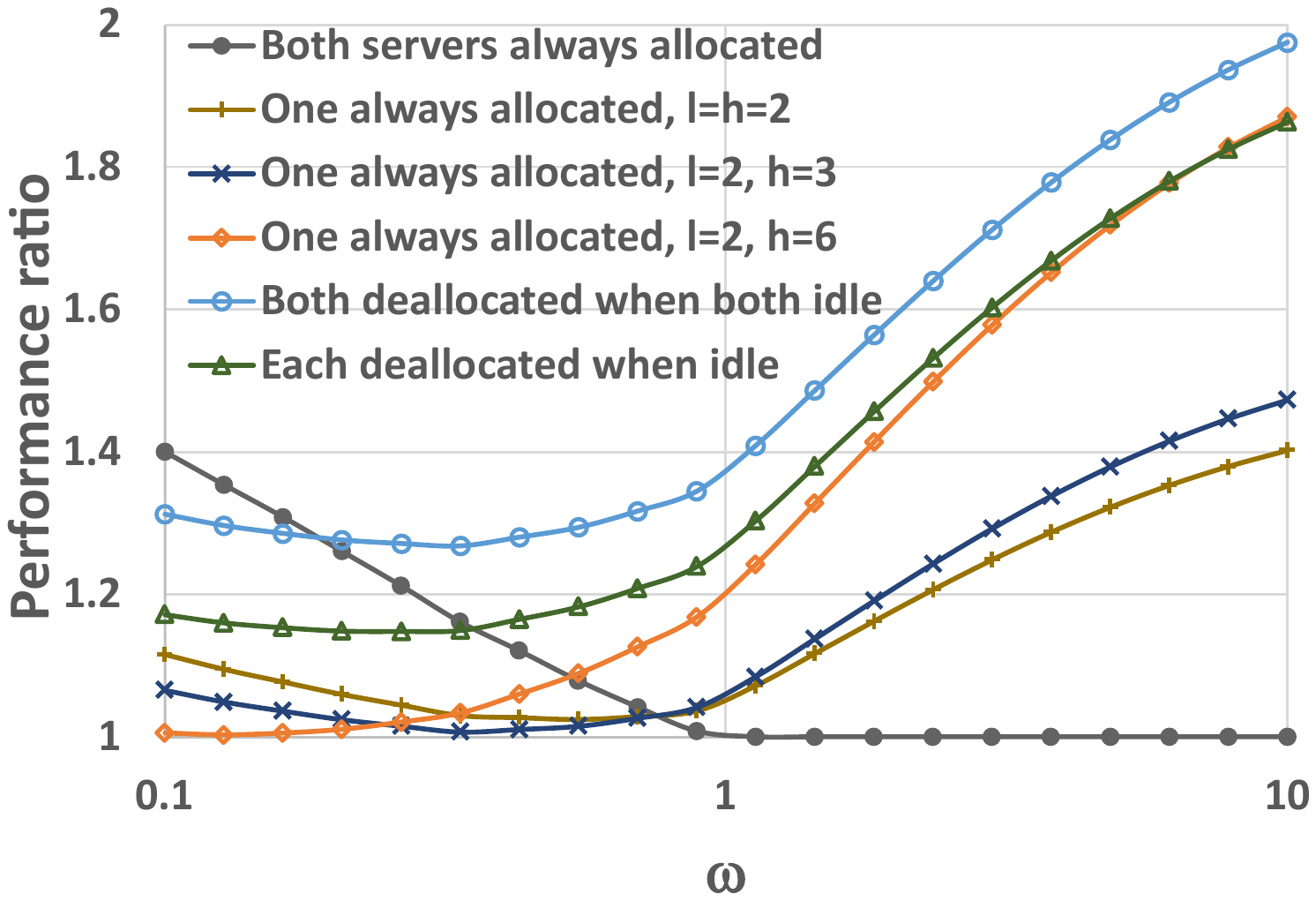}
\vspace{-10pt}
\caption{Impact of $\omega$ on the performance ratio for dual server allocation policies ($\mu = 1$, $\lambda=1$, $\Delta = 2$).}
\label{fig:dualserverpoliciesvaryingomega}
\vspace{-6pt}
\end{figure}

Figure~\ref{fig:dualserverpoliciesvaryingomega} illustrates the impact of the weight $\omega$ that is given to the $\lambda R$ term in the objective function given by Expression~\ref{eq:objfunction}.  The same dual server policies are considered as in Figure~\ref{fig:dualserverpolicies}, together with a ``Dual Servers, One Server Always Allocated'' policy with a higher allocation threshold ($h=6$).  The latter policy is noteworthy since for the particular parameters used in the figure, it yields close to optimal performance for small $\omega$.  This is because it becomes optimal to essentially batch up many requests before allocating a second server (which must periodically be allocated, since without a second server the single server utilization would be 100\% in this scenario), so as to reduce the frequency and therefore overhead cost of server allocation.  At the other end of the spectrum, for large $\omega$ it becomes optimal to simply always keep both servers  allocated. 

\section{Benefits of State-Dependent Routing in a Distributed Service System}\label{sec:routing}

In a system with multiple geographically distributed server sites a routing policy is needed to direct client requests to the most appropriate site.  Two broad classes of such policies are state-oblivious policies that use only information on average request rates, and state-dependent routing policies.  For systems using dynamic server allocation at each site, state-dependent routing might be particularly advantageous since routing could exploit knowledge of the current server allocation state at each site.   On the other hand, dynamic
server allocation allows each site to better manage variations in
load locally, which might reduce the benefits of state-dependent routing.

We study the potential benefits of state-dependent routing through consideration of a system with two sites.  At each site there is a population of clients that are ``local'' to the site, generating requests at some fixed rate.  The request rates at the two sites are such that it may be beneficial to route some or all of the requests generated by the client population local to site 1 to site 2 instead.
We evaluate how much performance improves by making the decision to route remotely state-dependent, rather than state-oblivious, when using optimal policies of each type together with optimal server allocation/deallocation.
\revOne{}{Comparing optimal state-dependent and state-oblivious routing provides more fundamental insight than comparing particular heuristic policies.
However, determining such optimal policies suffers from the well-known “curse of dimensionality”, and therefore our analysis focuses on the two-site case.}

We restrict attention to server allocation policies in which, at each site, at most one server allocation is allowed to be in progress at once.  Servers are assumed to homogeneous with processing rate $\mu$.  In the case of multiple servers at a site, there is assumed to be a shared queue so that when there are $n$ requests at the site and $m$ allocated servers, the total service rate at that site is min[$n,m$]$\mu$.
The overhead associated with processing a request remotely rather than locally is modelled by including a transfer time overhead in the request response time, with average value denoted by $D_R$.  We attribute this overhead to the extra time required to transfer the request response to the client, and assume that the time required to transfer the request itself to the remote site can be neglected.  This assumption is favorable for dynamic routing, since with no delay in transferring a request there can be no change in the states of the sites between when a request transfer decision is made and the request arrives at the remote site.

\subsection{Performance with Optimal Policies }\label{sec:optrouting}

\subsubsection{State-oblivious routing}\label{sec:optstateoblivious}
To determine the performance with optimal state-oblivious routing and optimal dynamic server provisioning, we apply policy iteration as described in Sec.~\ref{sec:optimal} considering only a single site in isolation, for each possible assigned request rate.  We then find the request rate split that yields the minimum value of a combined objective function. This objective function includes the
sum of the values of Expression~(\ref{eq:objfunction}) for the two sites, but also needs to incorporate a term for the transfer delay incurred by requests processed remotely (if any).  Treating this delay in the same way as the servicing delay, this latter term is given by $\omega$ times the rate at which requests are  routed remotely times $D_R$, yielding a combined objective function that can be written, using Little's Law, as $\omega Q + C_1 + C_2$, where $Q$ denotes the average total number of requests at one of the servers or with a response in transit back to the local site, and $C_1$ and $C_2$ denote the cost (as defined in Sec.~\ref{sec:metrics}) incurred at sites 1 and 2 respectively.   In our implementation we consider all possible request rate splits at a granularity of 0.01. 

\subsubsection{State-dependent routing}\label{sec:optstatedependent}
To determine the performance with optimal state-dependent routing and corresponding optimal dynamic server provisioning, we develop another semi-Markov decision model.
The set of system states is $\{ (n_1,m_1,a_1,n_2,m_2,a_2)\}$ such that $n_1, n_2, m_1, m_2 \geq 0$ and $a_1, a_2 \in\{0,1\}$, where $n_i$ gives the number of requests in the system (waiting or in service) at site $i$, $m_i$ gives the number of allocated servers at site $i$ prior to any action taken at the point of entering the state, and $a_i$ is 0 if no server allocation is in progress at site $i$ at the time the state is entered and 1 otherwise.
When applying policy iteration, we cap $n_1+n_2$.  As in Sec.~\ref{sec:performancecopmarisons},  we report results with a cap that is large enough to accommodate those states with non-negligible probability, and verify that larger caps do not yield different results.  The values of $m_1+a_1$ and $m_2+a_2$ are also capped, according to the desired number of potential servers at each site.

Each possible action in a state $(n_1,m_1,a_1,n_2,m_2,a_2)$ includes both a provisioning component and a routing component.
The provisioning action component is carried out at the time the state is entered.
We consider policies with possible provisioning action components as follows\footnote{To reduce the action space, we do not include some provisioning action components that would not be needed in an optimal policy, such as simultaneous server deallocations at both sites.}: (1) $sub$-$actions~IA_1, IA_2, IA_1/IA_2$: initiate allocation of a server at site 1 ($IA_1$, possible when $a_1=0$) or site 2 ($IA_2$, possible when $a_2=0$) or initiate allocations at both sites ($IA_1/IA_2$, possible when $a_1=a_2=0$); (2) $sub$-$actions~CA_1, CA_2, CA_1/CA_2$: cancel the in-progress server allocation at site 1 (possible when $a_1=1$) or site 2 (possible when $a_2=1$) or the in-progress server allocations at both sites (possible when $a_1=a_2=1$); (3) $sub$-$actions~D_1, D_2$: deallocate a server at site 1 (possible when $m_1 > 0$ and $a_1=0$) or site 2 (possible when $m_2 > 0$ and $a_2=0$); (4) $sub$-$actions~IA_1/CA_2, IA_2/CA_1$: initiate a server allocation at site 1 and cancel the in-progress server allocation at site 2 (possible when $a_1 = 0$ and $a_2=1$) or initiate a server allocation at site 2 and cancel the in-progress server allocation at site 1 (possible when $a_2 = 0$ and $a_1=1$); (5) $sub$-$actions~IA_1/D_2, IA_2/D_1$: initiate a server allocation at site 1 and deallocate a server at site 2 (possible when $a_1=a_2=0$ and $m_2 > 0$) or initiate a server allocation at site 2 and deallocate a server at site 1 (possible when $a_1=a_2=0$ and $m_1 > 0$); and (6) $sub$-$action~NC$: no changes.  The possible routing action components for a state are (1) $sub$-$action~RL$: set the policy for request routing when in the state to local routing; (2) $sub$-$action~R12$: set the policy for request routing when in the state to route an arrival at site 1 to site 2, and an arrival at site 2 locally.  Each provisioning action component can be paired with each routing action component, yielding a total of 26 potential actions.

The reward rates are defined such that the undiscounted average reward rate for a particular policy is equal to the negative of the same combined objective function as used for state-oblivious routing; i.e, 
$-(\omega Q + C_1 + C_2)$,
where $Q$ denotes the average total number of requests at one of the servers or with a response in transit back to the local site.  The state transition and reward rates, when no caps are placed on the values of the state variables, are given as follows for actions $IA_1 + RL$ and $IA_1 + R12$.  Transition rates and rewards for the other actions are straightforward to specify in an analogous fashion.  In the below, transition rates and rewards are superscripted by an action name as given by its two sub-action components, except in the case where a transition rate depends only on the provisioning action component of the action, in which case only that component is given.

For actions $IA_1 + RL$ and $IA_1 + R12$ (only possible when $a_1 = 0$):
{\footnotesize\begin{align}
& q^{IA_1+RL}[(n_1,m_1,0,n_2,m_2,a_2), (n_1+1,m_1,1,n_2,m_2,a_2)] \nonumber \\
& = q^{IA_1+R12}[(n_1,m_1,0,n_2,m_2,a_2), (n_1,m_1,1,n_2+1,m_2,a_2)]
= \lambda_1
\nonumber \\
& q^{IA_1}[(n_1,m_1,0,n_2,m_2,a_2), (n_1,m_1,1,n_2+1,m_2,a_2)]
= \lambda_2,
\nonumber \\
& q^{IA_1}[(n_1,m_1,0,n_2,m_2,a_2), (n_1-1,m_1,1,n_2,m_2,a_2)] \nonumber \\
&\hspace{20pt} = \textrm{min}[n_1,m_1]\mu,
\nonumber \\
& q^{IA_1}[(n_1,m_1,0,n_2,m_2,a_2), (n_1,m_1,1,n_2-1,m_2,a_2)] \nonumber \\
&\hspace{20pt} = \textrm{min}[n_2,m_2]\mu,
\nonumber \\
& q^{IA_1}[(n_1,m_1,0,n_2,m_2,a_2), (n_1,m_1+1,0,n_2,m_2,a_2)] \nonumber \\
& = q^{IA_1}[(n_1,m_1,0,n_2,m_2,1), (n_1,m_1,1,n_2,m_2+1,0)] 
= 1/\Delta,
\nonumber \\
& R^{IA_1+RL}[(n_1,m_1,0,n_2,m_2,a_2)] \nonumber \\
&\hspace{20pt} = - (\omega (n_1+n_2) + (m_1+m_2+a_2+1) \mu) \nonumber \\
& R^{IA_1+R12}[(n_1,m_1,0,n_2,m_2,a_2)] \nonumber \\
&\hspace{20pt} = - (\omega (n_1+n_2 + \lambda_1 D_R) + (m_1+m_2+a_2+1) \mu) \nonumber \\
\end{align}
}

\begin{figure*}[t]
\centering
\subfigure[Single each site ($\lambda_1 \! = \! \lambda_2 \! = \! 0.3$)]{\includegraphics[trim = 14mm 24mm 18mm 0mm, clip, width=0.23\textwidth]{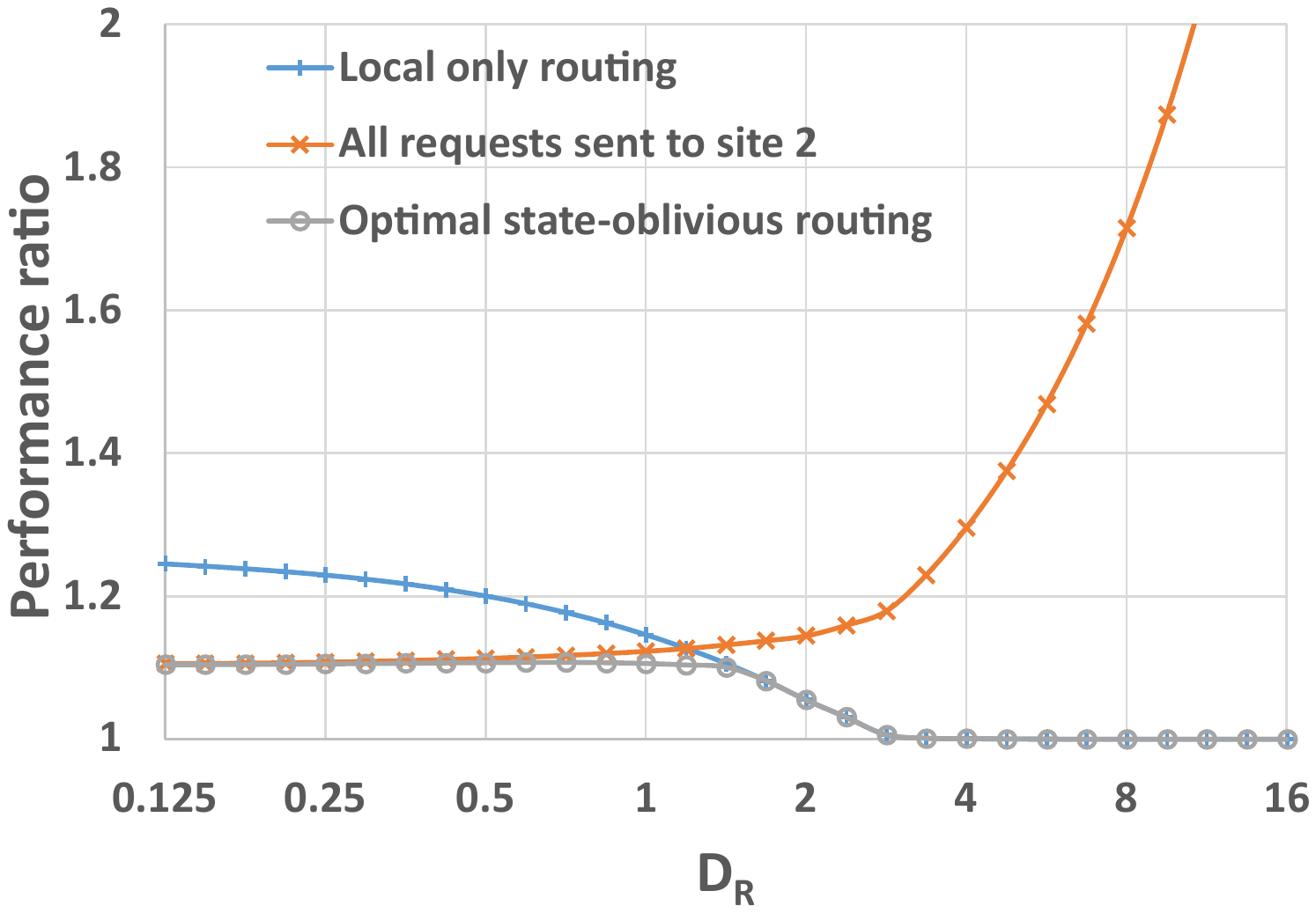}}
\subfigure[Dual each site ($\lambda_1 = \lambda_2 = 0.3$)]{\includegraphics[trim = 14mm 24mm 18mm 0mm, clip, width=0.23\textwidth]{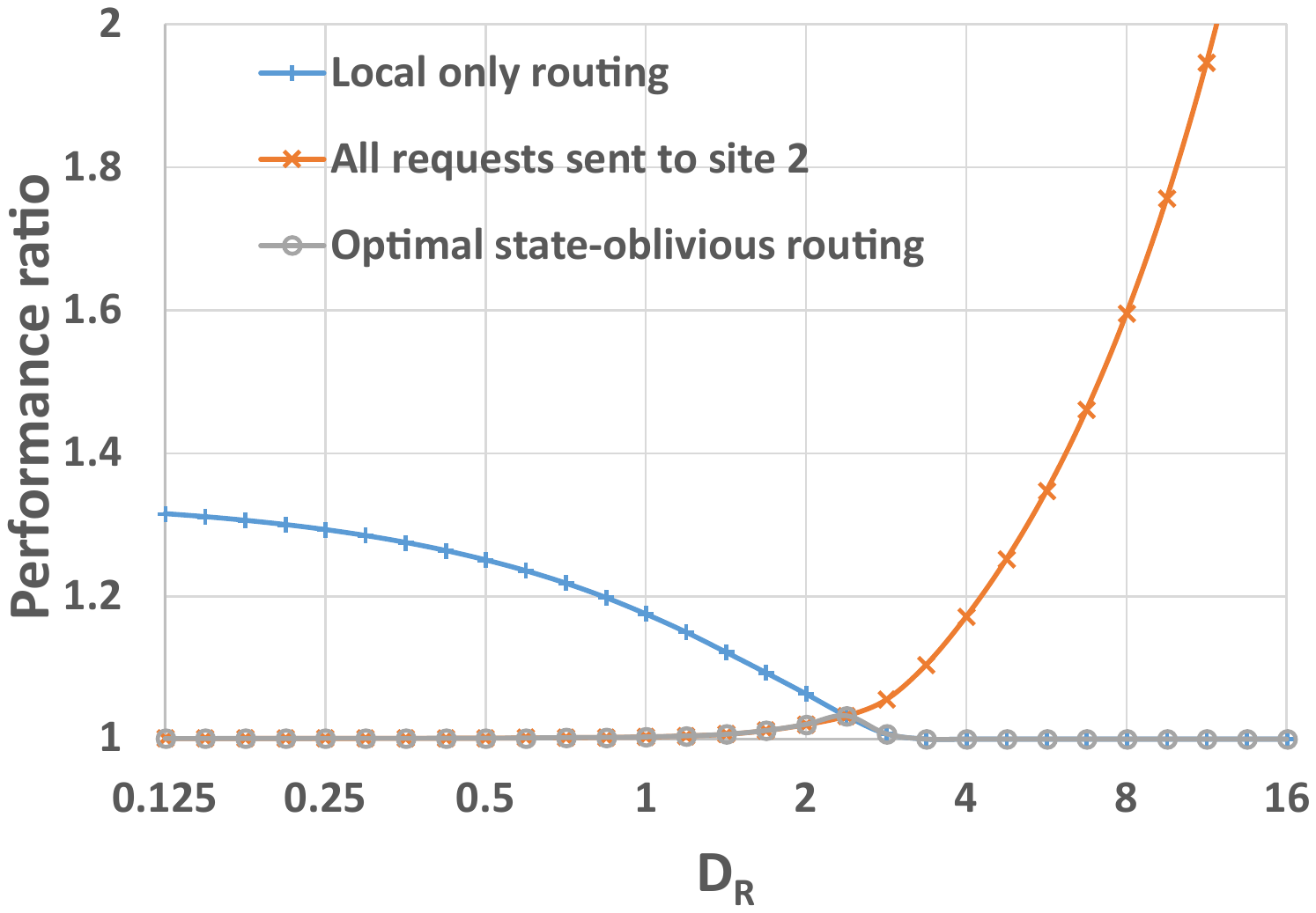}}
\subfigure[Single each ($\lambda_1 \! \! = \! 0.8$,\! $\lambda_2 \! \! = \! 0.04$)]{\includegraphics[trim = 14mm 24mm 18mm 0mm, clip, width=0.23\textwidth]{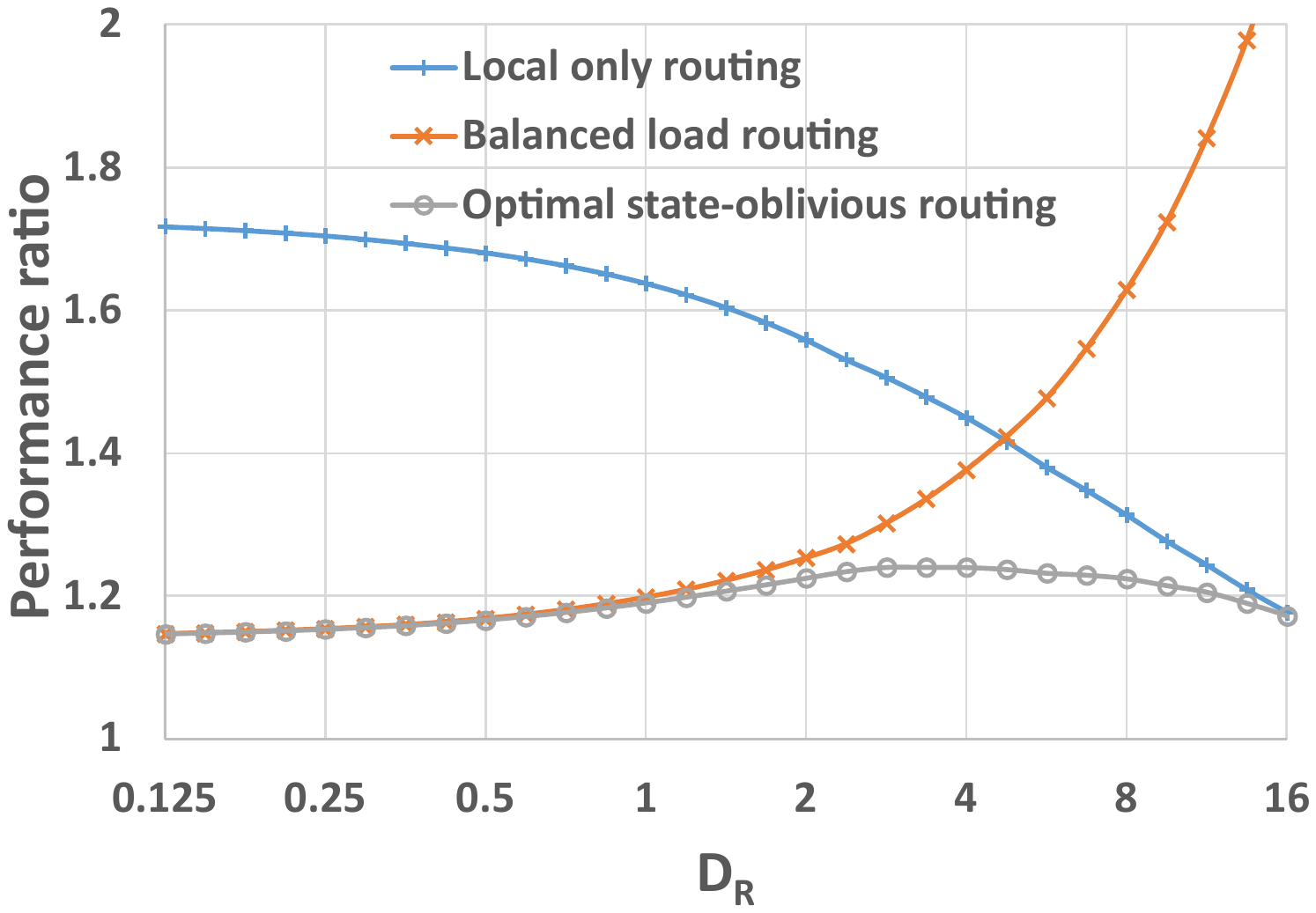}}
\subfigure[Dual each ($\lambda_1 \! = \! 1.7$, $\lambda_2 \! = \! 0.04$)]{\includegraphics[trim = 14mm 24mm 18mm 0mm, clip, width=0.23\textwidth]{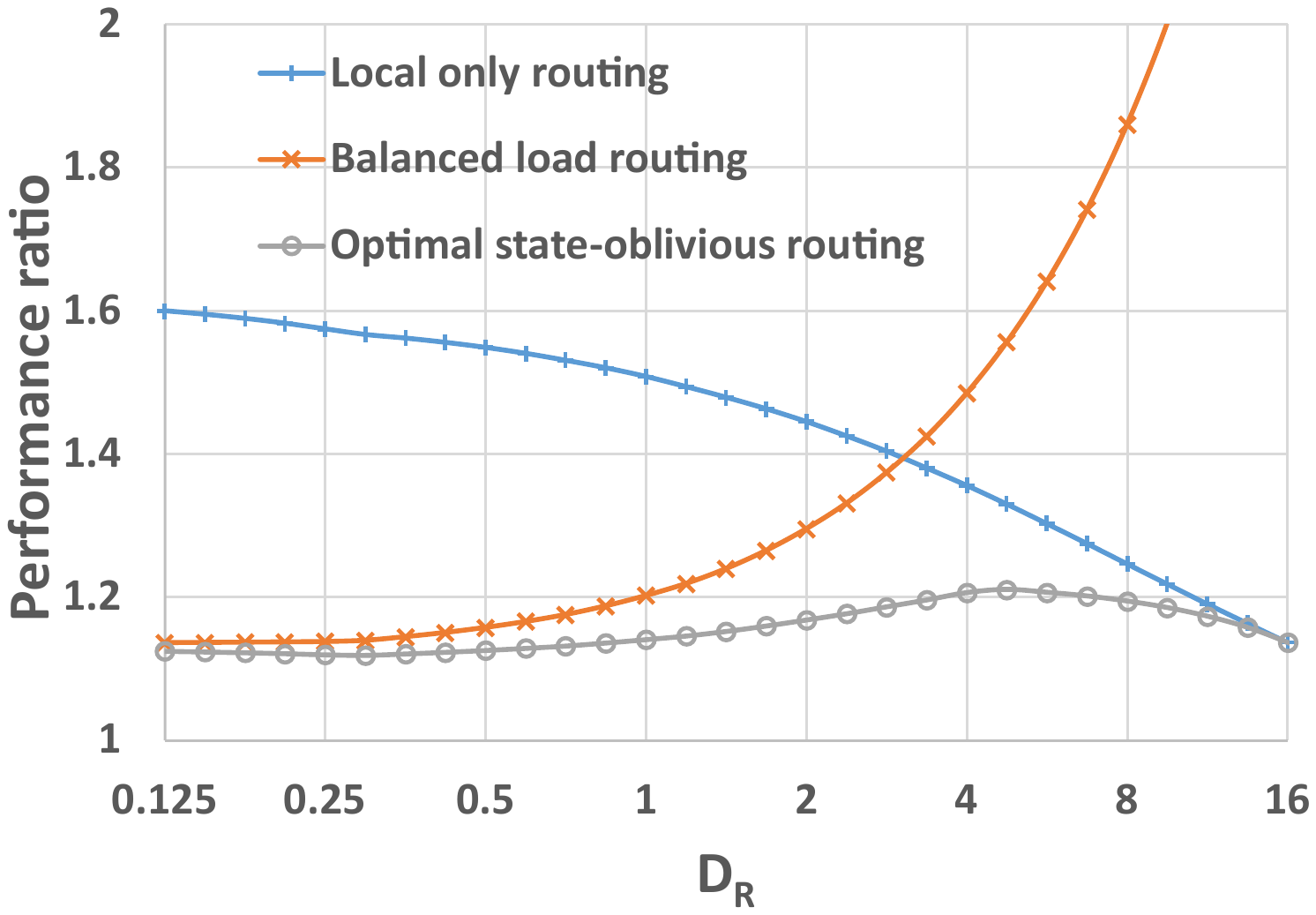}}
\caption{Example performance results for request routing policies ($\mu = 1$, $\omega=1$, $\Delta=2$).}
\label{fig:TwoSiteRouting}
\end{figure*}

\subsection{Example Results}

Numerical results are presented for two example scenarios.  In the first scenario, the request rates at the sites are equal and are low relative to the potential service capacities, and thus it may be advantageous to consolidate allocation of server resources at a single site (i.e. site 2) only.  In the second scenario, site 1 has a high request rate and site 2 has a low request rate, and it may be advantageous to process some of the requests from site 1 at site 2 for the purpose of alleviating the otherwise high load on site 1.  For each scenario, we consider both the case of a single allocatable server at each site and the case of dual potential servers at each site.
    
\subsubsection{Remote Routing for Server Consolidation}\label{sec:serverconsolidation}

Figures~\ref{fig:TwoSiteRouting}(a) (single server can be allocated at each site) and (b) (dual servers can be allocated at each site) show results for a scenario in which it may be advantageous to consolidate allocation of server resources at site 2 only.  The figures show the ratio of the performance with optimal state-oblivious routing to that with optimal state-dependent routing, in both cases with use of optimal dynamic server provisioning, as a function of the average transfer time $D_R$.  The figures also show the performance ratio for the policies of routing all requests locally, and routing all requests to site 2.

Optimal state-oblivious routing is seen to yield performance close to that of optimal state-dependent routing over the full range of average transfer times.  Almost always, the optimal state-oblivious routing is one of the two extremes; i.e., either route all requests locally, or route all requests to site 2.  In the case where dual servers can be allocated at each site (Figure~\ref{fig:TwoSiteRouting}(b)), this is also true for optimal state-dependent routing.  In the case where only a single server can be allocated at each site, however, for low values of average transfer time optimal state-dependent routing sends most but not all requests to site 2.  Not all requests are sent to site 2 because the appreciable total load at site 2 results in occasional queue buildups there, during which times it may be beneficial to route site 1 requests locally.  For lower request rates at each site, however, such as 0.2, the results become very similar in form to those in Figure~\ref{fig:TwoSiteRouting}(b).     

\subsubsection{Remote Routing to Reduce Load on Overloaded Site}\label{sec:overloadedsite}
Figures~\ref{fig:TwoSiteRouting} (c) (single server can be allocated at each site) and (d) (dual servers can be allocated at each site) show results for a scenario in which it may be advantageous to route some of the site 1 requests to site 2 so as to reduce the load on site 1. In addition to the performance ratio for optimal state-oblivious routing, the figures show the performance ratio for the policy of routing all requests locally, and for state-oblivious routing in which load is perfectly balanced across the two sites.

Optimal state-oblivious routing is, again, seen to yield performance close to that of optimal state-dependent routing over the full range of average transfer times, although the performance gap is larger (at most about 24\%) than is seen in Figures~\ref{fig:TwoSiteRouting}(a) and (b).  Also, a performance gap is present over the full range covered in each figure.  Similar results are seen with other choices of imbalanced request rates.  Finally, note that in each figure there is a significant region where the optimal state-oblivious routing policy is intermediate between the extremes of local only routing and balanced load routing.

One might speculate \emph{a priori} that there would be a larger performance gap between optimal state-oblivious and state-dependent routing\revTwo{, given the use of dynamic server allocation/deallocation}{}.  Note, however, that the optimal state-dependent routing policy is typically substantially different from a greedy routing policy, and less sensitive to current state.  For example, even if no server is currently allocated at site 2 and lower request delay could be achieved by processing a new site 1 request locally, it can still be better to route that request to site 2 and initiate the allocation of a server there, owing to the improved performance this may enable for future requests.
\revTwo{}{This reduced sensitivity to current state of optimal state-dependent routing, compared to with a greedy policy, is a consequence of the use of dynamic server allocation, and likely explains the relatively small gap between optimal state-oblivious and state-dependent routing.}

\section{Conclusions}\label{sec:conclusions}
In this paper we describe
policies, develop analytic models, and present performance comparisons and insights into how a service provide can best balance service costs and delays.  
First, we describe
several
simple dynamic server allocation 
policies and develop analytic models for their evaluation.  
Second, semi-Markov
decision models are developed and applied to quantify the
performance gaps between these simple 
policies and (often highly-complex) optimal allocation policies.  
\revTwo{We find that the simple policies we consider can often yield close to optimal
performance.  However, performance gaps widen
as the cost of server allocation increases, increasing the potential benefits of more complex policies.
Such policies could use}{}

\revTwo{}{Our results yield substantial insight into the design of dynamic server allocation policies, a topic of central importance in cloud computing systems.  Key design takeaways include the following:  (1) when the server setup delay is similar to or lower than the request service time, simple server allocation policies can yield close to optimal performance; (2) when the server setup delay is relatively high (e.g., twice the request service time or greater), performance gaps widen, increasing the potential benefit of more complex policies; and (3) in the unlimited server setting, there can be a substantial benefit to allowing multiple server allocations to be in progress at once when using a simple reactive policy, a capability that may not be needed if using a more complex policy.  With respect to the design of more complex policies, if needed, such policies require} 
more cautious rules with respect to initiating server allocations as well as  deallocating servers, through use of non-zero finite "holding-on" / "delayed-off" times and/or significantly higher load thresholds for initiating server allocations versus for server deallocations (e.g., $h$ versus $l$ in the simple dual server model of Figure~\ref{fig:statetransitiondiagramforMM1Scaling}).

Finally, we consider systems with multiple geographically distributed server sites.
\revTwo{}{Request routing in such a context is another central concern in cloud computing.} 
One might speculate that the performance
benefits of state-dependent request routing would increase when using dynamic server allocation, since routing
could exploit knowledge of the state of server allocation (as well as
of request queues) at each site.  We take a first look at this question in the context of a simple scenario with just two sites.  A semi-Markov decision model is developed and applied for the evaluation of optimal state-dependent routing.  Comparing optimal state-dependent and optimal state-oblivious routing, we find only modest performance
gaps.

\section*{Acknowledgements}
This work was supported by the Swedish Foundation for Strategic Research (SSF) and the Swedish Research Council (VR).

\bibliographystyle{IEEEtran}
\bibliography{paper}


\begin{IEEEbiography}[{\includegraphics[width=1in,height=1.55in,clip,keepaspectratio]{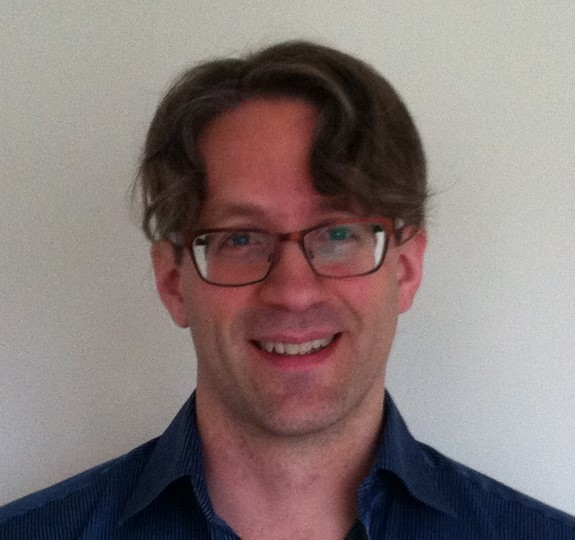}}]{Niklas Carlsson}
  is an Senior Associate Professor at Link\"oping
  University, Sweden. He received his M.Sc. degree in Engineering
  Physics from Ume\aa~University, Sweden, and his Ph.D. in Computer
  Science from the University of Saskatchewan, Canada. He has
  previously worked as a Postdoctoral Fellow at the University
  of Saskatchewan, Canada, and as a Research Associate at the
  University of Calgary, Canada. His research interests are in the
  areas of design, modeling, characterization, performance, and security of distributed systems and networks.
\end{IEEEbiography}

\begin{IEEEbiography}[{\includegraphics[trim=100mm 0mm 140mm 0mm, clip, width=1in, clip, keepaspectratio]{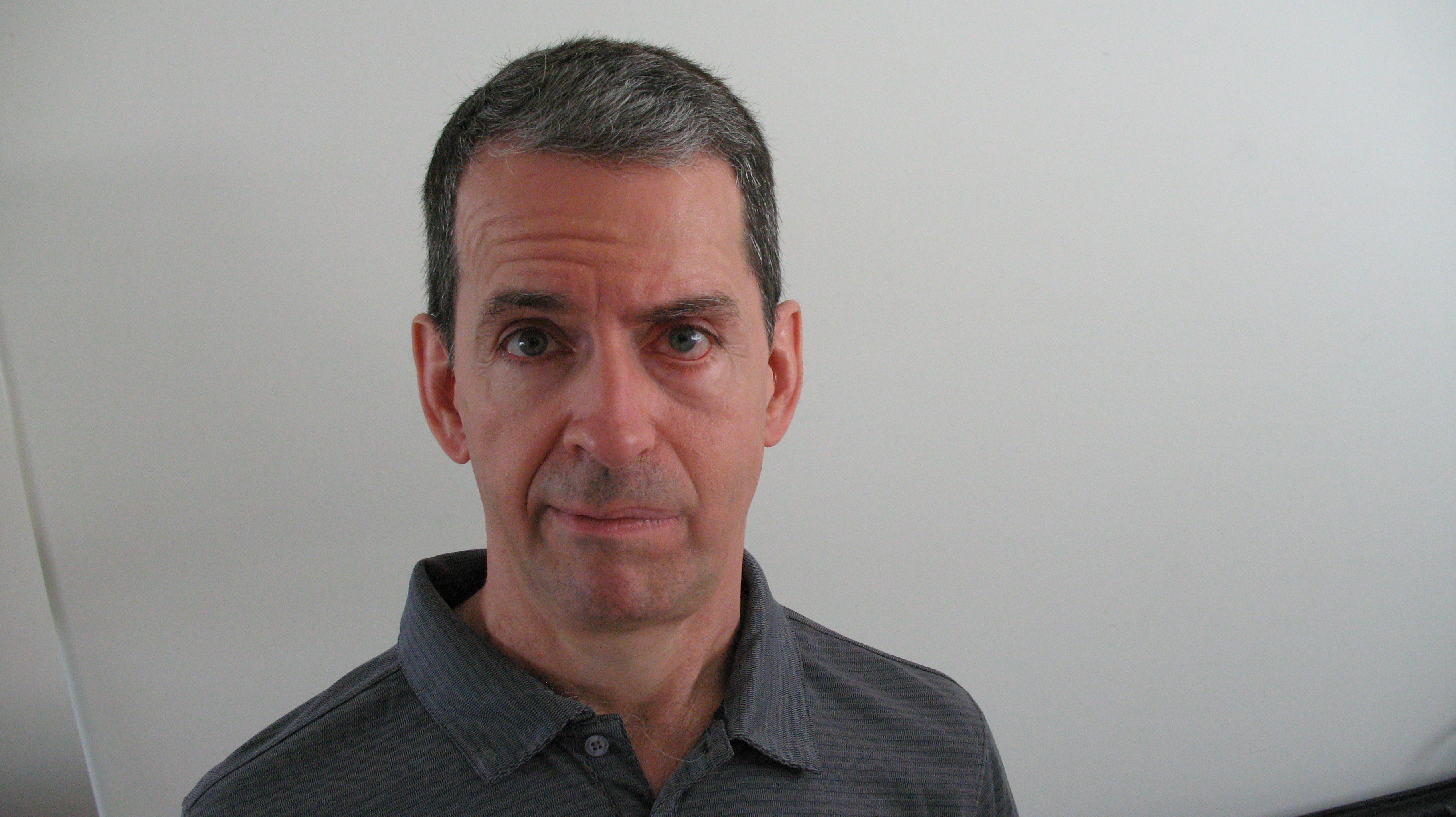}}]{Derek Eager}
  received the BSc degree in computer science from the University of Regina, Canada,
  and the MSc and PhD degrees in computer science from the University of Toronto, Canada.
  He is a professor in the Department of Computer Science, University of Saskatchewan, Canada.
  His research interests include the areas of performance evaluation, content distribution,
  and distributed systems and networks.
\end{IEEEbiography}

\clearpage

\appendix

\subsection{\revTwo{}{Extended Single Server Derivation}}\label{sec:singleserver-appendix}

Denote the steady-state probability of the state with label $s$ by $p_s$.
Assuming $\mu > \lambda$, we must have
\begin{align}\label{eq:util}
    \sum_{i=1}^{\infty} p_{i\textrm{A}} = \frac{\lambda}{\mu} \Leftrightarrow \sum_{j=1}^{k} p_{0\textrm{H}j} + p_{0\textrm{I}} + \sum_{i=1}^{\infty} p_{i\textrm{D}} = 1 - \frac{\lambda}{\mu}.
\end{align}
\revOne{}{The leftmost equality in Equation~(\ref{eq:util}) follows from the fact that the sum of the probabilities of the “A” states gives the fraction of time the server is actively processing requests.  This is the server utilization, which is equal to $\lambda/\mu$.}

We can express the $p_{0\textrm{H}j}$ and $p_{i\textrm{D}}$ probabilities in terms of $p_{0\textrm{I}}$, allowing the solution for $p_{0\textrm{I}}$ using Equation~(\ref{eq:util}).  From flow balance, we have
\begin{align}
p_{0\textrm{H}k} \left( \frac{k}{T} \right) = p_{0\textrm{I}} \lambda
\end{align}
and
\begin{align}
p_{0\textrm{H}j} \left( \frac{k}{T} \right) = p_{0\textrm{H}(j+1)} (\lambda + k/T) \;\;\;\;\;\; 1 \leq j < k,
\end{align}
yielding
\begin{align}\label{eq:p0Hj}
p_{0\textrm{H}j} = p_{0\textrm{I}} \left( \frac{\lambda T}{k} \right) \left( \frac{\lambda + k/T}{k/T} \right)^{k-j} \;\;\;\;\;\; 1 \leq j \leq k.
\end{align}
From Equation~(\ref{eq:p0Hj}) we get:
\begin{align}\label{eq:sump0Hj}
\sum_{j=1}^{k} p_{0\textrm{H}j} 
=  p_{0\textrm{I}} \left( (\lambda T / k + 1 )^k - 1 \right).
\end{align}
\revOne{}{Note that adding $p_{0I}$ to both sides of this equation and then dividing both sides by $p_{0I}$ yields the expected result that the ratio of the probability of being in a state with zero requests to the probability of being in state $0I$, is equal to the probability that no arrival occurs during a holding-on period.}
\revOne{We also have}{We also have, from flow balance,}
\begin{align}\label{eq:p1D}
p_{1\textrm{D}} ( \lambda + 1/\Delta ) = p_{0\textrm{I}} \lambda
\end{align}
and
\begin{align}
p_{(i+1)\textrm{D}} ( \lambda + 1/\Delta ) = p_{i\textrm{D}} \lambda  \;\;\;\;\;\; i \geq 1,
\end{align}
yielding
\begin{align}\label{eq:piDoffwhenidle}
p_{i\textrm{D}} = p_{0\textrm{I}} \left( \frac{\lambda}{\lambda + 1/\Delta} \right)^i \;\;\;\;\;\; i \geq 1.
\end{align}
From Equation~(\ref{eq:piDoffwhenidle}) we get:
\begin{align}\label{eq:sumpiD}
\sum_{i=1}^{\infty} p_{i\textrm{D}} =
p_{0\textrm{I}} \frac{\lambda / (\lambda + 1/\Delta)}{1 - \lambda / (\lambda + 1/\Delta)} = p_{0\textrm{I}}\lambda \Delta.
\end{align}
\revOne{}{Note that this equation yields the expected result that the ratio of the probability of being in a state in which the server is in setup delay to the probability of being in state $0I$, is equal to the ratio of the mean setup delay ($\Delta$) to the mean interarrival time ($1/\lambda$).}
Using Equations~(\ref{eq:sump0Hj}) and~(\ref{eq:sumpiD}) to substitute into~(\ref{eq:util}) gives:
\begin{align}\label{eq:p0I}
p_{0\textrm{I}} = \frac{1 - \lambda/\mu}{(\lambda T / k + 1 )^k + \lambda \Delta}. 
\end{align}

Denote by $p_n$ ($n \geq 1$) the steady-state probability of $n$ client requests being present at the server, i.e. the sum of $p_{n\textrm{A}}$ and $p_{n\textrm{D}}$.  From flow balance,
\begin{align}\label{eq:p1fb}
\mu (p_1 - p_{1\textrm{D}}) = \lambda \left( \sum_{j=1}^{k} p_{0\textrm{H}j} + p_{0I} \right).
\end{align}
Substitution from~(\ref{eq:sump0Hj}),~(\ref{eq:piDoffwhenidle}), and~(\ref{eq:p0I}) gives
\begin{align}\label{eq:p1}
p_1 =  \left( \frac{\lambda}{\mu} (\lambda T / k + 1 )^k + \frac{\lambda}{\lambda + 1/\Delta} \right) \left(\frac{1 - \lambda/\mu}{(\lambda T / k + 1 )^k + \lambda \Delta}\right).
\end{align}
Again applying flow balance,
\begin{align}\label{eq:pifb}
\mu (p_i - p_{i\textrm{D}}) = \lambda p_{i-1} \;\;\;\;\;\; i \geq 2,
\end{align}
yielding, for all $i \geq 1$,
\begin{align}\label{eq:pi}
p_i =  \left( \left( \frac{\lambda}{\mu} \right)^i (\frac{\lambda T}{k} + 1 )^k + \sum_{l=1}^{i} \left( \frac{\lambda}{\lambda + 1/\Delta} \right)^l \left( \frac{\lambda}{\mu}\right)^{i-l}  \right) & \times \nonumber \\ \left(\frac{1 - \lambda/\mu}{(\frac{\lambda T}{k} + 1 )^k + \lambda \Delta}\right) &.
\end{align}

Considering now the mean number of requests in the system $\sum_{i=1}^{\infty} i p_i$, where $p_i$ is given by Equation~(\ref{eq:pi}), note that
\begin{align}\label{eq:sumsoln}
\sum_{i=1}^{\infty} i (\lambda / \mu)^i = \frac{\lambda / \mu}{(1 - \lambda / \mu)^2}
\end{align}
and
{\footnotesize \begin{align}\label{eq:regroupingeq}
\sum_{i=1}^{\infty} i \sum_{l=1}^{i} \left( \frac{\lambda}{\lambda + 1/\Delta} \right)^l \left( \frac{\lambda}{\mu}\right)^{i-l} & = \sum_{i=1}^{\infty}  \left( \frac{\lambda}{\lambda + 1/\Delta} \right)^i \sum_{l=0}^{\infty} (l+i) \left( \frac{\lambda}{\mu} \right)^l \nonumber \\
& = \frac{\lambda / \mu}{(1 - \lambda / \mu)^2} (\lambda \Delta) + \frac{\lambda \Delta (1 + \lambda \Delta)}{1 - \lambda/\mu}.
\end{align}
}
Note that in the first line of (\ref{eq:regroupingeq}), the original double summation is rewritten to group together all of the resulting terms that include the same power of $\lambda / (\lambda + 1/\Delta)$ as one of the factors. Applying~(\ref{eq:sumsoln}) and~(\ref{eq:regroupingeq}) with~(\ref{eq:pi}), the mean number of requests in the system is given by
{\footnotesize\begin{align}
\sum_{i=1}^{\infty} i p_i & = \left( \frac{\lambda / \mu}{(1 - \lambda / \mu)^2} \left( (\frac{\lambda T}{k} + 1 )^k + \lambda \Delta \right) + \frac{\lambda \Delta (1 + \lambda \Delta)}{1 - \lambda/\mu} \right) \times \nonumber \\
& \;\;\;\;\;\; \;\;\;\;\;\; \;\;\;\;\;\; \;\;\;\;\;\; \;\;\;\;\;\; \;\;\;\;\;\; \;\;\;\;\;\; \;\;\;\;\;\; \;\;\;\;
\left(\frac{1 - \lambda/\mu}{(\frac{\lambda T}{k} + 1 )^k + \lambda \Delta}\right) \nonumber \\
& = \frac{\lambda / \mu}{1 -\lambda / \mu} + \frac{\lambda \Delta (1 + \lambda \Delta)} {(\frac{\lambda T}{k} + 1 )^k + \lambda \Delta}.
\end{align}
}
From Little's Law, the mean request response time $R$ is given by
\revOne{}{the mean number of requests in the system divided by $\lambda$:} 
\begin{align}
R = \frac{1/\mu}{1 -\lambda /\mu} + \frac{\Delta (1 + \lambda \Delta)} {(\frac{\lambda T}{k} + 1 )^k + \lambda \Delta}.
\end{align}
The cost $C$ is given by $\mu$ times the probability that the server is active or in setup or holding-on delay, i.e., by $\mu (1-p_{0I})$, yielding
\begin{align}
C = \mu  - \frac{\mu -\lambda }{(\lambda T/k + 1)^k + \lambda  \Delta} .
\end{align}


\subsection{Single Server Model with State-dependent Service Rates}\label{sec:MMmun}
\begin{figure}[t]
\centering
\includegraphics[width=0.46\textwidth]{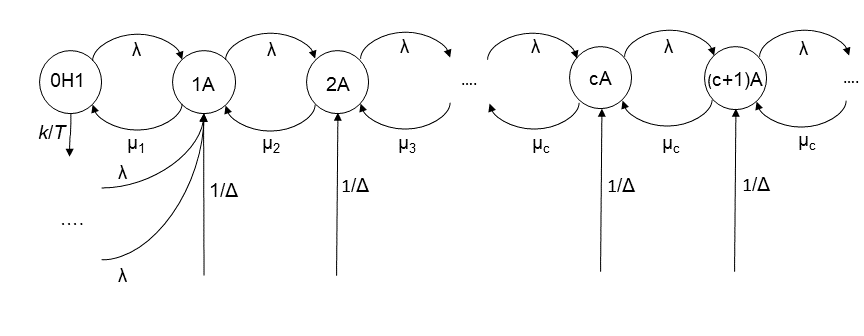}
\vspace{-14pt}
\caption{Top Row of State-Transition Diagram for Single Server Allocation/Deallocation Policy with State-dependent Service Rates.}
\label{fig:statetransitiondiagramformmun}
\vspace{-2pt}
\end{figure}

The state-transition diagram for this system is the same as that in 
\hardcoded{Figure~\ref{fig:mm1statetransitiondiagram}}{Figure 1 (in paper)} 
excepting for a modified top row as shown in 
\hardcoded{Figure~\ref{fig:statetransitiondiagramformmun}.}{Figure 1 (here).} 
As shown in the figure, the transition rate $u_i$ from state $i$A to state $(i-1)$A (or to state 0H1 for $i=1$) is dependent on $i$ for $1 \leq i < c$.  For $i \geq c$ the transition rate is constant and is equal to $u_c$.

As in the case of state-independent service rate, state probabilities can be expressed in terms of $p_{0\textrm{I}}$.  However, 
\hardcoded{Equation~(\ref{eq:util})}{Equation~(2 in paper)} 
does not hold in general, and we will solve for $p_{0\textrm{I}}$ using the constraint that the state probabilities must sum to one.  
\hardcoded{Equations~(\ref{eq:p0Hj}),~(\ref{eq:sump0Hj}), and~(\ref{eq:piDoffwhenidle})}{Equations~(5 in paper),~(6 in paper), and~(33 in paper)} 
also hold here, since the corresponding portion of the state-transition diagram is identical.  
\hardcoded{Equation~(\ref{eq:p1fb})}{Equation~(12 in paper)} 
holds here when $\mu$ in this equation is replaced by $\mu_1$, as does 
\hardcoded{Equation~(\ref{eq:pifb})}{Equation~(14 in paper)} 
when $\mu$ is replaced by $\mu_i$ for $2 \leq i \leq c$ and by $\mu_c$ for $i > c$.  In place of 
\hardcoded{Equation~(\ref{eq:p1})}{Equation~(13 in paper)} 
we get
\begin{align}\label{eq:unp1}
p_1 =  p_{0\textrm{I}}\left( \frac{\lambda}{\mu_1} (\lambda T / k + 1 )^k + \frac{\lambda}{\lambda + 1/\Delta} \right).
\end{align}
In place of 
\hardcoded{Equation~(\ref{eq:pi})}{Equation~(15 in paper)} 
we get
{\footnotesize\begin{align}\label{eq:unpi}
p_i =  p_{0\textrm{I}} \left(  \frac{\lambda^i}{\prod_{l=1}^i \mu_l} (\frac{\lambda T}{k} + 1 )^k + \sum_{l=1}^{i} \left( \frac{\lambda}{\lambda + 1/\Delta} \right)^l \left( \frac{\lambda^{i-l}}{\prod_{m=l+1}^{i}\mu_m}\right)  \right)  \nonumber \\
1 \leq i < c
\end{align}
}
and
{\footnotesize\begin{align}\label{eq:unpi2}
p_i =  p_{0\textrm{I}} & \left( \frac{\lambda^{c-1}}{\prod_{l=1}^{c-1} \mu_l} \left(\frac{\lambda}{\mu_c}\right)^{i-c+1} (\frac{\lambda T}{k} + 1 )^k  + \right. \nonumber \\
& \sum_{l=1}^{c-1} \left( \frac{\lambda}{\lambda + 1/\Delta} \right)^l \left( \frac{\lambda^{i-l}}{\mu_c^{i-c+1} \prod_{m=l+1}^{c-1}\mu_m}\right) + \nonumber \\
& \left. \sum_{l=c}^{i} \left( \frac{\lambda}{\lambda + 1/\Delta} \right)^l \left( \frac{\lambda}{\mu_c}\right)^{i-l} \right) \;\;\;\;\;\; i \geq c.
\end{align}
}

Considering now the sum over all $i \geq 1$ of $p_i$, note that
\begin{align}\label{eq:sumsoln2}
\sum_{i=1}^{c-1} \left( \frac{\lambda^i}{\prod_{l=1}^i \mu_l} \right) +
\sum_{i=c}^{\infty} \left( \frac{\lambda^{c-1}}{\prod_{l=1}^{c-1} \mu_l} \right) \left(\frac{\lambda}{\mu_c}\right)^{i-c+1} \nonumber \\
= \sum_{i=1}^{c-1} \left( \frac{\lambda^i}{\prod_{l=1}^i \mu_l} \right) +  \frac{\lambda^{c}}{(\prod_{l=1}^{c} \mu_l) (1 -\lambda/\mu_c)}
\end{align}
and
{\footnotesize\begin{align}\label{eq:regroupingeq2}
& \sum_{i=1}^{c-1} \sum_{l=1}^{i} \left( \frac{\lambda}{\lambda + \frac{1}{\Delta}} \right)^l \left( \frac{\lambda^{i-l}}{\prod_{m=l+1}^{i}\mu_m}\right) + \nonumber \\
& \sum_{i=c}^{\infty} \left( \sum_{l=1}^{c-1} \left( \frac{\lambda}{\lambda + \frac{1}{\Delta}} \right)^l \left( \frac{\lambda^{i-l}}{\mu_c^{i-c+1} \prod_{m=l+1}^{c-1}\mu_m}\right) + \right. \nonumber \\
& \left.
\sum_{l=c}^{i} \left( \frac{\lambda}{\lambda + \frac{1}{\Delta}} \right)^l \! \! \left( \frac{\lambda}{\mu_c}\right)^{i-l} \right) = \sum_{i=1}^{c-1}  \left( \frac{\lambda}{\lambda + \frac{1}{\Delta}} \! \!  \right)^i \left( \sum_{l=0}^{c-1-i}  \frac{\lambda^l}{\prod_{m=i+1}^{i+l}\mu_m} + \right. \nonumber \\
& \left. \frac{\lambda^{c-1-i}}{\prod_{m=i+1}^{c-1}\mu_m} \sum_{l=c-i}^{\infty} \left( \frac{\lambda}{\mu_c} \right)^{l-(c-1-i)}\right) + \sum_{i=c}^{\infty}  \left( \frac{\lambda}{\lambda + \frac{1}{\Delta}} \right)^i \sum_{l=0}^{\infty} \left( \frac{\lambda}{\mu_c} \right)^l \nonumber \\
& = \sum_{i=1}^{c-1} \! \left( \frac{\lambda}{\lambda + \frac{1}{\Delta}} \right)^i \! \! \left( \sum_{l=0}^{c-1-i}  \frac{\lambda^l}{\prod_{m=i+1}^{i+l}\mu_m} + \frac{\lambda^{c-i}}{(\prod_{m=i+1}^{c}\mu_m) (1 -\lambda/\mu_c)} \! \right) \nonumber \\
& + \left( \frac{\lambda}{\lambda + \frac{1}{\Delta}} \right)^c \frac{\lambda \Delta + 1}{1 - \lambda/\mu_c}.
\end{align}
}
As with the first line of 
\hardcoded{(\ref{eq:regroupingeq}),}{(17 in paper),} 
the original double summations in the first two lines of (\ref{eq:regroupingeq2}) are rewritten to group together all of the resulting terms that include the same power of $\lambda / (\lambda + 1/\Delta)$ as one of the factors.  Applying~(\ref{eq:sumsoln2}) and~(\ref{eq:regroupingeq2}) with~(\ref{eq:unpi}) and~(\ref{eq:unpi2}) yields
{\footnotesize\begin{align}\label{eq:unsumpi}
& \sum_{i=1}^{\infty} p_i = p_{0\textrm{I}} \left( \left( \sum_{i=1}^{c-1} \left( \frac{\lambda^i}{\prod_{l=1}^i \mu_l} \right) +  \frac{\lambda^{c}}{(\prod_{l=1}^{c} \mu_l) (1 -\lambda/\mu_c)} \right) (\frac{\lambda T}{k} + 1 )^k \right. \nonumber \\
& + \sum_{i=1}^{c-1}  \left( \frac{\lambda}{\lambda + \frac{1}{\Delta}} \right)^i \! \! \left( \sum_{l=0}^{c-1-i}  \frac{\lambda^l}{\prod_{m=i+1}^{i+l}\mu_m} + \frac{\lambda^{c-i}}{(\prod_{m=i+1}^{c}\mu_m) (1 -\lambda/\mu_c)}\right) \nonumber \\
& + \left. \left( \frac{\lambda}{\lambda + \frac{1}{\Delta}} \right)^c \frac{\lambda \Delta + 1}{1 - \lambda/\mu_c} \right).
\end{align}
}
Applying now the constraint that
\begin{align}
p_{0\textrm{I}} + \sum_{i=1}^{\infty}  p_i + \sum_{j=1}^{k} p_{0\textrm{H}j} = 1,
\end{align} 
\hardcoded{Equations~(\ref{eq:unsumpi}) and~(\ref{eq:sump0Hj})}{Equation~(\ref{eq:unsumpi}) and Equation (6 in paper)} 
yield
{\footnotesize\begin{align}\label{eq:unp0I}
& p_{0\textrm{I}} = 1 /
\left( \left( \sum_{i=1}^{c-1} \left( \frac{\lambda^i}{\prod_{l=1}^i \mu_l} \right) +  \frac{\lambda^{c}}{(\prod_{l=1}^{c} \mu_l) (1 -\lambda/\mu_c)} + 1 \right) (\frac{\lambda T}{k} + 1 )^k \right. \nonumber \\
& + \sum_{i=1}^{c-1}  \left( \frac{\lambda}{\lambda + \frac{1}{\Delta}} \right)^i \! \! \left( \sum_{l=0}^{c-1-i}  \frac{\lambda^l}{\prod_{m=i+1}^{i+l}\mu_m} + \frac{\lambda^{c-i}}{(\prod_{m=i+1}^{c}\mu_m) (1 -\lambda/\mu_c)}\right) \nonumber \\
& + \left. \left( \frac{\lambda}{\lambda + \frac{1}{\Delta}} \right)^c \frac{\lambda \Delta + 1}{1 - \lambda/\mu_c} \right).
\end{align}
}
Considering now the mean number of requests in the system $\sum_{i=1}^{\infty} i p_i$, where $p_i$ is given by Equations~(\ref{eq:unpi}) and~(\ref{eq:unpi2}), note that
{\footnotesize\begin{align}\label{eq:unsumsoln}
& \sum_{i=1}^{c-1} i \left( \frac{\lambda^i}{\prod_{l=1}^i \mu_l} \right) +
\sum_{i=c}^{\infty} i \left( \frac{\lambda^{c-1}}{\prod_{l=1}^{c-1} \mu_l} \right) \left(\frac{\lambda}{\mu_c}\right)^{i-c+1} \nonumber \\
& = \sum_{i=1}^{c-1} i \left( \frac{\lambda^i}{\prod_{l=1}^i \mu_l} \right) +
\left( \frac{\lambda^{c-1}}{\prod_{l=1}^{c-1} \mu_l} \right) \left( \sum_{i=c}^{\infty} (i-c+1)  \left(\frac{\lambda}{\mu_c}\right)^{i-c+1} \right. \nonumber \\
& + \left. (c-1) \sum_{i=c}^{\infty}  \left(\frac{\lambda}{\mu_c}\right)^{i-c+1} \right) \nonumber \\
& = \sum_{i=1}^{c-1} i \left( \frac{\lambda^i}{\prod_{l=1}^i \mu_l} \right) +  \left( \frac{\lambda^{c}}{\prod_{l=1}^{c} \mu_l} \right) \left( \frac{1}{(1 - \lambda/\mu_c)^2} + \frac{(c - 1)} {1 -\lambda/\mu_c} \right) 
\end{align}
}
and
{\footnotesize\begin{align}\label{eq:unregroupingeq}
& \sum_{i=1}^{c-1} i \sum_{l=1}^{i} \left( \frac{\lambda}{\lambda + \frac{1}{\Delta}} \right)^l \left( \frac{\lambda^{i-l}}{\prod_{m=l+1}^{i}\mu_m}\right) +  \sum_{i=c}^{\infty} \left( i \sum_{l=1}^{c-1} \left( \frac{\lambda}{\lambda + \frac{1}{\Delta}} \right)^l \times \right. \nonumber \\
& \left. \left( \frac{\lambda^{i-l}}{\mu_c^{i-c+1} \prod_{m=l+1}^{c-1}\mu_m}\right) + i
\sum_{l=c}^{i} \left( \frac{\lambda}{\lambda + \frac{1}{\Delta}} \right)^l \left( \frac{\lambda}{\mu_c}\right)^{i-l} \right) \nonumber \\
& = \sum_{i=1}^{c-1}  \left( \frac{\lambda}{\lambda + \frac{1}{\Delta}} \right)^i \left( \sum_{l=0}^{c-1-i} (i+l) \frac{\lambda^l}{\prod_{m=i+1}^{i+l}\mu_m} + \right. \nonumber \\
& \left. \frac{\lambda^{c-1-i}}{\prod_{m=i+1}^{c-1}\mu_m} \sum_{l=c-i}^{\infty} (i+l) \left( \frac{\lambda}{\mu_c} \right)^{l-(c-1-i)}\right) + \nonumber \\
& \sum_{i=c}^{\infty}  \left( \frac{\lambda}{\lambda + \frac{1}{\Delta}} \right)^i \sum_{l=0}^{\infty} (i+l) \left( \frac{\lambda}{\mu_c} \right)^l \nonumber \\
& = \sum_{i=1}^{c-1}  \left( \frac{\lambda}{\lambda + \frac{1}{\Delta}} \right)^i \left( \sum_{l=0}^{c-1-i} (i+l) \frac{\lambda^l}{\prod_{m=i+1}^{i+l}\mu_m} + \right. \nonumber \\
& \left. \frac{\lambda^{c-i}}{\prod_{m=i+1}^{c}\mu_m} \left(  \frac{1}{(1- \lambda/ \mu_c)^2} + \frac{c - 1}{(1- \lambda/ \mu_c)} \right) \right) + \nonumber \\
& \frac{(\lambda \Delta)^c}{(\lambda \Delta +1)^{c-1}} \left( \frac{\lambda / \mu_c}{(1 - \lambda / \mu_c)^2} + \frac{\lambda \Delta + c}{1 - \lambda / \mu_c} \right).
\end{align}
}
Applying~(\ref{eq:unsumsoln}) and~(\ref{eq:unregroupingeq}) with~(\ref{eq:unpi}) and~(\ref{eq:unpi2}), the mean number of requests in the system is given by
{\footnotesize\begin{align}
\sum_{i=1}^{\infty} i p_i & = p_{0\textrm{I}} \! \left((\frac{\lambda T}{k} + 1 )^k \! \left(\sum_{i=1}^{c-1} i \left( \frac{\lambda^i}{\prod_{l=1}^i \mu_l} \right)\! \! + \! \!\left( \frac{\lambda^{c}}{\prod_{l=1}^{c} \mu_l} \right) \! \!  \left( \frac{1}{(1 - \frac{\lambda}{\mu_c})^2} \right. \right. \right. \nonumber \\
& \left. \left. \left.  + \frac{(c - 1)} {1 -\lambda/\mu_c} \right) \right) + \sum_{i=1}^{c-1}  \left( \frac{\lambda}{\lambda + \frac{1}{\Delta}} \right)^i \! \! \left( \sum_{l=0}^{c-1-i} (i+l) \frac{\lambda^l}{\prod_{m=i+1}^{i+l}\mu_m} + \right. \right. \nonumber \\
& \left. \left. \frac{\lambda^{c-i}}{\prod_{m=i+1}^{c}\mu_m} \left(  \frac{1}{(1- \lambda/ \mu_c)^2} + \frac{c - 1}{1- \lambda/ \mu_c} \right) \right) \right. + \nonumber \\
& \left. \frac{(\lambda \Delta)^c}{(\lambda \Delta +1)^{c-1}} \left( \frac{\lambda / \mu_c}{(1 - \lambda / \mu_c)^2} + \frac{\lambda \Delta + c}{1 - \lambda / \mu_c} \right) \right)
\end{align}
}
where $p_{0\textrm{I}}$ is given by Equation~(\ref{eq:unp0I}).  From Little’s Law, division by $\lambda$ yields the mean request response time. In the case that the service provider is charged according to the highest service rate, the cost $C$ would be given by $\mu_c$ times the probability that the server is active or in setup or holding-on delay, i.e., by $\mu_c (1-p_{0I})$.


\subsection{Dual Servers, One Server Always Allocated, $1 \leq l \leq h$}\label{sec:dualoption2}

As for the case of $l = h$, the case of $1 \leq l \leq h$ can be analyzed by expressing all state probabilities in terms of $p_{h\textrm{B+}}$, and then solving for $p_{h\textrm{B+}}$ using the constraint that the state probabilities must sum to one.

As the corresponding portion of the state-transition diagram (refer to Figure~\ref{fig:statetransitiondiagramforMM1Scaling}) is unchanged, Equations~(\ref{eq:piD}),  (\ref{eq:sumofiD}), and~(\ref{eq:sumofipiD}) hold in this case as well.  Consider the state probabilities $p_{i\textrm{B+}}$ for $l \leq i \leq h$.  For $l+2 \leq i \leq h$  the flow balance Equation~(\ref{eq:scalingflowbalance}) is satisfied, and therefore, similarly as in Equation~(\ref{eq:solutionform}), we have 
\begin{align}\label{eq:solutionform2}
p_{(l+k)\textrm{B+}} = \alpha r_1^k + \beta r_2^k \;\;\;\;\;\; 0 \leq k \leq h - l,
\end{align}
where $\alpha$ and $\beta$ are independent of $k$, and 
$r_1$ and $r_2$ are given in~(\ref{eq:r1andr2}).
We have
\begin{align}
\alpha + \beta = p_{l\textrm{B+}}
\end{align}
and
\begin{align}
\alpha r_1 + \beta r_2 = p_{(l+1)\textrm{B+}} = p_{l\textrm{B+}} \frac{\lambda + \mu_1 + 1/\Delta}{\mu_1},
\end{align}
where the last equality follows from
flow balance.
Solving for $\alpha$ and $\beta$ yields
\begin{align}
\alpha & = p_{l\textrm{B+}} \frac{r_2 - (\lambda + \mu_1 + 1/\Delta)/\mu_1}{r_2 - r_1} = - p_{l\textrm{B+}} \frac{r_1}{r_2 - r_1}, \nonumber \\
\beta & = p_{l\textrm{B+}} \frac{(\lambda + \mu_1 + 1/\Delta)/\mu_1 - r_1}{r_2 - r_1} = p_{l\textrm{B+}} \frac{r_2}{r_2 - r_1} ,
\end{align}
where the last equalities follow from Equation~(\ref{eq:r1}) and the analogous equation for $r_2$, and since $r_1 r_2 = \lambda / \mu_1$.  Finally, applying Equation~(\ref{eq:solutionform2}) for $k=h-l$, we find that $p_{l\textrm{B+}}$ is given in terms of $p_{h\textrm{B+}}$ by
\begin{align}\label{eq:plD}
p_{l\textrm{B+}} = p_{h\textrm{B+}}\frac{ r_2 - r_1}{r_2^{h-l+1} - r_1^{h-l+1}  }.
\end{align}
This yields
\begin{align}\label{eq:piD2}
p_{i\textrm{B+}} = p_{h\textrm{B+}} \left(\frac{r_2^{i-l+1} - r_1^{i-l+1}}{r_2^{h-l+1} - r_1^{h-l+1}} \right) \;\;\; l \leq i \leq h.
\end{align}
Note that
\begin{align}\label{eq:sumofiD2}
\sum_{i=l}^{h-1} p_{i\textrm{B+}} = p_{h\textrm{B+}} \left( \frac{r_2\left(\frac{1 - r_2^{h-l}}{1 - r_2} \right) - r_1 \left(\frac{1 - r_1^{h-l}}{1 - r_1} \right)}{r_2^{h-l+1} - r_1^{h-l+1}} \right).
\end{align}

Consider now the state probabilities $p_{i\textrm{E}}$ for $l \leq i \leq h$.  The state probability $p_{l\textrm{E}}$ satisfies the flow balance equation
\begin{align}
p_{l\textrm{E}} \mu_2 = \left( \sum_{i=l}^{h-1} p_{i\textrm{B+}} + \sum_{i=h}^{\infty} p_{i\textrm{B+}} \right) (1/\Delta).
\end{align}
Applying Equations~(\ref{eq:sumofiD}) and~(\ref{eq:sumofiD2}) yields
{\footnotesize\begin{align}\label{eq:plE}
p_{l\textrm{E}} & = p_{h\textrm{B+}} \left( \frac{r_2\left(\frac{1 - r_2^{h-l}}{1 - r_2} \right) - r_1 \left(\frac{1 - r_1^{h-l}}{1 - r_1} \right)}{r_2^{h-l+1} - r_1^{h-l+1}} + 
\frac{1}{1 - r_1} \right)\left( \frac{1}{\Delta\mu_2} \right) \nonumber \\
& = p_{h\textrm{B+}} \left( \frac{(r_2 - r_1) (r_2^{h-l+1} - 1)}{r_2^{h-l+1} - r_1^{h-l+1}} \right)
\left( \frac{\mu_1}{\mu_2} \right),
\end{align}
}
where the last equality is obtained using $(1-r_1)(r_2-1) = 1/(\Delta\mu_1)$.  In the case of $l=h$, Equation~(\ref{eq:plE}) can be seen to match Equation~(\ref{eq:piE}), using the fact that $r_1r_2 = \lambda/\mu_1$.

Each state probability $p_{i\textrm{E}}$, $l < i \leq h$, satisfies the flow balance equation
{\footnotesize\begin{align}
& p_{i\textrm{E}} \mu_2 = p_{(i-1)\textrm{E}} \lambda + p_{l\textrm{E}} \mu_2 - \left( \sum_{m=l}^{i-1} p_{m\textrm{B+}} \right) (1/\Delta) \nonumber \\
& = p_{(i-1)\textrm{E}} \lambda + p_{l\textrm{E}} \mu_2 - 
p_{h\textrm{B+}}\left( \frac{r_2\left(\frac{1 - r_2^{i-l}}{1 - r_2} \right) - r_1 \left(\frac{1 - r_1^{i-l}}{1 - r_1} \right)}{r_2^{h-l+1} - r_1^{h-l+1}} \right) (1 / \Delta)
\nonumber \\
& = p_{(i-1)\textrm{E}} \lambda + p_{l\textrm{E}} \mu_2 - \nonumber \\
& \; \; \; \; \; \; \; \; \; \; p_{h\textrm{B+}}\left( \frac{(1 - r_1) r_2^{i-l+1} +(r_2-1)  r_1^{i-l+1} - (r_2 - r_1) }{r_2^{h-l+1} - r_1^{h-l+1}} \right) \mu_1,
\end{align}
}
yielding, for $l \leq i \leq h$, 
{\footnotesize\begin{align}\label{eq:piE2}
& p_{i\textrm{E}} = p_{l\textrm{E}} \left( \left(\frac{\lambda}{\mu_2}\right)^{i-l} +  \left( \sum_{m=0}^{i-l-1} \left(\frac{\lambda}{\mu_2}\right)^m \right) \right) - p_{h\textrm{B+}} \left(\frac{\mu_1}{\mu_2}\right) \times \nonumber \\
& \left( \frac{ \sum\limits_{m=l}^{i-1} (\frac{\lambda}{\mu_2})^{i-1-m} \! \!  \left(  (1 - r_1) r_2^{m-l+2} +(r_2-1)  r_1^{m-l+2} - (r_2 - r_1)   \right)}{r_2^{h-l+1} - r_1^{h-l+1}}  \right)  \nonumber \\
& = p_{h\textrm{B+}} \frac{\mu_1/\mu_2}{r_2^{h-l+1} - r_1^{h-l+1}} \times \nonumber \\
& \left(  (r_2 - r_1) \left(r_2^{h-l+1} \! \! \left((\lambda/\mu_2)^{i-l}  +  \textstyle{\sum_{m=0}^{i-l-1} (\lambda/\mu_2)^m} \right) - (\lambda/\mu_2)^{i-l} \right)   \right. - \nonumber \\ 
& \left.  \sum_{m=0}^{i-l-1} (\lambda/\mu_2)^m \left( (1 - r_1) r_2^{i-l-m+1} +(r_2-1)  r_1^{i-l-m+1} \right)  \right).
\end{align}
}

Each state probability $p_{i\textrm{E}}$, $i > h$, satisfies the flow balance equation~(\ref{eq:piEfb}), yielding
{\footnotesize\begin{align}
& p_{i\textrm{E}} = p_{h\textrm{E}} \left( \frac{\lambda}{\mu_2} \right)^{i-h} + p_{h\textrm{B+}} \left( \frac{\lambda - r_1 \mu_1}{\mu_2} \right) \sum_{m=0}^{i-h-1} r_1^m \left( \frac{\lambda}{\mu_2} \right)^{i-h-1-m} \nonumber \\
& = p_{h\textrm{B+}} \left\{ \frac{ (\mu_1/\mu_2) (\lambda/u_2)^{i-h}}{r_2^{h-l+1} - r_1^{h-l+1}} \times \right. \nonumber \\
& \left(  (r_2 - r_1) \! \left(r_2^{h-l+1} \! \! \left((\lambda/\mu_2)^{h-l} \! +  \textstyle{\sum_{m=0}^{h-l-1} (\lambda/\mu_2)^m} \right) \! - (\lambda/\mu_2)^{h-l} \right)   \right. - \nonumber \\ 
& \left.  \sum_{m=0}^{h-l-1} (\lambda/\mu_2)^m \left( (1 - r_1) r_2^{h-l-m+1} +(r_2-1)  r_1^{h-l-m+1} \right)  \right) + \nonumber \\
& \left. \left( \frac{\lambda - r_1 \mu_1}{\mu_2} \right) \sum_{m=0}^{i-h-1} r_1^m \left( \frac{\lambda}{\mu_2} \right)^{i-h-1-m} \right\}
\;\;\; i > h .
\end{align}
}
Similarly as for Equations~(\ref{eq:sumofiE}) and~(\ref{eq:sumofipiE}), we get 
{\footnotesize\begin{align}\label{eq:sumofiE2}
& \sum_{i=h}^{\infty} p_{i\textrm{E}} = \frac{p_{h\textrm{E}}}{1 - \lambda/\mu_2} + \frac{p_{h\textrm{B+}}}{\mu_2 - \lambda} \left( \frac{\lambda - r_1 \mu_1}{1 - r_1}  \right) \nonumber \\
& = p_{h\textrm{B+}} \left\{ \frac{\mu_1/\mu_2}{(r_2^{h-l+1} - r_1^{h-l+1})  (1 -\lambda/\mu_2) } \times \right. \nonumber \\
& \left(  (r_2 - r_1) \! \left(r_2^{h-l+1} \! \! \left((\lambda/\mu_2)^{h-l} \! + \! \! \textstyle{\sum_{m=0}^{h-l-1} (\lambda/\mu_2)^m} \right) \! - (\lambda/\mu_2)^{h-l} \right)   \right. - \nonumber \\ 
& \left.  \sum_{m=0}^{h-l-1} (\lambda/\mu_2)^m \left( (1 - r_1) r_2^{h-l-m+1} +(r_2-1)  r_1^{h-l-m+1} \right)  \right) + \nonumber \\
& \left. \frac{\lambda - r_1 \mu_1}{(\mu_2 - \lambda)( 1 - r_1)} \right\}
\end{align}
}
and
{\footnotesize\begin{align}\label{eq:sumofipiE2}
& \sum_{i=h}^{\infty} i p_{i\textrm{E}} =
p_{h\textrm{E}} \left( \frac{h - (h - 1) \lambda/\mu_2 }{(1 - \lambda/\mu_2 )^2} \right) +  \nonumber \\
& \; \; \; \; \; \; \; \; \; \; \; \; \; \; \; \frac{p_{h\textrm{B+}}}{\mu_2 - \lambda} \left( \frac{\lambda - r_1 \mu_1}{1 - r_1} \right) \left( h + \frac{1}{1 - \lambda/\mu_2} +  \frac{r_1}{1 - r_1} \right) \nonumber \\
& = p_{h\textrm{B+}} \left\{ \frac{ (\mu_1/\mu_2) (h - (h-1)\lambda/u_2)}{(r_2^{h-l+1} - r_1^{h-l+1}) (1 -\lambda/u_2)^2 } \times \right. \nonumber \\
& \left(  (r_2 - r_1) \! \left(r_2^{h-l+1} \! \!  \left((\lambda/\mu_2)^{h-l} \! +  \textstyle{\sum_{m=0}^{h-l-1} (\lambda/\mu_2)^m} \right) - (\lambda/\mu_2)^{h-l} \right)   \right. - \nonumber \\ 
& \left.  \sum_{m=0}^{h-l-1} (\lambda/\mu_2)^m \left( (1 - r_1) r_2^{h-l-m+1} +(r_2-1)  r_1^{h-l-m+1} \right)  \right) + \nonumber \\
& \left. \frac{\lambda - r_1 \mu_1}{(\mu_2 - \lambda)( 1 - r_1)} \left( h + \frac{1}{1 - \lambda/\mu_2} +  \frac{r_1}{1 - r_1} \right) \right\}.
\end{align}
}

Now, consider the state probabilities $p_{i\textrm{B}}$ for $l-1 \leq i \leq h-1$.
The state probability $p_{(h-1)\textrm{B}}$ satisfies the flow balance equation
\begin{align}
p_{(h-1)\textrm{B}} \lambda = p_{l\textrm{B+}} \mu_1 + p_{l\textrm{E}} \mu_2.
\end{align}
Applying Equations~(\ref{eq:plD}) and~(\ref{eq:plE}) yields
{\footnotesize\begin{align}\label{eq:p(h-1)B}
p_{(h-1)\textrm{B}} & = p_{h\textrm{B+}} \left( \frac{ r_2 - r_1}{r_2^{h-l+1} - r_1^{h-l+1}  }  +  \frac{(r_2 - r_1) (r_2^{h-l+1} - 1)}{r_2^{h-l+1} - r_1^{h-l+1}}
 \right)\left( \frac{\mu_1}{\lambda} \right) \nonumber \\
 & = p_{h\textrm{B+}} \left(\frac{(r_2 - r_1)r_2^{h-l+1}}{r_2^{h-l+1} - r_1^{h-l+1}}
 \right)\left( \frac{\mu_1}{\lambda} \right).
\end{align}
}
In the case of $l=h$, using  $r_1r_2 = \lambda/\mu_1$ Equation~(\ref{eq:p(h-1)B}) can be seen to match Equation~(\ref{eq:l=hp(h-1)B}).

Each state probability $p_{i\textrm{B}}$, $l-1 \leq i < h-1$, satisfies the flow balance equation
\begin{align}
p_{i\textrm{B}} \lambda = p_{(i+1)\textrm{B}}\mu_1 + p_{(h-1)\textrm{B}}\lambda,
\end{align}
yielding
{\footnotesize\begin{align}\label{eq:piB2}
& p_{i\textrm{B}} = p_{(h-1)\textrm{B}}\sum_{m=0}^{h-1-i} \left( \frac{\mu_1}{\lambda} \right)^m \nonumber \\
& = p_{h\textrm{B+}} \left(\frac{(r_2 - r_1)r_2^{h-l+1}}{r_2^{h-l+1} - r_1^{h-l+1}}
 \right) \sum_{m=1}^{h-i} \left( \frac{\mu_1}{\lambda} \right)^m \;\;\;\;\;\; l-1 \leq i \leq h - 1.
\end{align}
}

Finally, each state probability $p_{i\textrm{B}}$, $0 \leq i < l-1$, satisfies the flow balance equation
\begin{align}
p_{i\textrm{B}} \lambda = p_{(i+1)\textrm{B}}\mu_1,
\end{align}
yielding
{\footnotesize\begin{align}\label{eq:piB3}
& p_{i\textrm{B}} = p_{(l-1)\textrm{B}}\left( \frac{\mu_1}{\lambda} \right)^{l-1-i} \nonumber \\
& = p_{h\textrm{B+}} \left(\frac{(r_2 - r_1)r_2^{h-l+1}}{r_2^{h-l+1} - r_1^{h-l+1}}
 \right) \sum_{m=l-i}^{h-i} \left( \frac{\mu_1}{\lambda} \right)^m \;\;\;\;\;\; 0 \leq i \leq l - 1.
\end{align}
}

Applying the constraint that
\begin{align}
\sum_{i=l}^{h-1} p_{i\textrm{B+}} + \sum_{i=h}^{\infty} p_{i\textrm{B+}} + \sum_{i=l}^{h-1} p_{i\textrm{E}} + \sum_{i=h}^{\infty} p_{i\textrm{E}} + \sum_{i=0}^{h-1} p_{i\textrm{B}} = 1,
\end{align}
together with Equations~(\ref{eq:sumofiD}), (\ref{eq:sumofiD2}), (\ref{eq:piE2}),   (\ref{eq:sumofiE2}), (\ref{eq:piB2}), and~(\ref{eq:piB3}), the solution for $p_{h\textrm{B+}}$ is obtained, yielding the solution for all of the state probabilities.
The mean number of requests in the system is given by
\begin{align}
\sum_{i=l}^{h-1} i p_{i\textrm{B+}} + \sum_{i=h}^{\infty} i p_{i\textrm{B+}} + \sum_{i=l}^{h-1} i p_{i\textrm{E}} + \sum_{i=h}^{\infty} i p_{i\textrm{E}} + \sum_{i=0}^{h-1} i p_{i\textrm{B}},
\end{align}
which can evaluated using the solution for $p_{h\textrm{B+}}$ and Equations~(\ref{eq:sumofipiD}), (\ref{eq:piD2}), (\ref{eq:plE}), (\ref{eq:piE2}),   (\ref{eq:sumofipiE2}), (\ref{eq:piB2}), and~(\ref{eq:piB3}).  From Little’s Law, division by $\lambda$ yields the mean request response time.  The cost $C$ is given by $\mu_1$ times $\sum_{i=0}^{h-1} p_{i\textrm{B}}$ (evaluated using the solution for $p_{h\textrm{B+}}$ and Equations~(\ref{eq:piB2}) and~(\ref{eq:piB3})) plus $\mu_2$ times $\sum_{i=l}^\infty ( p_{i\textrm{B+}} + p_{i\textrm{E}} )$ (evaluated using the solution for $p_{h\textrm{B+}}$ and Equations~(\ref{eq:sumofiD}), ~(\ref{eq:sumofiD2}), ~(\ref{eq:piE2}) and~(\ref{eq:sumofiE2})).

\subsection{Dual Servers, both Dynamically Allocated/Deallocated}\label{sec:dualoption3}
 
The analysis of this system (refer to Figure~\ref{fig:statetransitiondiagramforDualServer}) is carried out by expressing all state probabilities in terms of $p_{0\textrm{I}}$, and then solving for $p_{0\textrm{I}}$ using the constraint that the state probabilities must sum to one.  In the following we assume $2 \mu > \lambda$.

Considering first the state probabilities $p_{i\textrm{D}}$, from flow balance we have
\begin{align}\label{eq:Dualp1D}
p_{1\textrm{D}} ( \lambda + 1/\Delta ) = p_{0\textrm{I}} \lambda
\end{align}
and
\begin{align}
p_{(i+1)\textrm{D}} ( \lambda + 2/\Delta ) = p_{i\textrm{D}} \lambda  \;\;\;\;\;\; i \geq 1,
\end{align}
yielding
\begin{align}\label{eq:DualpiDoffwhenidle}
p_{i\textrm{D}} = p_{0\textrm{I}} \left( \frac{\lambda}{\lambda + 1/\Delta} \right) \left( \frac{\lambda}{\lambda + 2/\Delta} \right)^{i-1} \;\;\;\;\;\; i \geq 1.
\end{align}

Consider now the state probabilities $p_{i\textrm{B}}$.  We seek expressions for these probabilities, in terms of $p_{0\textrm{I}}$, such that the flow balance equation
{\footnotesize\begin{align}\label{eq:Dualflowbalance}
& p_{i\textrm{B}} \mu = p_{(i-1)\textrm{B}} (\lambda + \mu + 1/\Delta) - p_{(i-2)\textrm{B}} \lambda - 2 p_{(i-1)\textrm{D}} /\Delta \nonumber \\
& = p_{(i-1)\textrm{B}} (\lambda + \mu + 1/\Delta) - p_{(i-2)\textrm{B}} \lambda - 2 \left(\frac{p_{0\textrm{I}} \lambda}{\lambda \Delta + 1} \right) \left( \frac{\lambda}{\lambda + 2/\Delta} \right)^{i-2} 
\end{align}
}
is satisfied for all $i \geq 3$.  In the case that $\lambda/(\lambda + 2/\Delta)$ does not equal $r_1$ as defined below, the general form of solution of this recurrence relation is given by
\begin{align}\label{eq:Dualsolutionform}
p_{i\textrm{B}} = \alpha r_1^i + \beta r_2^i + \gamma \left( \frac{\lambda}{\lambda + 2/\Delta} \right)^i \;\;\;\;\;\; i \geq 1,
\end{align}
where $\alpha$ and $\beta$ are independent of $i$, and similarly as in Equation ~(\ref{eq:r1andr2}), $r_1$ and $r_2$ are given by 
\begin{align}\label{eq:Dualr1andr2}
r_1 = \frac{ (\lambda + \mu + 1/\Delta)/\mu - \sqrt{( (\lambda + \mu + 1/\Delta)/\mu)^2  - 4 \lambda/\mu}}{2},
\nonumber \\
r_2 = \frac{ (\lambda + \mu + 1/\Delta)/\mu + \sqrt{( (\lambda + \mu + 1/\Delta)/\mu)^2  - 4 \lambda/\mu}}{2}.
\end{align}The last term in Equation~(\ref{eq:Dualsolutionform}) is a particular solution of this recurrence relation, implying that
{\footnotesize\begin{align}
\gamma \left( \frac{\lambda}{\lambda + 2/\Delta} \right)^i \mu  & = \gamma \left( \frac{\lambda}{\lambda + 2/\Delta} \right)^{i-1} (\lambda + \mu + 1/\Delta) - \nonumber \\
& \gamma \left( \frac{\lambda}{\lambda + 2/\Delta} \right)^{i-2} \lambda - 2 \left(\frac{p_{0\textrm{I}} \lambda}{\lambda \Delta + 1} \right) \left( \frac{\lambda}{\lambda + 2/\Delta} \right)^{i-2} 
\end{align}
}
which gives
\begin{align}
\gamma =  \frac{2 p_{0\textrm{I}} \left( \lambda + 2/\Delta \right)^2}{(\lambda + 1/\Delta )(2\mu - \lambda - 2/\Delta)}  .
\end{align}
Note that $\lambda/(\lambda + 2/\Delta) = r_1$ when $\mu = \lambda/2 + 1/\Delta$, and so for the case under consideration when $\lambda/(\lambda + 2/\Delta) \neq r_1$,  $2\mu - \lambda - 2/\Delta \neq 0$.

Since $r_2 > 1$ and $0 < r_1 < \textrm{min}[\lambda/\mu, 1]$ (assuming $1/\Delta > 0$), $\beta$ must be zero if valid state probabilities are to be obtained. Thus, we get
\begin{align}\label{eq:DualpiBinit}
p_{i\textrm{B}} = \alpha r_1^i + \frac{2 p_{0\textrm{I}} \left( \lambda + 2/\Delta \right)^2}{(\lambda + 1/\Delta )(2\mu - \lambda - 2/\Delta)} \! \left( \frac{\lambda}{\lambda + 2/\Delta} \right)^i\; i \geq 1.
\end{align}
From flow balance, $p_{1\textrm{B}} \mu = p_{0\textrm{I}} \lambda$, and so we have
\begin{align}
p_{0\textrm{I}} \lambda / \mu = \alpha r_1 + \frac{2 p_{0\textrm{I}} \left( \lambda + 2/\Delta \right) \lambda }{(\lambda + 1/\Delta)(2\mu - \lambda - 2/\Delta)}.
\end{align}
Solving for $\alpha$, substituting into 
Equation~(\ref{eq:DualpiBinit}), and rearranging terms gives
{\footnotesize\begin{align}\label{eq:DualpiB}
p_{i\textrm{B}} = p_{0\textrm{I}} \! \left( \! \frac{ \lambda r_1^{i-1}}{\mu} + \frac{2 \left( \lambda + 2/\Delta \right) \lambda }{\lambda + 1/\Delta} \! \left( \! \frac{ \left( \frac{\lambda}{\lambda + 2/\Delta} \right)^{i-1} - r_1^{i-1}}{2\mu - \lambda - 2/\Delta} \right) \right) \;\; i \geq 1.
\end{align}
}
For the case of $\lambda/(\lambda + 2/\Delta) = r_1$ (and therefore $2\mu - \lambda - 2/\Delta = 0$), we get
{\footnotesize\begin{align}
p_{i\textrm{B}} = p_{0\textrm{I}} \left( \frac{ 2 \lambda}{\lambda + 2/\Delta} + \frac{4 (i-1) \lambda }{(\lambda + 4/\Delta) (\lambda \Delta + 1)} \right) \left( \frac{\lambda}{\lambda + 2/\Delta} \right)^{i-1} \;\; i \geq 1.
\end{align}
}

Consider now the state probabilities $p_{i\textrm{E}}$ for $i \geq 2$.  The state probability $p_{2\textrm{E}}$ satisfies the flow balance equation
\begin{align}
p_{2\textrm{E}} (2 \mu)  = p_{1\textrm{B}} (\lambda + \mu) - p_{2\textrm{B}} \mu - p_{1\textrm{D}} /\Delta
\end{align}
yielding
{\footnotesize\begin{align}\label{eq:Dualp2E}
p_{2\textrm{E}} & = p_{0\textrm{I}} \! \! \left( \! \! \frac{\lambda(\lambda + \mu)}{2 \mu^2} - \! \! \left( \! \! \frac{\lambda r_1}{2 \mu} + \frac{\left( \lambda + \frac{2}{\Delta} \right) \lambda }{\lambda + \frac{1}{\Delta}} \! \! \left( \! \! \frac{ \frac{\lambda}{\lambda + \frac{2}{\Delta}} - r_1}{2\mu - \lambda - \frac{2}{\Delta}} \right) \! \right) \! -  \frac{\lambda / (2 \mu)}{\lambda \Delta + 1} \! \right) \nonumber \\
& = p_{0\textrm{I}} \frac{\lambda}{2\mu}\left(\frac{\lambda}{\mu} + \frac{r_1\left((\lambda+1/\Delta)(\lambda+2/\Delta)+2\mu/\Delta \right) - \lambda(\lambda+2/\Delta)}{(\lambda+1/\Delta)(2\mu - \lambda-2/\Delta)} \right) \nonumber \\
& = p_{0\textrm{I}} \lambda \left(\frac{\mu \lambda^2 + (\mu r_1 - \lambda) (\lambda + 1/\Delta)(\lambda+2/\Delta) + 2\mu^2r_1/\Delta }{2\mu^2(\lambda+1/\Delta)(2\mu - \lambda-2/\Delta)} \right)
\end{align}
}
where we have assumed the case of $\lambda/(\lambda + 2/\Delta) \neq r_1$.  For the case of $\lambda/(\lambda + 2/\Delta) = r_1$ (for which $\mu = \lambda/2 + 1/\Delta$), we get instead
{\footnotesize\begin{align}
& p_{2\textrm{E}} = p_{0\textrm{I}} \left( \lambda \frac{3 \lambda + 2/\Delta}{(\lambda + 2/\Delta)^2} - \right. \nonumber \\
& \left. \left( \frac{ \lambda}{\lambda + 2/\Delta} + \frac{2 \lambda }{(\lambda + 4/\Delta) (\lambda \Delta + 1)} \right) \left( \frac{\lambda}{\lambda + 2/\Delta} \right) - \frac{\lambda / (\lambda + 2/\Delta)}{\lambda \Delta + 1} \right) \nonumber \\
& = p_{0\textrm{I}} \lambda^2 \left( \frac{2 \lambda^2 \Delta + 9 \lambda + 8/\Delta }{(\lambda + 2/\Delta)^2 (\lambda + 4/\Delta)(\lambda\Delta+1)}\right).
\end{align}
}

In the following, we assume the case of $\lambda/(\lambda + 2/\Delta) \neq r_1$; the analysis for $\lambda/(\lambda + 2/\Delta) = r_1$ is carried out analogously. The state probability $p_{i\textrm{E}}$ for $i \geq 4$ satisfies the flow balance equation
{\footnotesize\begin{align}
p_{i\textrm{E}} (2 \mu) & = p_{(i-1)\textrm{E}} (\lambda + 2\mu) - p_{(i-2)\textrm{E}} \lambda - p_{(i-1)\textrm{B}}/\Delta \nonumber \\
& = p_{(i-1)\textrm{E}} (\lambda + 2\mu) - p_{(i-2)\textrm{E}} \lambda - \nonumber \\
& \frac{p_{0\textrm{I}}}{\Delta} \left( \frac{ \lambda r_1^{i-2}}{\mu} + \frac{2 \left( \lambda + 2/\Delta \right) \lambda }{\lambda + 1/\Delta} \left(\frac{ \left( \frac{\lambda}{\lambda + 2/\Delta} \right)^{i-2} - r_1^{i-2}}{2\mu - \lambda - 2/\Delta} \right) \right),
\end{align}
}
The general form of solution of this recurrence relation is given by
\begin{align}
p_{i\textrm{E}} = \alpha \left(\frac{\lambda}{2\mu} \right)^i  + \phi \left(\frac{\lambda}{\lambda+2/\Delta}\right)^i+ \gamma r_1^i
\end{align}
where $\gamma$ and $\phi$ are such that
\begin{align}
\gamma r_1^i  & = \gamma r_1^{i-1} \left( \frac{\lambda}{2\mu} + 1 \right) - \gamma r_1^{i-2} \left( \frac{\lambda}{2\mu} \right) - \frac{p_{0\textrm{I}}}{\Delta \mu} \left( \frac{ \lambda r_1^{i-2}}{2 \mu} \right) + \nonumber \\
& \frac{p_{0\textrm{I}}}{\Delta \mu} \left( \frac{\left( \lambda + 2/\Delta \right) \lambda }{(\lambda + 1/\Delta)(2\mu - \lambda - 2/\Delta)}  \right)r_1^{i-2}
\end{align}
and
{\footnotesize\begin{align}
\phi \! \left(\frac{\lambda}{\lambda+2/\Delta}\right)^i \! \! & = \phi \! \left(\frac{\lambda}{\lambda+2/\Delta}\right)^{i-1} \! \! \left( \frac{\lambda}{2\mu} + 1 \right) \! \! - \phi \! \left(\frac{\lambda}{\lambda+2/\Delta}\right)^{i-2} \! \! \frac{\lambda}{2\mu} - \nonumber \\
& \frac{p_{0\textrm{I}}}{\Delta \mu} \left( \frac{\left( \lambda + 2/\Delta \right) \lambda }{(\lambda + 1/\Delta)(2\mu - \lambda - 2/\Delta)}  \right)\left( \frac{\lambda}{\lambda + 2/\Delta} \right)^{i-2},
\end{align}
}
implying that
{\footnotesize\begin{align}
\gamma = p_{0\textrm{I}} \lambda \left( \frac{2\mu/\Delta + (\lambda + 1/\Delta)(\lambda+2/\Delta)}{2 \mu (\lambda + 1/\Delta)(2\mu-\lambda-2/\Delta)\left( r_1 - \frac{\lambda\Delta}{2} (1 - r_1) \right)}  \right)
\end{align}
}
(note that $\lambda/(\lambda + 2/\Delta) \neq r_1$ implies that $r_1 - \frac{\lambda\Delta}{2} (1 - r_1) \neq 0$) and
\begin{align}
\phi = p_{0\textrm{I}} \left( \frac{\left( \lambda + 2/\Delta \right)^3 }{(\lambda + 1/\Delta)(2\mu - \lambda - 2/\Delta)^2}  \right).
\end{align}
Thus we get, for $i \geq 2$,
{\footnotesize\begin{align}\label{eq:DualinitialpiE}
p_{i\textrm{E}} & = \alpha \! \left(\frac{\lambda}{2\mu} \right)^i  \! \! + p_{0\textrm{I}} \! \left( \frac{\left( \lambda + 2/\Delta \right)^3 }{(\lambda + 1/\Delta)(2\mu - \lambda - 2/\Delta)^2}  \right) \! \left(\frac{\lambda}{\lambda+2/\Delta}\right)^i  + \nonumber \\
& p_{0\textrm{I}} \lambda \left( \frac{2\mu/\Delta + (\lambda + 1/\Delta)(\lambda+2/\Delta)}{2 \mu (\lambda + 1/\Delta)(2\mu-\lambda-2/\Delta)\left( r_1 - \frac{\lambda\Delta}{2} (1 - r_1) \right)}  \right) r_1^i \nonumber \\
& = \alpha \! \left(\frac{\lambda}{2\mu} \right)^i \! \!  + p_{0\textrm{I}} \! \left( \frac{\left( \lambda + 2/\Delta \right) \lambda^2 }{(\lambda + 1/\Delta)(2\mu - \lambda - 2/\Delta)^2}  \right) \! \left(\frac{\lambda}{\lambda+2/\Delta}\right)^{i-2} \! \! \! + \nonumber \\
& p_{0\textrm{I}} \lambda \left( \frac{\left(2\mu/\Delta + (\lambda + 1/\Delta)(\lambda+2/\Delta) \right) r_1^2 }{2 \mu (\lambda + 1/\Delta)(2\mu-\lambda-2/\Delta)\left( r_1 - \frac{\lambda\Delta}{2} (1 - r_1) \right)}  \right) r_1^{i-2}.
\end{align}
}
Applying this equation for $i=2$ with Equation~(\ref{eq:Dualp2E}) gives
{\footnotesize\begin{align}
& \alpha \left(\frac{\lambda}{2\mu} \right)^2  + p_{0\textrm{I}} \left( \frac{\left( \lambda + 2/\Delta \right) \lambda^2 }{(\lambda + 1/\Delta)(2\mu - \lambda - 2/\Delta)^2}  \right) + \nonumber \\
& p_{0\textrm{I}} \lambda \left( \frac{\left(2\mu/\Delta + (\lambda + 1/\Delta)(\lambda+2/\Delta) \right) r_1^2}{2 \mu (\lambda + 1/\Delta)(2\mu-\lambda-2/\Delta)\left( r_1 - \frac{\lambda\Delta}{2} (1 - r_1) \right)}  \right) = \nonumber \\
& p_{0\textrm{I}} \lambda \left(\frac{\mu \lambda^2 + (\mu r_1 - \lambda) (\lambda + 1/\Delta)(\lambda+2/\Delta) + 2\mu^2r_1/\Delta }{2\mu^2(\lambda+1/\Delta)(2\mu - \lambda-2/\Delta)} \right).
\end{align}
}
Solving for $\alpha$ yields
{\footnotesize\begin{align}
\alpha = p_{0\textrm{I}} \left( \frac{ \splitfrac{ \lambda^3 (r_1 + \mu\Delta (1 - r_1)) + \lambda^2 (4 \mu +  5 r_1 / \Delta - 8 r_1 \mu)}{ \splitfrac{ +
(4 \lambda/\Delta) (\mu + 2 r_1 / \Delta - 4 r_1 \mu)}{ + (4 r_1/\Delta) ( 1/\Delta^2 - 2\mu/\Delta - \mu^2)}}}{(\lambda+1/\Delta)(2\mu-\lambda-2/\Delta)^2(r_1-\frac{\lambda\Delta}{2}(1-r_1))} \right) . 
\end{align}
}
Substitution into Equation~(\ref{eq:DualinitialpiE}) then yields
{\footnotesize\begin{align}\label{eq:DualpiE}
& p_{i\textrm{E}} = p_{0\textrm{I}} \left( \left( \frac{ \splitfrac{ \lambda^3 (r_1 + \mu\Delta (1 - r_1)) + \lambda^2 (4 \mu +  5 r_1 / \Delta - 8 r_1 \mu)}{ \splitfrac{ +
(4 \lambda/\Delta) (\mu + 2 r_1 / \Delta - 4 r_1 \mu)}{ + (4 r_1/\Delta) ( 1/\Delta^2 - 2\mu/\Delta - \mu^2)}}}{(\lambda+1/\Delta)(2\mu-\lambda-2/\Delta)^2(r_1-\frac{\lambda\Delta}{2}(1-r_1))} \right) \times \right. \nonumber \\
& \left. \left(\frac{\lambda}{2\mu} \right)^i  + \left( \frac{\left( \lambda + 2/\Delta \right) \lambda^2 }{(\lambda + 1/\Delta)(2\mu - \lambda - 2/\Delta)^2}  \right) \left(\frac{\lambda}{\lambda+2/\Delta}\right)^{i-2}  + \right. \nonumber \\
&  \left.\left( \frac{\lambda \left(2\mu/\Delta + (\lambda + 1/\Delta)(\lambda+2/\Delta) \right) r_1^2 }{2 \mu (\lambda + 1/\Delta)(2\mu-\lambda-2/\Delta)\left( r_1 - \frac{\lambda\Delta}{2} (1 - r_1) \right)}  \right) r_1^{i-2} \right) 
\;\;\ i \geq 2.
\end{align}
}

From Equations~(\ref{eq:DualpiDoffwhenidle}), (\ref{eq:DualpiB}), and (\ref{eq:DualpiE}), we have
{\footnotesize\begin{align}\label{eq:DualsumpiDoffwhenidle}
\sum_{i=1}^\infty p_{i\textrm{D}} & = p_{0\textrm{I}} \sum_{i=1}^\infty \left( \frac{\lambda}{\lambda + 1/\Delta} \right) \left( \frac{\lambda}{\lambda + 2/\Delta} \right)^{i-1} \nonumber \\
& = p_{0\textrm{I}} \left( \frac{\lambda}{\lambda + 1/\Delta} \right) \left( 1 + \frac{\lambda\Delta}{2} \right),
\end{align}
\begin{align}\label{eq:DualsumpiB}
\sum_{i=1}^\infty p_{i\textrm{B}} & = p_{0\textrm{I}} \sum_{i=1}^\infty \left( \! \frac{ \lambda r_1^{i-1}}{\mu} + \frac{2 \left( \lambda + 2/\Delta \right) \lambda }{\lambda + 1/\Delta} \left(\frac{ \left( \frac{\lambda}{\lambda + 2/\Delta} \right)^{i-1} - r_1^{i-1}}{2\mu - \lambda - 2/\Delta} \right) \! \right) \nonumber \\
& = p_{0\textrm{I}} \left( \frac{\lambda}{\mu(1-r_1)} + \frac{\left( \lambda + 2/\Delta \right) \lambda }{\lambda + 1/\Delta} \left(\frac{ \lambda \Delta - 2 r_1/(1-r_1)}{2\mu - \lambda - 2/\Delta} \right) \right),
\end{align}
}
and
{\footnotesize\begin{align}\label{eq:DualsumpiE}
& \sum_{i=2}^\infty p_{i\textrm{E}} = \nonumber \\
& p_{0\textrm{I}} \! \! \sum_{i=2}^\infty \! \left( \! \left( \! \frac{ \splitfrac{ \lambda^3 (r_1 + \mu\Delta (1 - r_1)) + \lambda^2 (4 \mu +  5 \frac{r_1}{\Delta} - 8 r_1 \mu)}{ \splitfrac{ +
\frac{4 \lambda}{\Delta} (\mu + 2 \frac{r_1 }{\Delta} - 4 r_1 \mu)}{ + \frac{4 r_1}{\Delta} ( \frac{1}{\Delta^2} - 2\frac{\mu}{\Delta} - \mu^2)}}}{(\lambda+\frac{1}{\Delta})(2\mu-\lambda-\frac{2}{\Delta})^2(r_1-\frac{\lambda\Delta}{2}(1-r_1))} \! \! \right) \! \! \times \right. \nonumber \\
& \left. \left(\frac{\lambda}{2\mu} \right)^i  + \left( \frac{\left( \lambda + 2/\Delta \right) \lambda^2 }{(\lambda + 1/\Delta)(2\mu - \lambda - 2/\Delta)^2}  \right) \left(\frac{\lambda}{\lambda+2/\Delta}\right)^{i-2}  + \right. \nonumber \\
&  \left.\left( \frac{\lambda \left(2\mu/\Delta + (\lambda + 1/\Delta)(\lambda+2/\Delta) \right) r_1^2 }{2 \mu (\lambda + 1/\Delta)(2\mu-\lambda-2/\Delta)\left( r_1 - \frac{\lambda\Delta}{2} (1 - r_1) \right)}  \right) r_1^{i-2} \right) \nonumber \\
& = p_{0\textrm{I}} \! \left( \! \left( \frac{ \splitfrac{ \lambda^3 (r_1 + \mu\Delta (1 - r_1)) + \lambda^2 (4 \mu +  5 r_1 / \Delta - 8 r_1 \mu)}{ \splitfrac{ +
(4 \lambda/\Delta) (\mu + 2 r_1 / \Delta - 4 r_1 \mu)}{ + (4 r_1/\Delta) ( 1/\Delta^2 - 2\mu/\Delta - \mu^2)}}}{(\lambda+1/\Delta)(2\mu-\lambda-2/\Delta)^2(r_1-\frac{\lambda\Delta}{2}(1-r_1))} \right) \times \right. \nonumber \\
& \left.\left(\frac{\lambda^2}{2\mu (2\mu-\lambda)} \right)  + \left( \frac{\left( \lambda + 2/\Delta \right) \lambda^2 }{(\lambda + 1/\Delta)(2\mu - \lambda - 2/\Delta)^2}  \right) \left( 1 + \frac{\lambda\Delta}{2} \right)  + \right. \nonumber \\
&  \left. \frac{\lambda \left(2\mu/\Delta + (\lambda + 1/\Delta)(\lambda+2/\Delta) \right) r_1^2 }{2 \mu (\lambda + 1/\Delta)(2\mu-\lambda-2/\Delta)\left( r_1 - \frac{\lambda\Delta}{2} (1 - r_1) \right)(1 - r_1)}  \right).
\end{align}
}
Applying the constraint that
\begin{align}
p_{0\textrm{I}} + \sum_{i=1}^{\infty} p_{i\textrm{D}} + \sum_{i=1}^{\infty} p_{i\textrm{B}} + \sum_{i=2}^{\infty} p_{i\textrm{E}} = 1,
\end{align}
together with Equations~(\ref{eq:DualsumpiDoffwhenidle}), (\ref{eq:DualsumpiB}), and~(\ref{eq:DualsumpiE}), the solution for $p_{0\textrm{I}}$ is obtained, yielding the solution for all of the state probabilities.

The mean number of requests in the system is given by
\begin{align}
\sum_{i=1}^{\infty} i p_{i\textrm{D}} + \sum_{i=l}^{\infty} i p_{i\textrm{B}} + \sum_{i=2}^{\infty} i p_{i\textrm{E}}.
\end{align}
From Equations~(\ref{eq:DualpiDoffwhenidle}), (\ref{eq:DualpiB}), and (\ref{eq:DualpiE}), we have
{\footnotesize{\begin{align}\label{eq:DualsumipiDoffwhenidle}
\sum_{i=1}^\infty i p_{i\textrm{D}} & = p_{0\textrm{I}} \sum_{i=1}^\infty i \left( \frac{\lambda}{\lambda + 1/\Delta} \right) \left( \frac{\lambda}{\lambda + 2/\Delta} \right)^{i-1} \nonumber \\
& = p_{0\textrm{I}} \left( \frac{\lambda}{\lambda + 1/\Delta} \right) \left( 1 + \frac{\lambda\Delta}{2} \right)^2,
\end{align}
\begin{align}\label{eq:DualsumipiB}
& \sum_{i=1}^\infty i p_{i\textrm{B}} = p_{0\textrm{I}} \! \sum_{i=1}^\infty i \left( \! \! \frac{ \lambda r_1^{i-1}}{\mu} + \frac{2 \left( \lambda + \frac{2}{\Delta} \right) \lambda }{\lambda + \frac{1}{\Delta}} \! \left(\frac{ \left( \frac{\lambda}{\lambda + \frac{2}{\Delta}} \right)^{i-1} - r_1^{i-1}}{2\mu - \lambda - \frac{2}{\Delta}} \right) \! \right) \nonumber \\
& = p_{0\textrm{I}} \left( \frac{\lambda}{\mu(1-r_1)^2} + \frac{2 \left( \lambda + \frac{2}{\Delta} \right) \lambda }{\lambda + \frac{1}{\Delta}} \left(\frac{ (1 + \frac{\lambda \Delta}{2})^2 - 1/(1-r_1)^2 }{2\mu - \lambda - \frac{2}{\Delta}} \right) \right),
\end{align}
}
and
\footnotesize\begin{align}\label{eq:DualsumipiE}
& \sum_{i=2}^\infty i p_{i\textrm{E}} = \nonumber \\
& p_{0\textrm{I}} \! \sum_{i=2}^\infty i \left( \! \left( \frac{ \splitfrac{ \lambda^3 (r_1 + \mu\Delta (1 - r_1)) + \lambda^2 (4 \mu +  5 r_1 / \Delta - 8 r_1 \mu)}{ \splitfrac{ +
(4 \lambda/\Delta) (\mu + 2 r_1 / \Delta - 4 r_1 \mu)}{ + (4 r_1/\Delta) ( 1/\Delta^2 - 2\mu/\Delta - \mu^2)}}}{(\lambda+1/\Delta)(2\mu-\lambda-2/\Delta)^2(r_1-\frac{\lambda\Delta}{2}(1-r_1))} \right) \! \! \times \right. \nonumber \\
& \left. \left(\frac{\lambda}{2\mu} \right)^i  + \left( \frac{\left( \lambda + 2/\Delta \right) \lambda^2 }{(\lambda + 1/\Delta)(2\mu - \lambda - 2/\Delta)^2}  \right) \left(\frac{\lambda}{\lambda+2/\Delta}\right)^{i-2}  + \right. \nonumber \\
&  \left.\left( \frac{\lambda \left(2\mu/\Delta + (\lambda + 1/\Delta)(\lambda+2/\Delta) \right) r_1^2 }{2 \mu (\lambda + 1/\Delta)(2\mu-\lambda-2/\Delta)\left( r_1 - \frac{\lambda\Delta}{2} (1 - r_1) \right)}  \right) r_1^{i-2} \right) \nonumber \\
& = p_{0\textrm{I}} \! \left( \! \left( \frac{ \splitfrac{ \lambda^3 (r_1 + \mu\Delta (1 - r_1)) + \lambda^2 (4 \mu +  5 r_1 / \Delta - 8 r_1 \mu)}{ \splitfrac{ +
(4 \lambda/\Delta) (\mu + 2 r_1 / \Delta - 4 r_1 \mu)}{ + (4 r_1/\Delta) ( 1/\Delta^2 - 2\mu/\Delta - \mu^2)}}}{(\lambda+1/\Delta)(2\mu-\lambda-2/\Delta)^2(r_1-\frac{\lambda\Delta}{2}(1-r_1))} \right) \! \times \right. \nonumber \\
& \left. \left(\frac{\lambda^2(2-\frac{\lambda}{2\mu})}{(2\mu-\lambda)^2} \right)  + \left( \! \frac{\left( \lambda + \frac{2}{\Delta} \right) \lambda^2 }{(\lambda + \frac{1}{\Delta})(2\mu - \lambda - \frac{2}{\Delta})^2} \! \right)\!\! \left( \! 1 + \frac{\lambda\Delta}{2} \!  \right) \!\!\left( \! 2 + \frac{\lambda \Delta}{2} \! \right) + \right. \nonumber \\
&  \left. \frac{\lambda \left(2\mu/\Delta + (\lambda + 1/\Delta)(\lambda+2/\Delta) \right) r_1^2 (2 - r_1)}{2 \mu (\lambda + 1/\Delta)(2\mu-\lambda-2/\Delta)\left( r_1 - \frac{\lambda\Delta}{2} (1 - r_1) \right)(1 - r_1)^2}  \right).
\end{align}
}
From Little’s Law, division of the mean number of requests in the system by $\lambda$ yields the mean request response time.  The cost $C$ is given by $\mu( p_{1\textrm{D}}+p_{1\textrm{B}})$ (evaluated using the solution for $p_{0\textrm{I}}$ and Equations~(\ref{eq:Dualp1D}) and~(\ref{eq:DualpiB})) plus $2 \mu \sum_{i=2}^\infty ( p_{i\textrm{D}} + p_{i\textrm{B}} + p_{i\textrm{E}} )$ (evaluated using the solution for $p_{0\textrm{I}}$ and Equations~(\ref{eq:Dualp1D}), ~(\ref{eq:DualsumpiDoffwhenidle}),  ~(\ref{eq:DualpiB}), ~(\ref{eq:DualsumpiB}), and~(\ref{eq:DualsumpiE})).


\end{document}